\documentclass[aps,pra,twocolumn,superscriptaddress]{revtex4-2}
\usepackage[encapsulated]{CJK}
\usepackage{graphicx}
\usepackage{float}
\usepackage{siunitx}
\usepackage{fixmath}
\usepackage{titlesec}
\usepackage{amsmath,amsfonts,amssymb}
\usepackage[colorlinks=true,linkcolor=blue,citecolor=blue]{hyperref}
\usepackage{color,soul}

\usepackage{amsmath}  \usepackage{amssymb}  \usepackage{amsfonts}  \usepackage{bm}  \usepackage{bbm}   \usepackage{bbold}   \usepackage{braket}  \usepackage{color}  \usepackage{comment}  \usepackage{dcolumn}  \usepackage{enumerate}  \usepackage{epsfig}  \usepackage{gensymb}  \usepackage{graphicx}  \usepackage{indentfirst}  \usepackage{lmodern}  \usepackage{mathrsfs}  \usepackage{mathtools}  \usepackage{psfrag}  \usepackage{pst-all}  \usepackage{soul}  \usepackage{units}  \usepackage{xcolor}

\def\ii{{\rm i}}  \def\ee{{\rm e}}
\def\me{m_{\rm e}}  \def\kB{{k_{\rm B}}}
  
                \def\Gb{{\bf G}}          \def\jb{{\bf j}}  \def\Kb{{\bf K}}            \def\Qb{{\bf Q}}  \def\qb{{\bf q}}  \def\Rb{{\bf R}}  \def\rb{{\bf r}}       
\def\xx{\hat{\bf x}}                
\def\kpar{k_\parallel}  \def\kparb{{\bf k}_\parallel} 
        
  \def\qparb{{\bf q}}
\def\EFg{{E_{\rm F}^{\rm g}}}  \def\EFAu{{E_{\rm F}^{\rm Au}}}  \def\DEFg{{\Delta E_{\rm F}^{\rm g}}}
\def\kFg{{k_{\rm F}^{\rm g}}}  \def\kFAu{{k_{\rm F}^{\rm Au}}}  \def\vFg{{v_{\rm F}^{\rm g}}}
\def\VVb{V_{\rm b}}

\begin{document}
\title{Exciton-assisted electron tunneling in van der Waals heterostructures}
\author{Lujun Wang} 
\thanks{These authors contributed equally.}
\affiliation{Photonics Laboratory, ETH Z\"urich, 8093 Z\"urich, Switzerland}

\author{Sotirios Papadopoulos}
\thanks{These authors contributed equally.}
\affiliation{Photonics Laboratory, ETH Z\"urich, 8093 Z\"urich, Switzerland}

\author{Fadil Iyikanat}
\thanks{These authors contributed equally.}
\affiliation{ICFO-Institut de Ciencies Fotoniques, The Barcelona Institute of Science and Technology, 08860 Castelldefels (Barcelona), Spain}

\author{Jian Zhang}
\affiliation{Transport at Nanoscale Interfaces Laboratory, Empa, Swiss Federal Laboratories for Materials Science and Technology, 8600 D\"ubendorf, Switzerland}

\author{Jing Huang}
\affiliation{Photonics Laboratory, ETH Z\"urich, 8093 Z\"urich, Switzerland}

\author{Kenji Watanabe}
\affiliation{National Institute for Material Science, 1-1 Namiki, Tsukuba, 305-0044, Japan}

\author{Takashi Taniguchi}
\affiliation{National Institute for Material Science, 1-1 Namiki, Tsukuba, 305-0044, Japan}

\author{Michel Calame}
\affiliation{Transport at Nanoscale Interfaces Laboratory, Empa, Swiss Federal Laboratories for Materials Science and Technology, 8600 D\"ubendorf, Switzerland}
\affiliation{Department of Physics, University of Basel, 4056 Basel, Switzerland}
\affiliation{Swiss Nanoscience Institute, University of Basel, 4056 Basel, Switzerland}

\author{Mickael L. Perrin}
\affiliation{Transport at Nanoscale Interfaces Laboratory, Empa, Swiss Federal Laboratories for Materials Science and Technology, 8600 D\"ubendorf, Switzerland}
\affiliation{Department of Information Technology and Electrical Engineering, ETH Z\"urich, 8092 Z\"urich, Switzerland}
\affiliation{Quantum Center, ETH Z\"urich, 8093 Z\"urich, Switzerland}

\author{F. Javier Garc\'{\i}a de Abajo}
\email{javier.garciadeabajo@nanophotonics.es}
\affiliation{ICFO-Institut de Ciencies Fotoniques, The Barcelona Institute of Science and Technology, 08860 Castelldefels (Barcelona), Spain}
\affiliation{ICREA-Instituci\'o Catalana de Recerca i Estudis Avan\c{c}ats, Passeig Llu\'{\i}s Companys 23, 08010 Barcelona, Spain}

\author{Lukas Novotny}
\email{lnovotny@ethz.ch}
\affiliation{Photonics Laboratory, ETH Z\"urich, 8093 Z\"urich, Switzerland}

\begin{abstract}
The control of elastic and inelastic electron tunneling relies on materials with well defined interfaces. Van der Waals materials made of two-dimensional constituents form an ideal platform for such studies. Signatures of acoustic phonons and defect states have been observed in current-to-voltage ($I \textendash V$) measurements. These features can be explained by direct electron-phonon or electron-defect interactions. Here, we use a novel tunneling process that involves excitons in transition metal dichalcogenides (TMDs). We study tunnel junctions consisting of graphene and gold electrodes separated by hexagonal boron nitride (hBN) with an adjacent TMD monolayer and observe prominent resonant features in $I \textendash V$ measurements. These resonances appear at bias voltages that correspond to TMD exciton energies. By placing the TMD outside of the tunneling pathway, we demonstrate that this phonon-exciton mediated tunneling process does not require any charge injection into the TMD. This work demonstrates the appearance of optical modes in electrical transport measurements and introduces a new functionality for optoelectronic devices based on van der Waals materials.
\end{abstract}

\maketitle

\section{INTRODUCTION}

The isolation of two-dimensional (2D) crystals combined with advances in fabrication techniques have enabled the realization of a new type of material, known as van der Waals (vdW) heterostructures, in which different atomic layers are assembled together in a desired sequence~\cite{Geim2013}. Tailored heterostructures comprising graphene, hexagonal boron nitride (hBN), transition metal dichalcogenides (TMDs), and other 2D materials are currently designed to display  properties that are absent in the individual constituents, thus providing a  platform for fundamental studies~\cite{Wallbank2016, Ma2016, Yankowitz2012, Wang2019, Cao2018, Rivera2015} and novel device applications~\cite{Koppens2014, Massicotte2016, Withers2015}. In this respect, tunnel junctions with different material combinations form an interesting system for investigating electron tunneling processes. Previous experiments have shown phonon-assisted resonant electron tunneling in  metal-insulator junctions~\cite{Chynoweth1962}, in conventional semiconductor heterostructures~\cite{Eaves1985}, and in graphene-based systems~\cite{Brar2007, Zhang2008, Wehling2008, Vdovin2016}. Similarly, exciton-assisted resonant tunneling has been observed in conventional semiconductor quantum wells~\cite{Cao1995, Cao1997}. Plasmon-assisted resonant tunneling has been investigated in metallic quantum well structures hosting silver nanorods~\cite{Qian2021} and graphene-based structures~\cite{Enaldiev2017}. Furthermore, evidence for defect-assisted resonant tunneling has been observed in hBN-based junctions~\cite{Chandni2015}. 

Here, we demonstrate exciton-assisted resonant electron tunneling in vdW tunnel junctions. Our electron transport measurements reveal distinct resonant peaks that coincide in energy with TMD excitons. We investigate the $I \textendash V$ characteristics of TMD/graphene/hBN/Au tunnel junctions and compare them with TMD-free reference structures. Resonances observed in differential conductance ($\mathrm{d}I/\mathrm{d}V$) measurements agree with TMD exciton energies despite the fact that the TMD is placed outside the electron tunneling pathway. These resonances can be the explained by a one-step process involving indirect excitons and a two-step process involving both phonons and direct excitons. Both of these processes conserve energy and in-plane momentum. Owing to the large exciton binding energies of TMD monolayers~\cite{Wang2018, Goryca2019}, such as WS$_\mathrm{2}$, MoS$_\mathrm{2}$, WSe$_\mathrm{2}$, and MoSe$_\mathrm{2}$, the resonant features can be observed at room temperature. While exciton-phonon interactions in TMDs have been investigated by optical methods \cite{Barati2022,Funk2021,Meneghini2022,Du2019}, in our study we observe  this interaction directly in electronic transport measurements, shedding light on the ways excitons are involved in the conservation of momentum during tunneling. Beyond its fundamental interest, our work establishes a new platform for the investigation of the physical processes involved in the electrical generation of excitons in TMDs.

\begin{figure*}
\begin{centering} \includegraphics[width=0.95\textwidth]{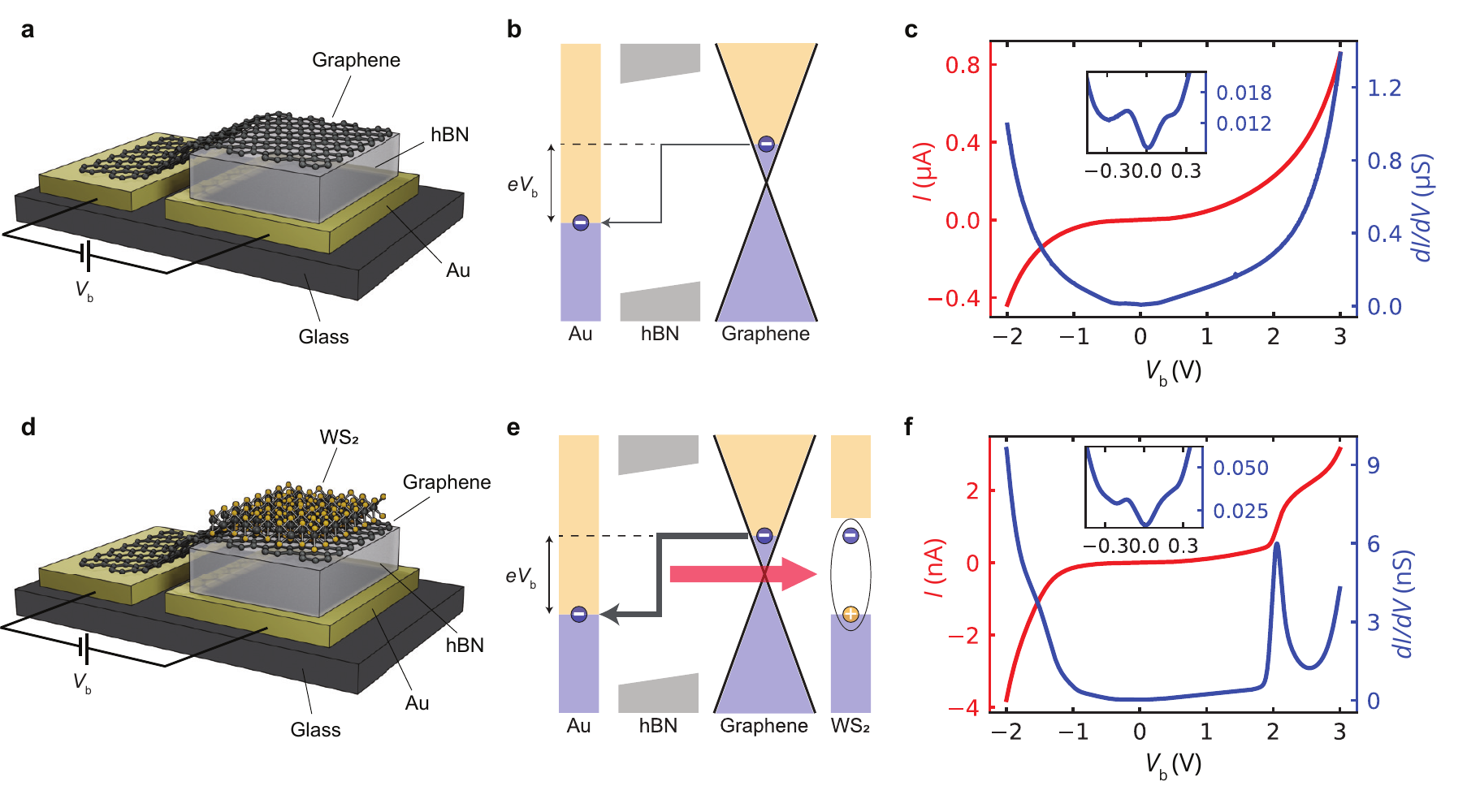} \par\end{centering}
	\caption{{\bf Device schematics, band diagrams, and $I \textendash V$ characteristics.} \textbf{a,} Illustration of a graphene/hBN/Au tunneling device. The device is protected by a top hBN layer (not shown for better visibility). A bias voltage $V_\mathrm{b}$ applied between the graphene and gold electrodes gives rise to a tunneling current through the hBN spacer. \textbf{b,} Band diagram of the device for positive $V_\mathrm{b}$. Electrons tunnel from graphene to Au both elastically (not shown) and inelastically (kinked arrow). \textbf{c,} Recorded $I \textendash V$ (red) and $\mathrm{d}I/\mathrm{d}V$ (blue) curves at room temperature from the device in \textbf{a}. No distinctive features are observed besides an overall asymmetry due to the electronic DOS in Au.
Inset: Zoom into the low-bias region displaying phonon-assisted resonances. \textbf{d,} Illustration of a WS$_\mathrm{2}$/graphene/hBN/Au tunnel device, including a protecting top hBN layer (not shown). \textbf{e,} Band diagram of the device at a positive bias. Electron tunneling can be mediated by the creation of excitons (encircled electron-hole pair) in WS$_\mathrm{2}$, as indicated by the red arrow. \textbf{f,} Recorded $I \textendash V$ (red) and $\mathrm{d}I/\mathrm{d}V$ (blue) curves at room temperature from the device in \textbf{d}. A new feature appears near $V_\mathrm{b}=\SI{2.05}{\volt}$. Inset: Zoom into the low-bias region displaying phonon-assisted resonances.}
	\label{fig:fig1}
\end{figure*}

\section{RESULTS AND DISCUSSION}

Our reference device is illustrated in Fig.~\ref{fig:fig1}a, where the graphene and Au electrodes are separated by a 3-4~nm layer of insulating hBN. Applying a bias voltage between the two electrodes generates a tunnel current through the hBN barrier. The band diagram is depicted in Fig.~\ref{fig:fig1}b for a positive bias voltage $V_\mathrm{b}$. The inelastic electron tunneling process, indicated by the kinked arrow, can be mediated by different modes of the structure, including phonons, defects, photons, and surface plasmons~\cite{Lambe1976, Parzefall2015, Parzefall2019, Kuzmina2021}. The measured $I \textendash V$ curve of such a device is plotted in red in Fig.~\ref{fig:fig1}c, which features a nearly exponential dependence on $V_\mathrm{b}$ for both polarities, in agreement with previous reports~\cite{Lee2011, Britnell2012}. To gain further insight, we evaluate the differential conductance $\mathrm{d}I/\mathrm{d}V$, shown in blue in Fig.~\ref{fig:fig1}c. This plot reveals an asymmetry in bias voltage (i.e., the differential conductance increases more rapidly for negative $V_\mathrm{b}$). This can be understood by the abrupt increase of the electronic density of states (DOS) in Au for negative bias voltages (see Supplementary Fig.~\ref{FigS9}). In addition, some minute features can be observed near the zero bias region, as shown in the inset of Fig.~\ref{fig:fig1}c. The minimum appearing around $V_\mathrm{b}=0$~V is a signature of inelastic electron tunneling assisted by graphene phonons \cite{Zhang2008}. The latter mediate the in-plane momentum mismatch between the electronic states in Au and graphene~\cite{Brar2007, Zhang2008, Wehling2008, Parzefall2019}.

As shown in Fig.~\ref{fig:fig1}d, we place a TMD monolayer on top of the graphene electrode and investigate its influence on electron tunneling.  To avoid direct tunneling between the TMD and Au we place the TMD flake fully inside the graphene area. The band diagram of a  WS$_\mathrm{2}$ based device is sketched in Fig.~\ref{fig:fig1}e for positive bias. Once $eV_\mathrm{b}$ reaches the exciton energy of WS$_\mathrm{2}$, a new inelastic tunneling channel opens up. It arises from the coupling of tunneling electrons to  WS$_\mathrm{2}$ excitons, as indicated by the red arrow. Figure~\ref{fig:fig1}f shows the measured $I \textendash V$ dependence (red curve) of the WS$_\mathrm{2}$ device. It exhibits a characteristic feature near $V_\mathrm{b}=\SI{2}{\volt}$, which is absent in the $I \textendash V$ curve of the reference device (Fig.~\ref{fig:fig1}c). We attribute this sudden increase in the tunneling current to the onset of exciton-assisted resonant tunneling. Note that this feature is not visible for negative $V_\mathrm{b}$ because it is masked by the high current arising from the large electronic DOS associated with $5d$  electrons in Au (see Supplementary Fig.~\ref{FigS9}). For the same reason, the breakdown voltage of the device is considerably lower for negative $V_\mathrm{b}$ and we cannot extend our measurements beyond \SI{-2}{\volt}.
Fig.~\ref{fig:fig1}f shows the corresponding $\mathrm{d}I/\mathrm{d}V$ curve (blue). For near-zero bias, it shows phonon-assisted resonances (inset), as in the case of the reference device. In contrast to the reference device, however, we now observe a pronounced resonance peak at $\sim$\SI{2.05}{\volt}. The energy of \SI{2.05}{\electronvolt} matches well that of excitonic excitations in monolayer WS$_\mathrm{2}$~\cite{Wang2018, Goryca2019}, which provides strong support in favor of the exciton-assisted tunneling mechanism.

\begin{figure}
\begin{centering} \includegraphics[width=0.45\textwidth]{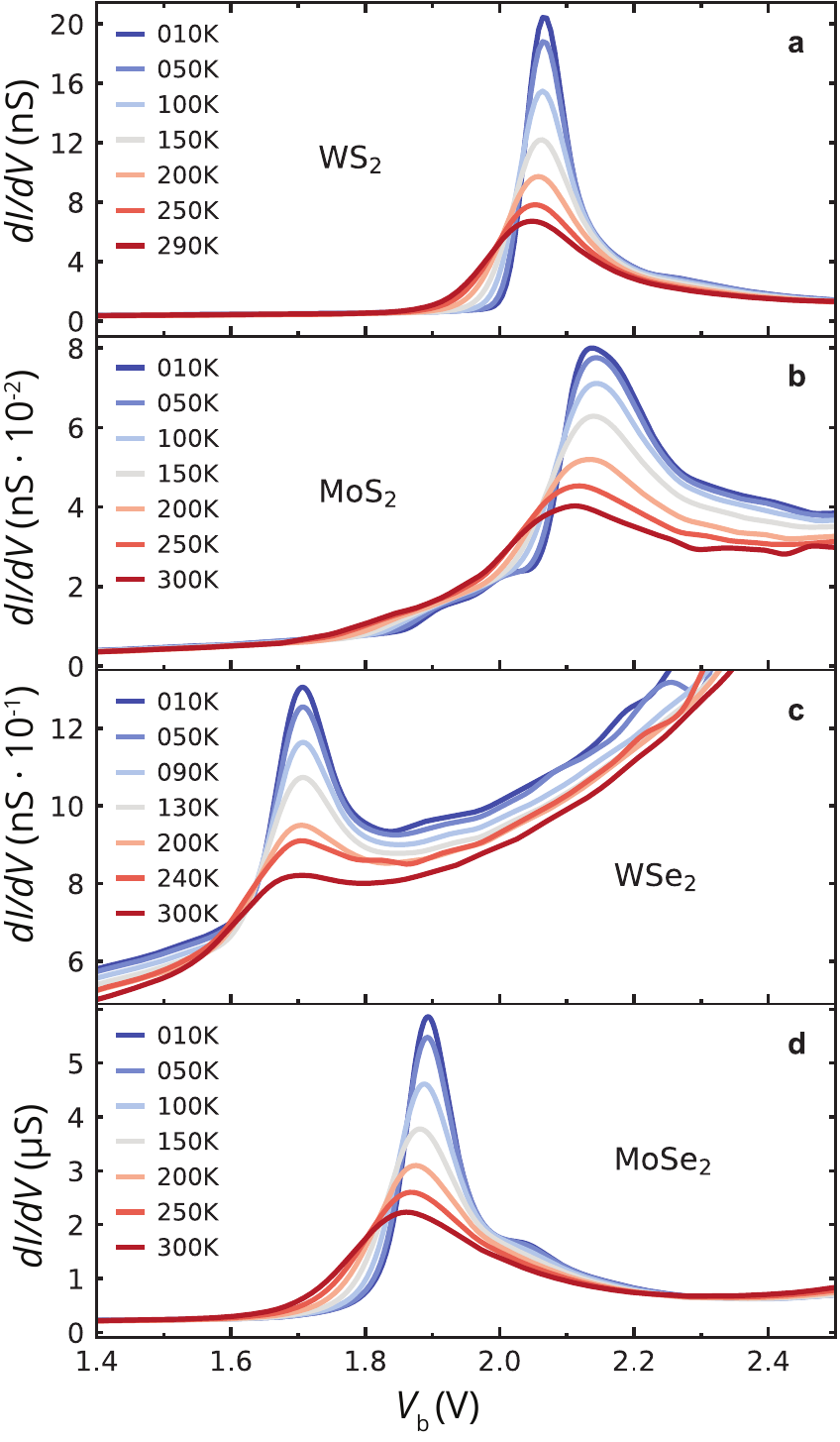} \par\end{centering}
	\caption{{\bf Temperature dependence of the tunneling spectra.} We show $\mathrm{d}I/\mathrm{d}V$ measurements for WS$_\mathrm{2}$ (\textbf{a}), MoS$_\mathrm{2}$ (\textbf{b}), WSe$_\mathrm{2}$ (\textbf{c}), and MoSe$_\mathrm{2}$ (\textbf{d}) devices at temperatures indicated by the legends.}
	\label{fig:fig2}
\end{figure} 

\begin{figure*}
\begin{centering} \includegraphics[width=0.85\textwidth]{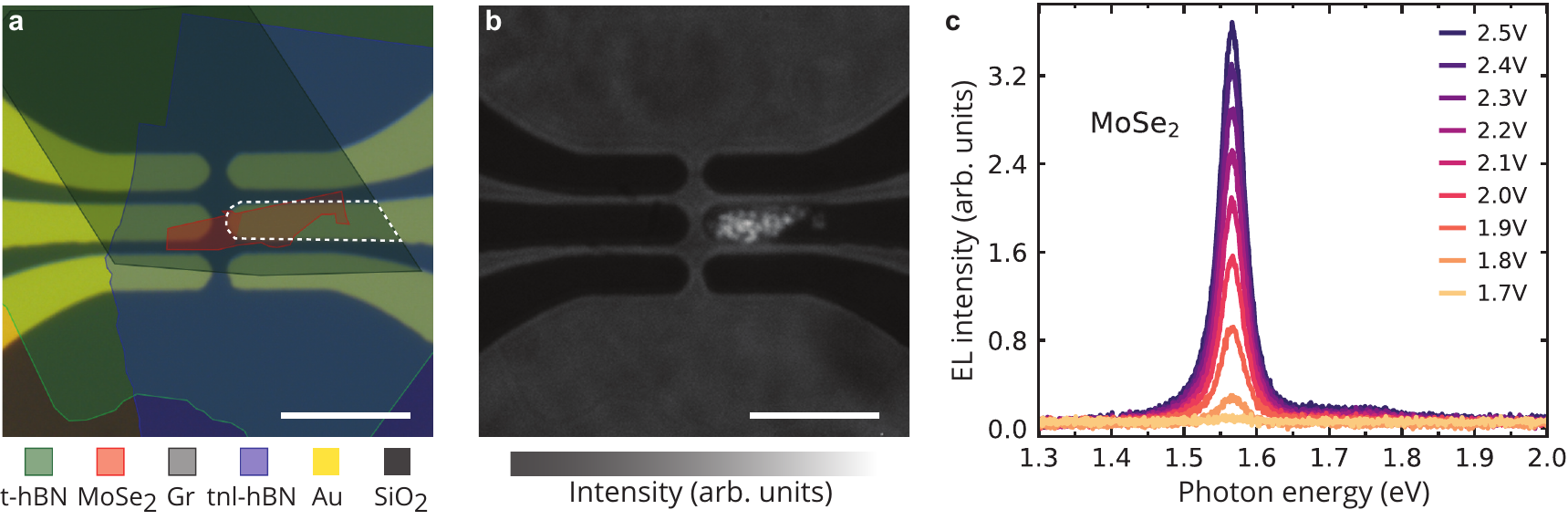} \par\end{centering}
	\caption{{\bf Radiative decay of tunneling-induced excitons.} \textbf{a}, Optical microscope image of a MoSe$_\mathrm{2}$ device fabricated on a glass (SiO$_\mathrm{2}$) substrate. The device consists of a vertical stack of a top hBN (t-hBN) protection layer, a MoSe$_\mathrm{2}$ monolayer, a graphene (Gr) monolayer, a \SI{\sim3.3}{nm} thick tunnel hBN layer (tnl-hBN), and a Au electrode. The graphene/hBN/Au tunnel junction is indicated by the white dashed line, where the graphene is partially covered by a MoSe$_\mathrm{2}$ monolayer (red polygon) on top. The upper two electrodes on the left serve as electrical contacts to the graphene sheet. Scale bar: \SI{10}{\micro\meter}. \textbf{b}, Image of light emitted from the device at an applied voltage of \SI{2.5}{\volt}, superimposed on the reference image of the device taken with back illumination. 
	Scale bar: \SI{10}{\micro\meter}. \textbf{c}, Electroluminescence (EL) spectra for different bias voltages. The exciton peak at \SI{\sim1.57}{\electronvolt} becomes more pronounced with increasing bias voltage.}
	\label{fig:fig3}
\end{figure*}

To substantiate our findings, we perform temperature dependent measurements for devices with different types of TMD monolayers. Our results are shown in Fig.~\ref{fig:fig2}\mbox{a-d} for WS$_\mathrm{2}$, MoS$_\mathrm{2}$, WSe$_\mathrm{2}$, and MoSe$_\mathrm{2}$. For all of these TMDs, we observe that the resonance becomes sharper and stronger with decreasing temperature, an effect that we attribute to the lower thermal broadening of both the electron distribution in the electrodes~\cite{Lambe1976} and the exciton resonance.

For the WS$_\mathrm{2}$ device, we observe a shift in peak position to larger bias voltages, from \SI{\sim2.05}{\volt} (\SI{290}{\kelvin}) to \SI{\sim2.07}{\volt} (\SI{10}{\kelvin}), consistent with the temperature-dependent measurements of excitonic resonances in WS$_\mathrm{2}$~\cite{Hanbicki2015, Jadczak2017}. Upon closer inspection, we find another resonance near \SI{2.26}{\volt}, which becomes visible at low temperatures for the WS$_\mathrm{2}$ device (Fig.~\ref{fig:fig2}a and Fig.~S6a) suggesting coupling to excitons of higher energy.

Measurements for a MoS$_\mathrm{2}$ device are plotted in Fig.~\ref{fig:fig2}b, where a main peak at $V_\mathrm{b} = \SI{2.1}{\volt}$ is observed, exhibiting a similar temperature dependence as the WS$_\mathrm{2}$ device. A weak feature appears at slightly lower $V_\mathrm{b}$ and develops into two distinct shoulders at lower temperatures, one at \SI{\sim1.92}{\volt} and another one at \SI{\sim2.02}{\volt} (see also Fig.~S6). The irregular features at bias voltages beyond \SI{2.2}{\volt}, especially at high temperatures, can be attributed to measurement instabilities.

For the WSe$_\mathrm{2}$ device shown in Fig.~\ref{fig:fig2}c, we observe a main resonance at $V_\mathrm{b} = \SI{1.7}{\volt}$ and the temperature dependence is similar to the previous two devices. The wiggles appearing at higher bias voltages can be assigned to measurement instabilities (see Supplementary Sec.~\ref{sec:instabilities} and Figs.~\ref{FigS6} and \ref{FigS7}). The MoSe$_\mathrm{2}$ device plotted in Fig.~\ref{fig:fig2}d presents the same temperature behaviour. The resonance in this case appears at $V_\mathrm{b} = \SI{1.9}{\volt}$. The smaller feature at \SI{\sim2.04}{\volt}, which becomes more pronounced at low temperatures, is likely  arising from higher energy excitons. The resonances appear at different bias voltages for every TMD, an observation that hints to exciton coupling. 

Further evidence for exciton-assisted electron tunneling is provided by our electroluminescence (EL) measurements. Excitons generated by resonant electron tunneling can partially decay through radiative electron-hole recombination. As shown in Fig.~\ref{fig:fig3}c, this radiative decay gives rise to a distinctive peak in the measured EL spectrum. Figure~\ref{fig:fig3}a illustrates the layout of a MoSe$_\mathrm{2}$-based device. Electrons tunnel in the region where graphene, hBN, and the Au electrode overlap (the area enclosed by a white dashed line). The MoSe$_\mathrm{2}$ monolayer indicated by the red polygon is placed directly above the graphene layer. The upper two electrodes on the left-hand side serve as electrical contacts to  graphene. The photograph in Fig.~\ref{fig:fig3}b shows that EL is observed when a voltage $V_\mathrm{b}=2.5$\,V is applied. The emission is restricted to the region where the MoSe$_\mathrm{2}$ monolayer overlaps with the tunneling device. Spectra of the emitted light for different bias voltages are plotted in Fig.~\ref{fig:fig3}c. The peak centered at \SI{\sim1.57}{\electronvolt} agrees with previous studies~\cite{Tongay2012, Tonndorf2013} and can be assigned to the $A$ exciton of MoSe$_\mathrm{2}$ according also to our photoluminescence (PL) spectroscopy measurements (see Supplementary Table~\ref{TableS2}). Note, however, that the corresponding peak in the differential conductivity measurements shown in Fig.~\ref{fig:fig2}d appears at higher voltages than the $A$ exciton energy, suggesting coupling to higher-order excitons. The EL intensity of the other three TMD devices in Fig.~\ref{fig:fig2} is below our detection threshold because we used a considerably thicker hBN spacer to prevent breakdown, so the resulting current densities are more than two orders of magnitude lower than those of the MoSe$_\mathrm{2}$ device (see Supplementary Fig.~\ref{FigS1}). Excitonic light emission from a WSe$_\mathrm{2}$ device with a thinner hBN tunnel barrier is shown in Supplementary Fig.~\ref{FigS5}.

\begin{figure*}
\begin{centering} \includegraphics[width=0.65\textwidth]{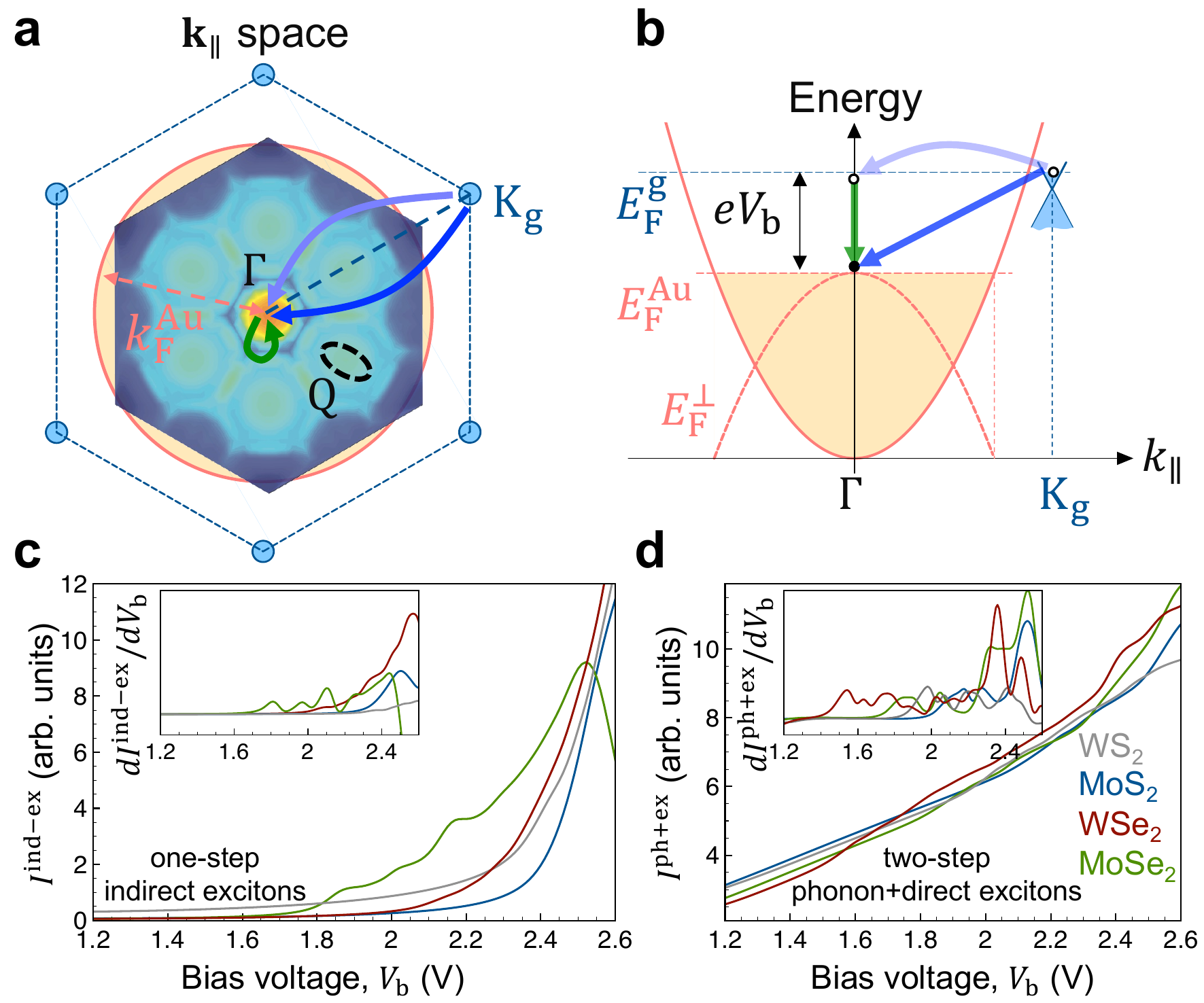} \par\end{centering}
\caption{{\bf Exciton-assisted electron tunneling pathways.}
\textbf{a}, We consider two possible tunneling channels under the configuration of Fig.~\ref{fig:fig1}d, illustrated here by an on-scale representation of the graphene first Brillouin zone (1BZ, dashed hexagon in the space of parallel wave vector $\kparb$), the surface-projected gold Fermi surface (orange circle), and the TMD 1BZ (color plot, showing the nonlocal surface conductivity for MoSe$_2$ at 2.1~eV photon energy). The two channels are (1) one-step tunneling assisted by the creation of indirect TMD excitons (blue arrow) and (2) two-step tunneling associated with phonon creation (purple arrow) followed by direct-exciton creation (green arrow). Direct excitons produce the intense feature at the ${\rm\Gamma}$ point in the TMD conductivity, while indirect excitons show up as maxima at special regions of the TMD 1BZ, such as the Q point. \textbf{b}, Gold conduction band dispersion diagram (orange), along with the energy-momentum region occupied by graphene electrons (light blue), involving a bias energy $e\VVb$ as well as the gold and graphene Fermi energies $E_{\rm F}^{\rm Au}$ and $E_{\rm F}^{\rm g}$. As a function of $k_\parallel$, the out-of-plane gold Fermi surface (dashed orange parabola) has a minimum energy mismatch (i.e., a maximum spill-out of gold electrons towards the hBN barrier) at the ${\rm\Gamma}$ point. The two considered tunneling channels bridge the energy-momentum mismatch between the graphene K point and gold Fermi energy at $\kpar=0$.
\textbf{c},\textbf{d}, Calculated voltage-dependent tunneling current associated with one-step indirect TMD exciton creation (\textbf{c}) and two-step phonon plus direct-exciton creation (\textbf{d}) for a hBN tunnel barrier of 3~nm and different TMDs. The insets show the corresponding $dI/dV$ curves.}
\label{fig:fig4}
\end{figure*}

We continue by theoretically exploring the mechanisms responsible for resonant tunneling and EL. The tunneling process involves conservation of both energy and in-plane momentum, for which there is a clear mismatch between graphene and gold, as illustrated in Fig.~\ref{fig:fig4}a,b (see also Supplementary Fig.~\ref{FigS10}). In addition, tunneling favors final electron gold states with large out-of-plane energy, which reside close to the conduction band of the hBN spacer. Consequently, the momentum of final states lies near the surface-projected $\Gamma$ point of the involved materials, as illustrated in Fig.~\ref{fig:fig4}b, and also in Supplementary Fig.~\ref{FigS10}c, where we observe a dramatic depletion of electron spill-out in gold-bound electrons when moving away from the $\Gamma$ point, corresponding to $\kparb=0$ in the space of parallel wave vector $\kparb$. Tunneling is therefore expected to be dominated by transitions from graphene electrons near the K$_{\rm g}$ point in this material to gold states near its Fermi level at the surface-projected ${\rm\Gamma}$ point, involving a large in-plane wave vector transfer given by ${\rm\Gamma}$K$_{\rm g}\approx17$~nm$^{-1}$. Phonons can provide such large momentum, and in fact, graphene and hBN have similar in-plane lattice parameters (2\% mismatch), so phonons (frequency $\omega_p$) in both materials can assist quasi-elastic tunneling, giving rise to the features observed at low bias voltages $\VVb\sim\hbar\omega_p/e\sim65$~mV in the insets of Fig.~\ref{fig:fig1}c,f and in agreement with previous studies \cite{Zhang2008}.

We identify two tunneling mechanisms that can bridge the graphene-gold momentum mismatch, as illustrated by thick arrows in Fig.~\ref{fig:fig4}a,b: (1) a single-step processes ($I^{\rm ind\mbox{-}ex}$ current) involving the creation of indirect TMD excitons (blue arrow); and (2) a two-step process ($I^{\rm ph+ex}$ current) in which phonons provide the required large in-plane momentum (purple arrow) and direct excitons the energy difference between initial and final states (green arrow). A detailed analysis of these channels is provided in Supplementary Secs.~\ref{theorysection}--\ref{TMDsigma}, from which we conclude that the associated currents bear a dependence on $\VVb$ given by
\begin{subequations}
\label{Itheory}
\begin{align}
I^{\rm ind \mbox{-} ex}\propto\;&{\rm Im}\big\{-W_{\Gb\Gb}(\kparb,d,d,eV_{\rm b}/\hbar)\big\}, \label{Itheory1}\\
I^{\rm ph+ex}\propto\;&{\rm Im}\big\{-W_{00}(0,d,d,eV_{\rm b}/\hbar-\omega_p)\big\}, \label{Itheory2}
\end{align}
\end{subequations}
expressed in terms of the screened interaction $W_{\Gb\Gb'}(\kparb,z,z',\omega)$ \cite{paper149}. The latter is defined as the potential created at a distance $z$ from the gold surface by a unit charge placed at a distance $z'$, oscillating with frequency $\omega$, and decomposed in parallel wave vectors $\kparb$ within the first Brillouin zone (1BZ) of the TMD, as well as reciprocal lattice vectors $\Gb$ and $\Gb'$. We set $z=z'=d$ at the graphene plane, where the initial source of electrons is located. Also, the frequency $\omega$ is determined by the associated bias frequency $e\VVb/\hbar$ in $I^{\rm ind\mbox{-}ex}$, a value that needs to be corrected by the emitted phonon frequency in $I^{\rm ph+ex}$. The actual expressions that we use to compute the contributions in Fig.~\ref{fig:fig4}c,d are slightly more complex, as detailed in the Sec.~\ref{SupplementaryInformation}, but essentially captured by Eqs.~(\ref{Itheory}).

The single-step current $I^{\rm ind\mbox{-}ex}$ [Eq.~(\ref{Itheory1})] receives equal contributions from each of the six smallest non-vanishing reciprocal lattice vectors $\Gb$ of the TMD, involving a dominant $\kparb$ value in the 1BZ where the optical conductivity reaches a maximum (color plot in Fig.~\ref{fig:fig4}a). These features are associated with indirect excitons connecting the K point of the TMD with the Q point. Although the relative lattice orientations of TMD and graphene are undefined in our samples, we note that $|\kparb+\Gb|$ provides the required $\sim17$~nm$^{-1}$ in-plane wave vector, and in fact, even after averaging over lattice orientations, indirect excitons produce discernible shoulders in the $\VVb$-dependent profile of the resulting current (Fig.~\ref{fig:fig4}c), which are better visualized in the $dI/dV$ curves (inset). The single-step indirect-exciton-assisted current thus displays features similar to experiment, although a determination of their detailed energies demands further computations exceeding our current resources. In the two-step tunneling process [Eq.~(\ref{Itheory2})], the screened interaction is evaluated at $\kparb=0$ and $\Gb=0$, and the resulting current displays direct-exciton features (Fig.~\ref{fig:fig4}d), but their strengths relative to the background are small compared to those of the indirect-exciton mechanism.

Since it is revealed that not only phonons but also indirect excitons between K point valence and Q point conduction band states can provide the missing in-plane momentum for tunneling in our devices, we finalize our study by comparing the peak positions of our $\mathrm{d}I/\mathrm{d}V$ curves with direct and indirect exciton energies. Supplementary Table~\ref{TableS2} includes experimental and calculated values of exciton energies and in-plane momentum values for various TMDs. The Mo-based TMDs present peaks that are higher in energy than $A$ excitons and they fit well with the reported values of K-Q indirect excitons, suggesting that mostly indirect exciton transitions contribute to the observed resonant tunneling behavior. In contrast, K-Q indirect-exciton energies in the W-based TMDs are closer to the $A$ exciton energies and both of them appear near their corresponding $\mathrm{d}I/\mathrm{d}V$ peak energies, making the distinction between them more difficult. Although a one-step indirect-exciton transition is likely to be more efficient, the EL emission from our MoSe$_2$ devices suggests the involvement of direct excitons. Such radiative exciton decay can happen either from the phonon+direct-exciton transitions or by phonon-assisted luminescence from indirect-excitons, as reported in previous studies \cite{Funk2021}.

\section{CONCLUSION}

In conclusion, we have demonstrated exciton-assisted resonant electron tunneling in graphene/hBN/Au tunnel junctions with a TMD monolayer placed in proximity to the graphene layer. This process manifests as an abrupt increase in the tunnel current when the bias electron energy $eV_\mathrm{b}$ matches an exciton energy, resulting in a resonance peak in the $\mathrm{d}I/\mathrm{d}V$ curve. An observed blueshift of the resonant peak with decreasing temperature is consistent with that of the corresponding excitons revealed by optical spectroscopy~\cite{Hanbicki2015, Jadczak2017}. We find that the exciton states giving rise to the main resonance peaks in the $\mathrm{d}I/\mathrm{d}V$ curves are different for the four studied TMD devices. We theoretically explain our measurements by electron tunneling mechanisms that involve either indirect or direct excitons. Indirect excitons can supply both the energy and in-plane momentum required to tunnel from graphene to gold, while direct excitons require additional in-plane momentum supplied by phonons. Our findings are further substantiated by optical measurements, which reveal excitonic light emission driven by inelastic electron tunneling. In our devices, the TMD layer is placed outside of the electron tunneling path, which allows us to suppress exciton generation by direct charge injection. This device structure provides a platform not only for studying fundamental aspects of tunneling processes, but also for exploring novel device functionalities for optoelectronics, all-electrical sensing and spectroscopy.

\section{METHODS}

\textbf{Sample fabrication.} All flakes are mechanically exfoliated from bulk crystals in air (hBN) or in an argon-based glovebox (TMD and graphene). The heterostructures are first assembled using a standard pick-up technique with a PDMS/PC stamp~\cite{Zomer2014}, then transferred onto the prefabricated Au electrodes on a glass substrate in the glovebox.

\textbf{Electrical and optical measurements.} Room temperature electrical measurements are performed using a Keithley 2602B source meter. Low-temperature measurements are performed in a variable temperature probe station. Currents are measured using a Femto DDPCA-300 current amplifier. An ADwin Pro II DAC is used to apply the bias voltage and read the output voltage of the current amplifier. For optical measurements, the samples are mounted on a Nikon TE300 inverted microscope under ambient conditions. The emitted light is collected by a $\times100$ objective (NA=0.9) and analysed using an Andor iXon Ultra camera and a Princeton Instruments Acton SpectraPro 300i spectrometer.

\textbf{Theoretical calculations.} The screened interaction in Eqs.~(\ref{Itheory}) is obtained by combining first-principles calculations of the nonlocal conductivity of TMD monolayers, the random-phase-approximation nonlocal response of graphene, and the specular-reflection model for the nonlocal response of the gold surface based on the Lindhard permittivity for the bulk metal. The anisotropic response of hBN is accounted for through a local permittivity tensor. To deal with indirect TMD excitons, the response of the heterostructure is calculated with inclusion of umklapp processes for the reflection and transmission coefficients of the TMD layer, which permeate the screened interaction, such that it becomes a tensor labeled by TMD reciprocal lattice vectors and wave vectors within the 1BZ. The derivation of Eqs.~(\ref{Itheory}) further involves the analysis of the electron potential landscape in the heterostructure, which is incorporated through the corresponding electron Green function. A detailed self-contained analysis of these elements is presented in the Supplementary Sec.~\ref{theorysection}--\ref{TMDsigma}, along with graphical information on the TMD nonlocal conductivities and the screened interaction of our heterostructures.

\section*{}
\subsection*{Author contributions}
L.W., S.P. and L.N. conceived the experiment. L.W. fabricated the devices. L.W. and S.P. performed the measurements. L.W. analysed the data. S.P. and J.H. supported in the device fabrication. J.Z. performed the low temperature measurements with the support from M.L.P and M.C.. S.P. and L.N. helped with the data interpretation. F.I. and F.J.G.d.A. developed the theory, performed numerical calculations, and wrote the theoretical discussion. K.W. and T.T. provided the high-quality hBN crystals. L.N. initiated and supervised the project. L.W. and S.P. wrote the paper and all authors discussed the results and worked on the manuscript.

\subsection*{Acknowledgments}
The authors would like to thank Mathieu Luisier, Achint Jain, Ronja Khelifa, Shengyu Shan and Paritosh Karnatak for fruitful discussions. This study was supported by funding from ETH Zurich under ETH Grant No. ETH-15 19-1 SYNEMA, the ETH Zurich Foundation project number 2013-08 (11) with a donation from the Stavros Niarchos Foundation, and the Swiss National Science Fund under grant number $200020\_192362$. K.W. and T.T. acknowledge support from the JSPS KAKENHI (Grant Numbers 19H05790, 20H00354 and 21H05233). M.L.P. acknowledges funding by the Swiss National Science Foundation (SNSF) under Eccellenza Professorial Fellowship no. $\mathrm{PCEFP2\_203663}$ and funding by the Swiss State Secretariat for Education, Research and Innovation (SERI) under contract number MB22.00076. M.C. acknowledges funding by the Swiss National Science Foundation under the Sinergia grant no. 189924 (Hydronics). F.J.G.A. acknowledges funding by the Spanish MCINN (PID2020-112625GB-I00 and CEX2019-000910-S), Generalitat de Catalunya (CERCA and AGAUR), and Fundaci\'os Cellex and Mir-Puig.

\section{SUPPLEMENTARY INFORMATION}
\label{SupplementaryInformation}

\renewcommand{\thetable}{S\arabic{table}}%
\renewcommand{\thefigure}{S\arabic{figure}}%
\renewcommand{\theequation}{S\arabic{equation}}%
\renewcommand{\thesection}{S\arabic{section}}%

\begin{figure}
\begin{centering} \includegraphics[width=0.5\textwidth]{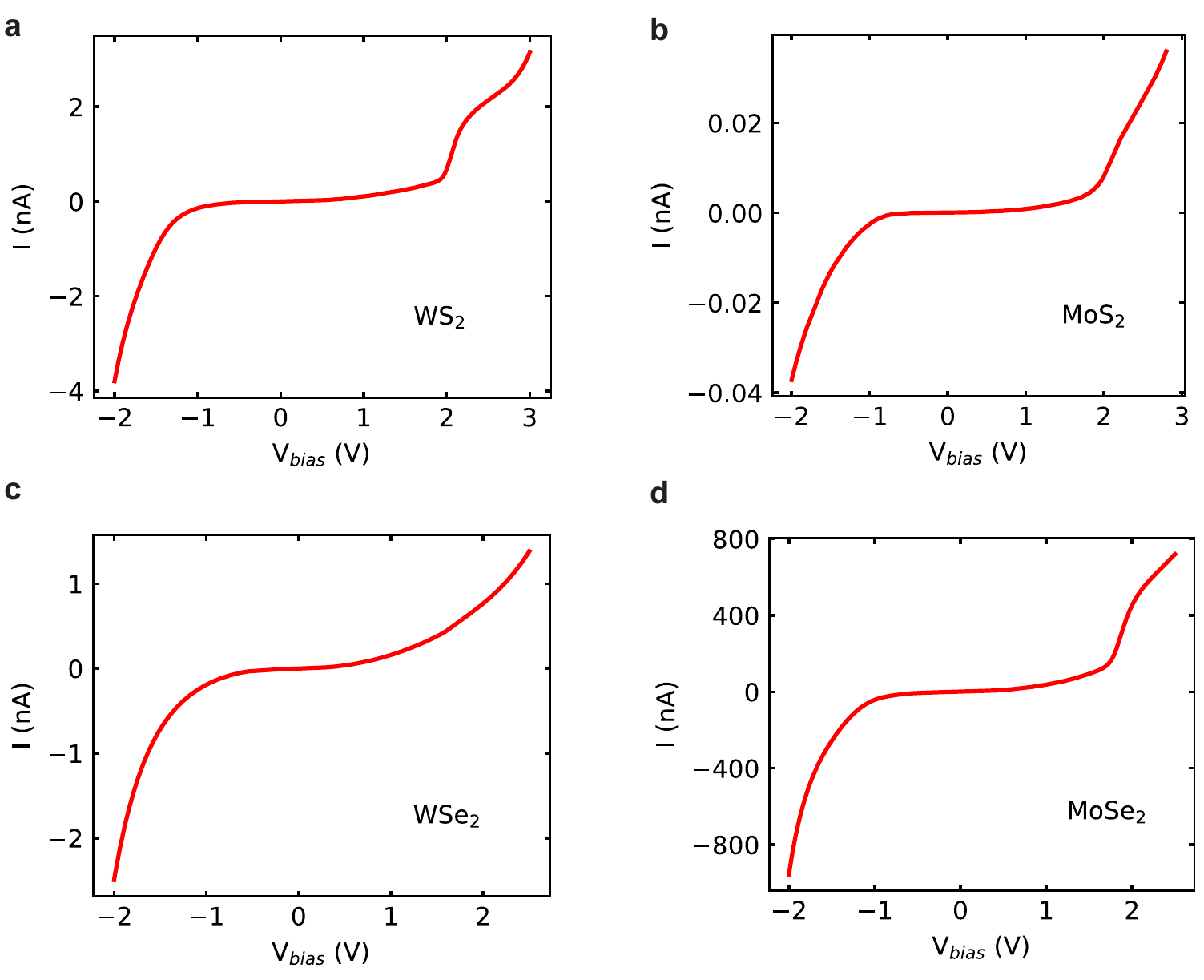} \par\end{centering}
    \caption{Measured $I \textendash V$ characteristics of the four TMD devices discussed in the main text at room temperature.}
    \label{FigS1}
\end{figure}

\subsection{Device details and $I \textendash V$ characteristics}\label{sec:IVs_main}
Geometrical details of the four devices discussed in the main text are provided in Table~\ref{TableS1}. The device size is determined by the overlap region of graphene, the hBN tunnel barrier, and the Au electrode. The TMD coverage is defined as the percentage of the device area under a TMD monolayer. The thickness of the hBN tunnel barrier is $\sim$3--4~nm, as estimated from the combination of the optical contrast and AFM measurements. However, the exact number of hBN atomic layers is difficult to determine with our AFM setup.

\begin{table*}
	\begin{center}
		\setlength{\tabcolsep}{10pt}
		\begin{tabular}{l|c|c|c|c}
		\hline
		Device name & WS$_2$ & MoS$_2$ & WSe$_2$ & MoSe$_2$ \\ \hline
		Device size ($\SI{}{\micro\meter}\times\SI{}{\micro\meter}$) & $\sim2.5\times5$ & $\sim2.5\times4$ & $\sim2.5\times10$ & $\sim2.5\times12$ \\ \hline
		TMD coverage (\%) & $\sim73$ & $\sim70$ & $\sim35$ & $\sim50$ \\ \hline
		\end{tabular}
		\label{TableS1}
		\caption{Sizes and TMD coverages of the devices.}
	\end{center}
\end{table*}

The measured $I \textendash V$ characteristics of the devices are plotted in Fig.~\ref{FigS1}. The exciton-assisted tunneling feature is less visible in the WSe$_2$ device than in the other three devices due to a smaller TMD coverage. Nevertheless, the exciton signature is clearly observed in the $\mathrm{d}I/\mathrm{d}V$ spectrum of this device, as shown in the main text. The current of the MoSe$_2$ device at a given bias voltage is more than two orders of magnitude larger than that of the other three devices due to a thinner hBN tunnel barrier. An exciton-assisted light emission can be measured for this device and is shown in the main text, while no light emission is detected for the other three devices. Assuming a similar electron-to-photon efficiency for all of our TMD devices, the number of emitted photons is at least two orders of magnitude smaller in the other three devices than in the MoSe$_2$ device at a given bias voltage, and thus, the electroluminescence signal lies below the detection limit of our setup.

\begin{figure}
\begin{centering} \includegraphics[width=0.5\textwidth]{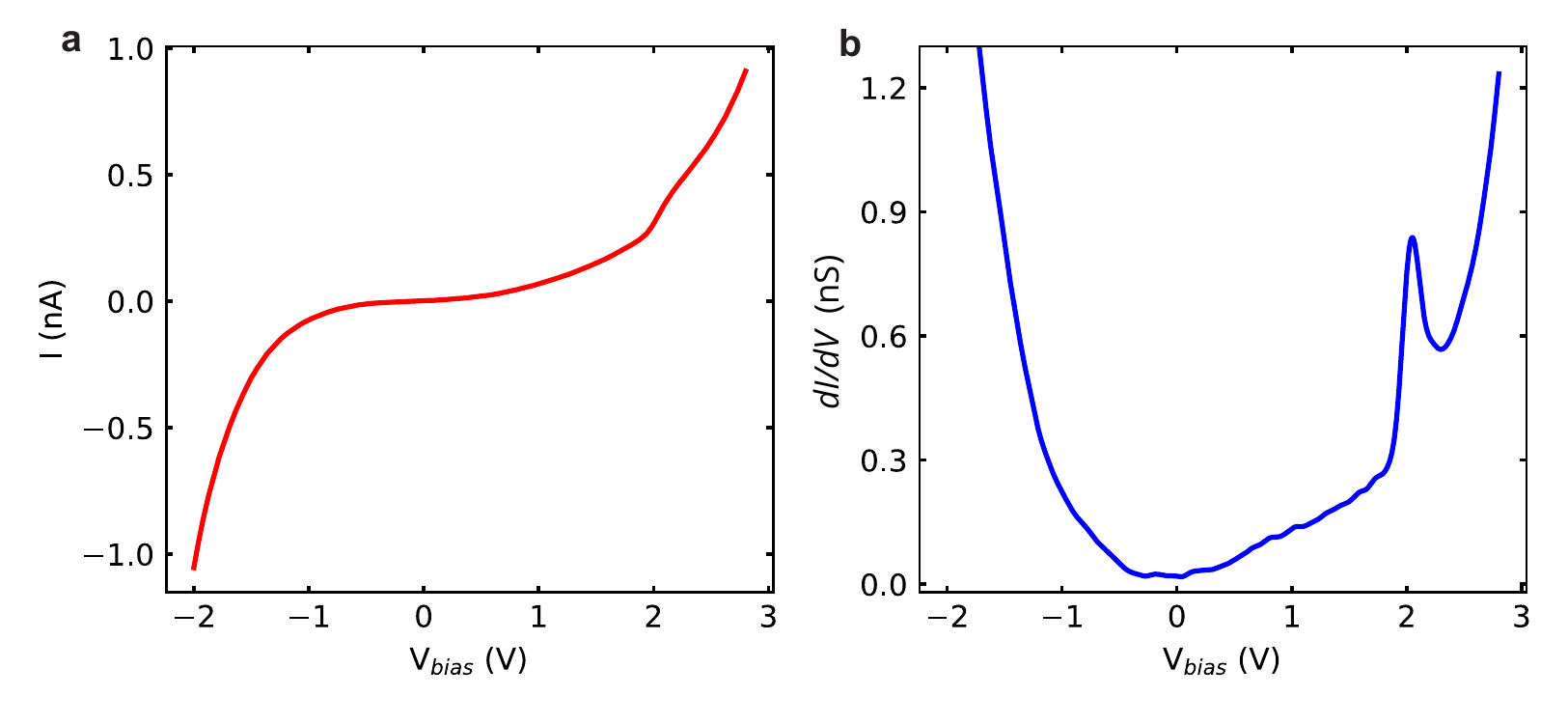} \par\end{centering}
    \caption{Measured $I \textendash V$ characteristics (\textbf{a}) and corresponding $\mathrm{d}I/\mathrm{d}V$ curve (\textbf{b}) of a second WS$_2$ device at room temperature.}
    \label{FigS2}
\end{figure}

\begin{figure}
\begin{centering} \includegraphics[width=0.5\textwidth]{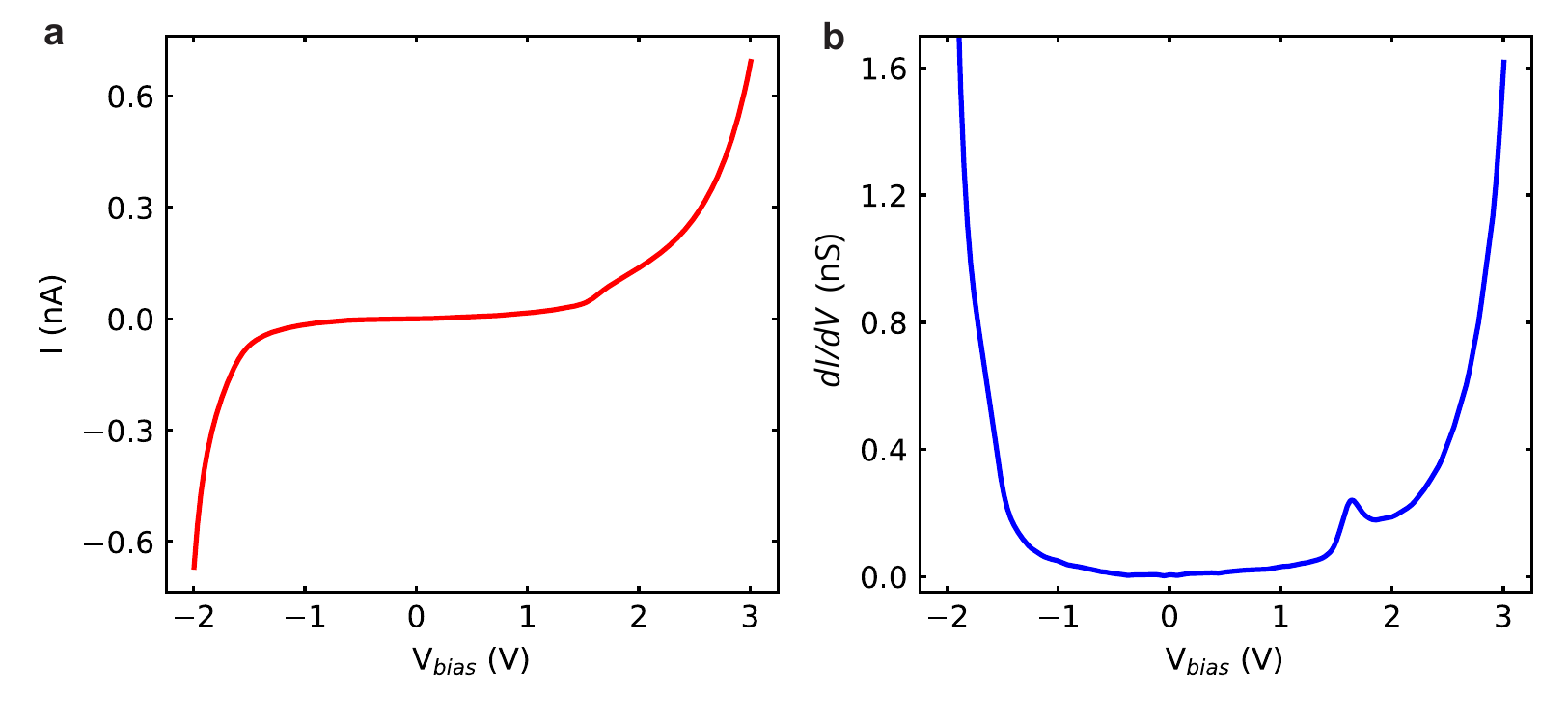} \par\end{centering}
    \caption{Measured $I \textendash V$ characteristics (\textbf{a}) and corresponding $\mathrm{d}I/\mathrm{d}V$ curve (\textbf{b}) of a second WSe$_2$ device at room temperature.}
    \label{FigS3}
\end{figure}

\begin{figure}
\begin{centering} \includegraphics[width=0.5\textwidth]{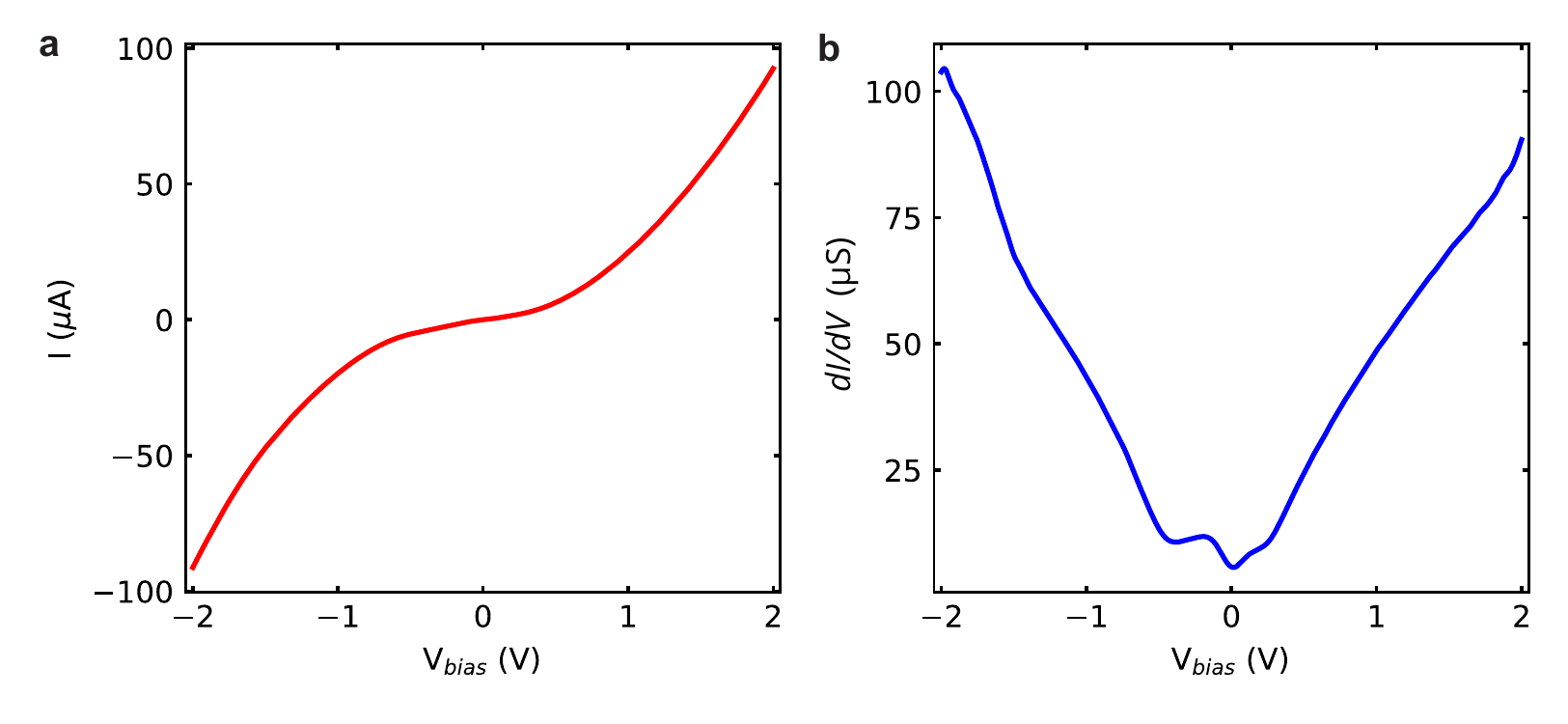} \par\end{centering}
    \caption{Measured $I \textendash V$ characteristics (\textbf{a}) and corresponding $\mathrm{d}I/\mathrm{d}V$ curve (\textbf{b}) of a WSe$_2$ device with a thinner ($\sim2.3$~nm) hBN tunnel barrier at room temperature.}
    \label{FigS4}
\end{figure}

\begin{figure}
\begin{centering} \includegraphics[width=0.5\textwidth]{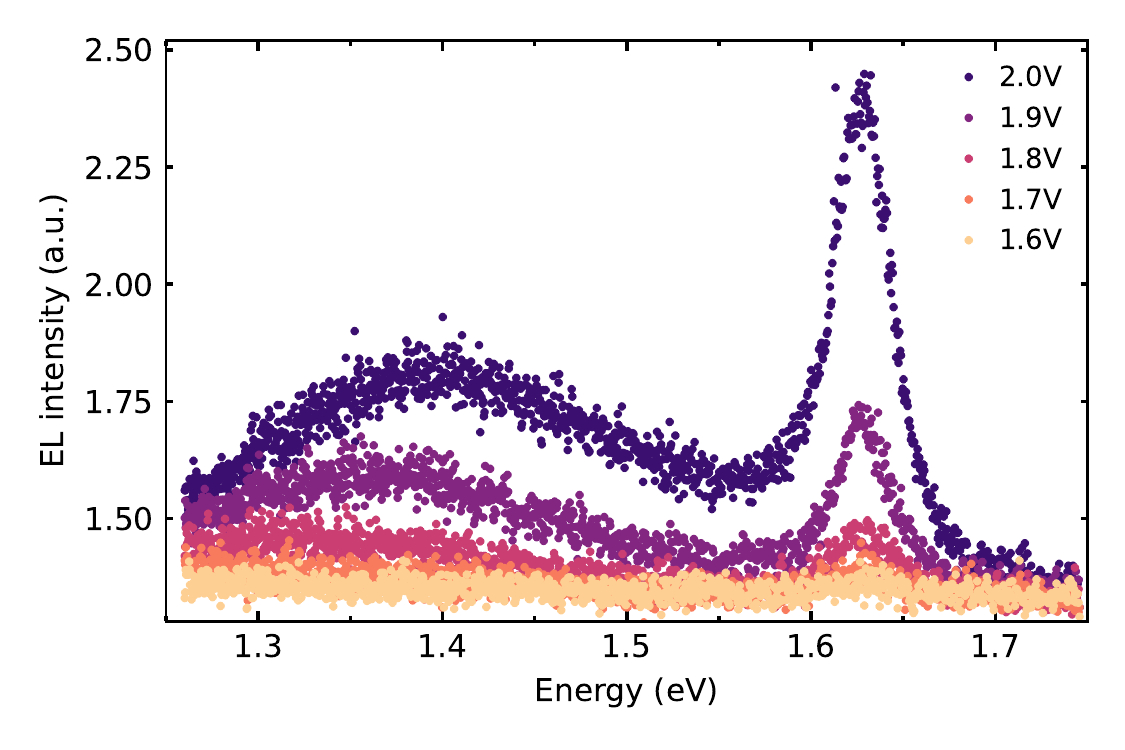} \par\end{centering}
    \caption{Measured spectra of the light emitted by the WSe$_2$ device of Fig.~\ref{FigS4} at different bias voltages.}
    \label{FigS5}
\end{figure}

\begin{figure*}
	\begin{centering} \includegraphics[width=0.9\textwidth]{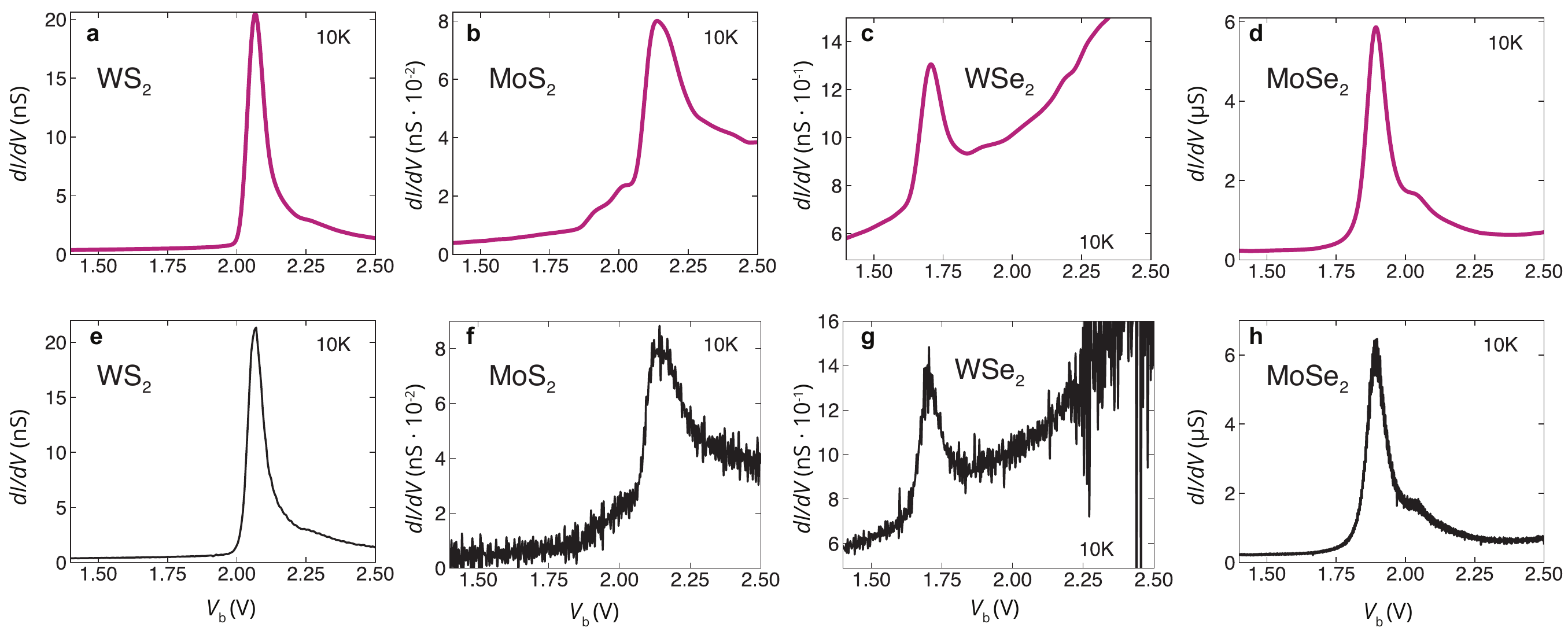} \par\end{centering}
	\caption{{\bf Measured $\mathrm{d}I/\mathrm{d}V$ curves at 10~K for the various TMDs studied in this work.} Curves in panels (\textbf{a}-\textbf{d}) are treated with a Savitzky–Golay filter, whereas the unfiltered data are presented in (\textbf{e}-\textbf{h}). Features observed at $V_{b}>$\SI{2.2}{\volt} appear to be artificial due to the excess noise at those voltages, as shown in (\textbf{g}).}
	\label{FigS6}
\end{figure*}

\begin{figure*}
	\begin{centering} \includegraphics[width=0.65\textwidth]{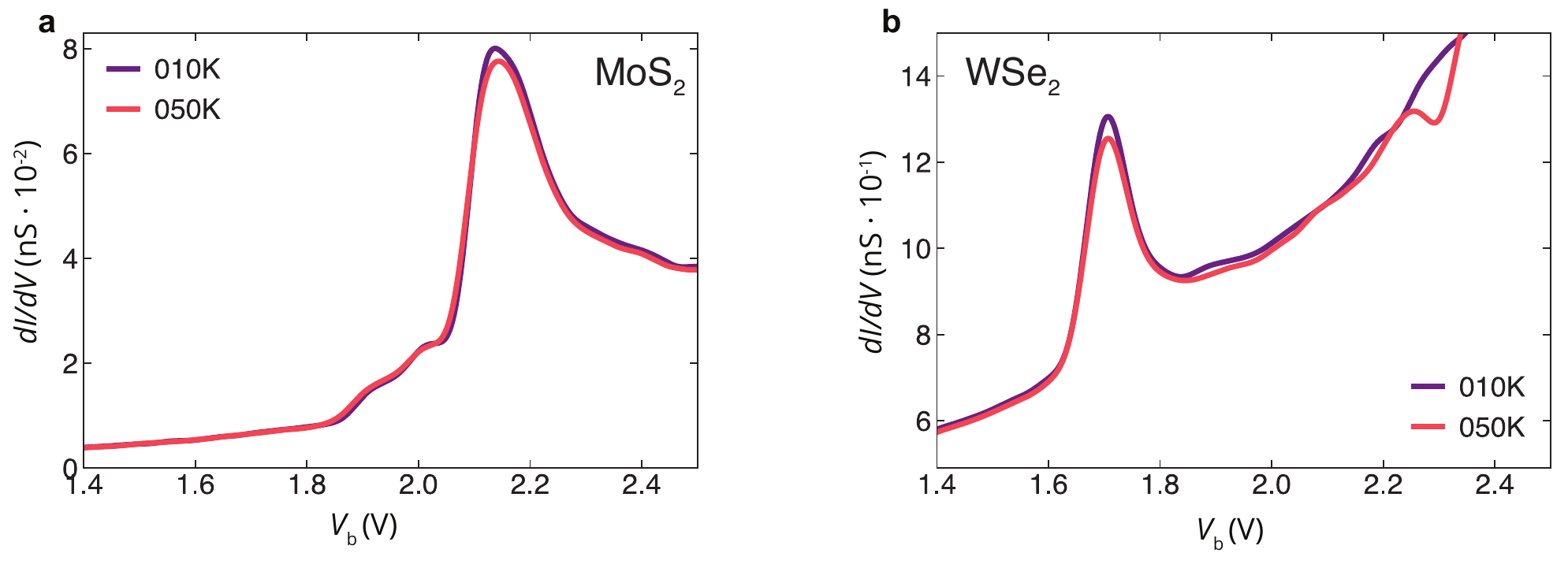} \par\end{centering}
	\caption{{\bf Comparison between 50~K and 10~K measurements for the $\mathrm{d}I/\mathrm{d}V$ curves of (\textbf{a}) MoS$_2$ and (\textbf{b}) WSe$_2$ devices.} In (\textbf{a}), some features appear consistently in both measurements close to the exciton resonance, whereas in (\textbf{b}), features at higher energies fluctuate, suggesting that they could originate in measurement instabilities.}
	\label{FigS7}
\end{figure*}

\begin{figure}
	\begin{centering} \includegraphics[width=0.4\textwidth]{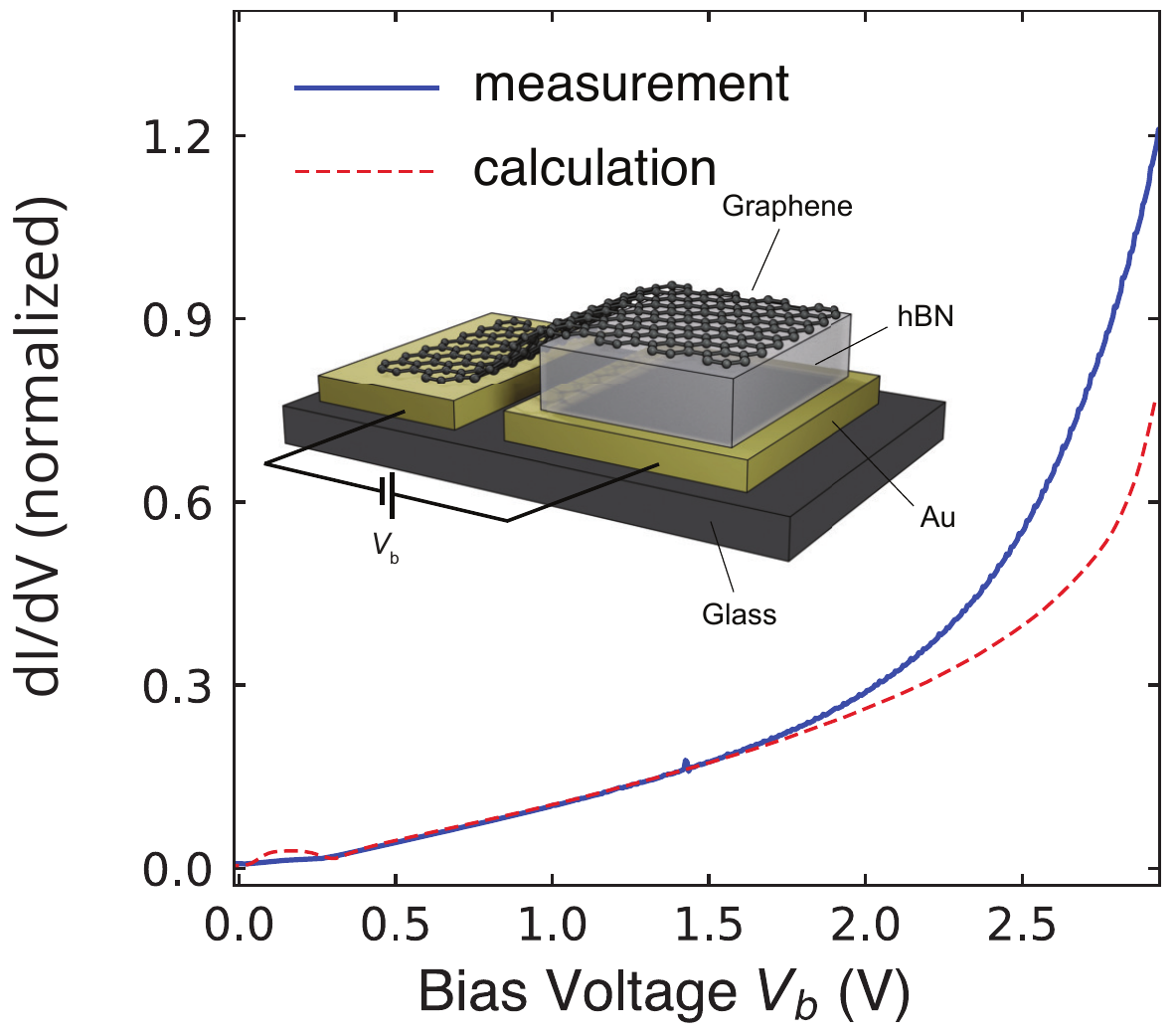} \par\end{centering}
	\caption{{\bf Comparison between $\mathrm{d}I/\mathrm{d}V$ measurements and a simple tunneling model with an infinitely high tunneling barrier.} The model follows the measurement up to 2~V, with significant deviations only appearing above 2.5~V. The device structure under consideration is illustrated in the inset.}
	\label{FigS8}
\end{figure}

\subsection{Data for additional devices}\label{sec:add_devices}
To further corroborate our findings, we fabricated two more devices, from which the results shown in the main text are reproduced. The $I \textendash V$ and $\mathrm{d}I/\mathrm{d}V$ curves of a second WS$_2$ device and a second WSe$_2$ device are displayed in Figs.~\ref{FigS2} and \ref{FigS3}, respectively. Resonance features can be clearly observed in the $\mathrm{d}I/\mathrm{d}V$ spectra, at \SI{\sim2.05}{\volt} for WS$_2$ and \SI{\sim1.64}{\volt} for WSe$_2$, which are essentially the same results as we find in the devices discussed in the main text. The reproducibility of the results further supports our exciton-assisted tunneling interpretation.

\subsection{Discrimization between actual features and measurement instabilities}\label{sec:instabilities} 
In this section, we are presenting $\mathrm{d}I/\mathrm{d}V$ curves for the four TMD devices (WS$_2$, MoS$_2$, WSe$_2$, and MoSe$_2$) measured at 10~K after (Fig.~\ref{FigS6}a-d) and before (Fig.~\ref{FigS6}e-h) filtering noise by means of a Savitzky–Golay filter. We conclude that the features that appear at high voltages in the $\mathrm{d}I/\mathrm{d}V$ curve of the WSe$_2$ device in Fig.~\ref{FigS6}c can be attributed to measurement instabilities and increased noise. This becomes apparent in the unfiltered version of the measurements in Fig.~\ref{FigS6}g, where increased noise and fluctuations are observed. Finally, the features appearing close to the exciton resonances for the rest of the devices (WS$_2$, MoS$_2$, and MoSe$_2$) are still observed in the unfiltered data. To further support this interpretation, we show in Fig.~\ref{FigS7}a the MoS$_2$ $\mathrm{d}I/\mathrm{d}V$ curve at both 50~K and 10~K, with features appearing consistently in both of them, suggesting that they are physical. This is in contrast to the $\mathrm{d}I/\mathrm{d}V$ curve of the MoS$_2$ device in Fig.~\ref{FigS7}b, where the high-energy features vary for different measurements at 50~K and 10~K, suggesting that they are due to instabilities.

\subsection{Graphene/hBN/Au tunnel junction: comparison between theory and experiment}\label{sec:Simple_tunnel_junction}
In the tunnel junctions presented in the main text, we explore a range of voltages up to 3~V. To further our understanding on whether we measure contributions from Fowler-Nordheim (F-N) tunneling or thermionic effects, we use a simple tunneling model with an infinitely high tunneling barrier to compare to our measurements. The phonon-induced tunneling current between graphene (Gr) and Au is calculated as \cite{Parzefall2019}
\begin{align}
		I \propto \int_{\hslash\Omega}^{eV_b} \mathcal{T}^2 \cdot \rho_\mathrm{{Au}} \cdot \rho_\mathrm{{Gr}}(E)~dE, 
\end{align}
\noindent
where $\hslash\Omega$ is the phonon energy, $\mathcal{T}$ is the transmission matrix element (assumed to be constant in accordance with the DFT calculations presented in the supplementary material of Ref.~\cite{Parzefall2019}), $\rho_\mathrm{{Au}}$ is the DOS of Au (also assumed to be constant for $V_b>0$ according to the DFT calculations in Fig.~\ref{FigS9}), and $\rho_\mathrm{{Gr}}$ is the DFT-calculated DOS of graphene taken from Ref.~\cite{Joucken2019}  and shifted in energy according to the level of graphene doping ($\sim0.3$~eV \cite{Parzefall2019}). By taking the derivative of the calculated current as a function of voltage, we obtain the $\mathrm{d}I/\mathrm{d}V$ curve shown in Fig.~\ref{FigS8} (red dashed curve). We scale the calculation to fit the low-voltage region of the measured $\mathrm{d}I/\mathrm{d}V$ curve. This is the region in which contributions of F-N or thermionic effects should be limited. Our comparison reveals that the calculated $\mathrm{d}I/\mathrm{d}V$ curve closely follows our measurement up to 2~V, with significant deviations only above 2.5~V. These high-voltage contributions, which may be the result of F-N or thermionic effects, seem to become relevant at voltages outside the range in which we explore exciton energies ($1.5-2.25$~V).

\subsection{Excitonic light emission from a WSe$_2$ device with a thinner hBN tunnel barrier}\label{sec:WSe2_thin_IV}
Here, we investigate another WSe$_2$ device that has a thinner hBN tunnel barrier ($2-3$~nm). The measured $I \textendash V$ profile is plotted in Fig.~\ref{FigS4}, showing a much larger tunnel current than the devices with thicker tunnel barriers discussed above. In the $\mathrm{d}I/\mathrm{d}V$ spectrum, the phonon-related feature shows up near zero bias voltage~\cite{Zhang2008}, but no clear exciton signatures can be observed. We attribute this result to the fact that the background arising from the larger total current overshadows any signature from the much smaller exciton-assisted tunneling current~\cite{Chandni2015}. However, exciton-assisted light emission is observed from this device, as shown in Fig.~\ref{FigS5}. The peak at \SI{\sim1.63}{\electronvolt} in the electroluminescence spectra matches well the WSe$_2$ A exciton ground state~\cite{He2014}, thus pointing to a coupling of the tunneling electrons to excitons in the WSe$_2$ monolayer. The bump feature around \SI{1.4}{\electronvolt} in the spectra can be attributed to the broad light emission resulting from the coupling of tunneling electrons to optical modes and surface-plasmon polaritons in the device~\cite{Parzefall2019}.

\subsection{Theory of inelastic electron tunneling between graphene and gold}
\label{theorysection}

\subsubsection{Preliminary considerations regarding energy--momentum conservation}
\begin{figure}
\begin{centering} \includegraphics[width=0.4\textwidth]{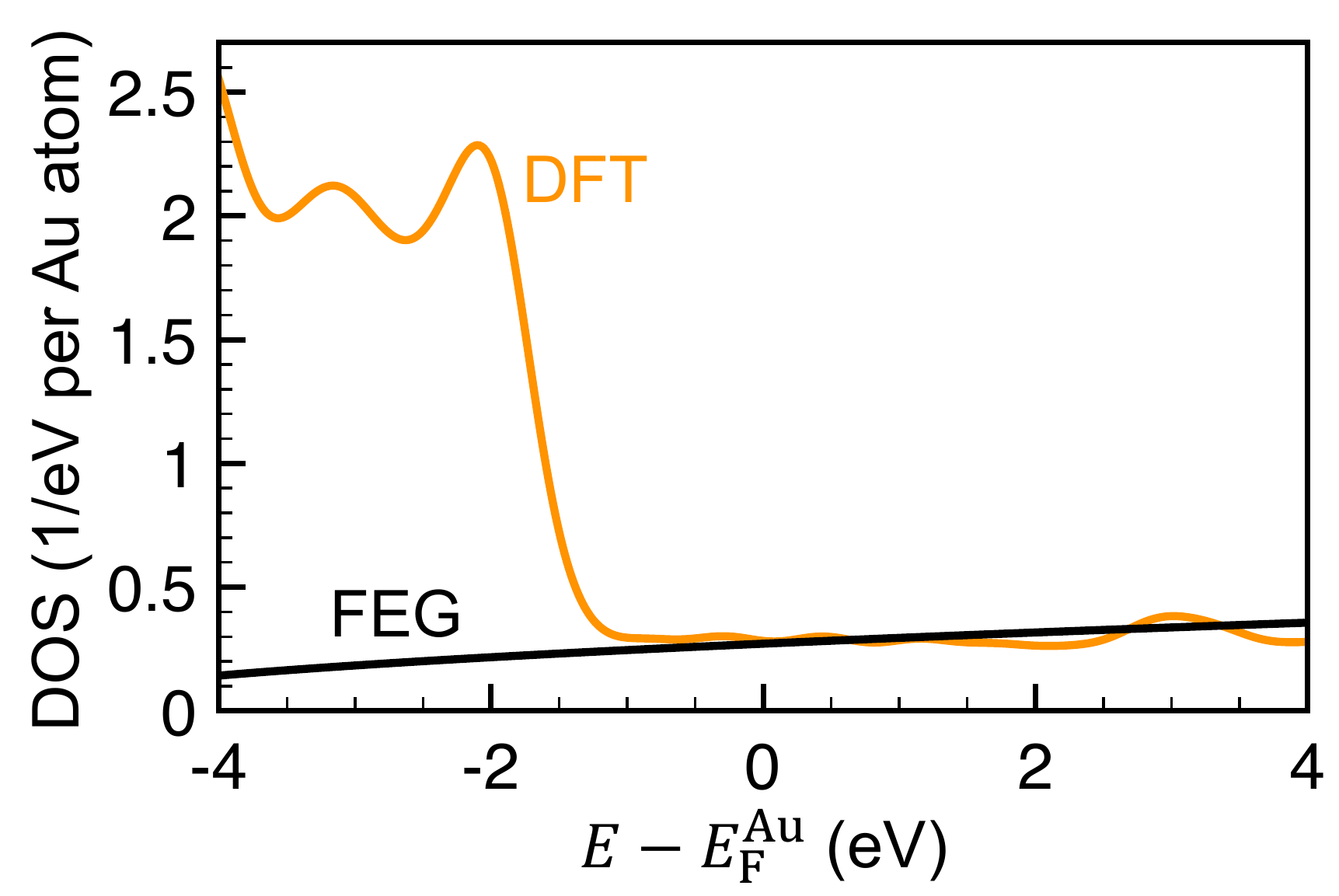} \par\end{centering}
\caption{{\bf Density of states (DOS) of gold.} We compare density-functional-theory (DFT) calculations (including a Gaussian energy smearing of full width at half maximum $\sigma=0.3$~eV) with the free-electron gas (FEG) model. In the latter, the DOS is $\rho_E=(m^*_{\rm Au}/\pi^2\hbar^3)\sqrt{2m^*_{\rm Au}E}$ and the Fermi energy is $\EFAu=(12\pi^2)^{2/3}(\hbar^2/2m^*_{\rm Au}a^2)\approx5.52$~eV, where $a=4.08$~{\AA} is the lattice constant and $m^*_{\rm Au}\approx\me$ is the effective mass.}
\label{FigS9}
\end{figure}

We first note that the electronic density of states (DOS) of gold near the Fermi level is dominated by the 6sp-band, which is well-described by the free-electron gas (FEG) model up to energies $\sim3$~eV above the Fermi level (see Fig.~\ref{FigS9}). The contribution of the 5d band shows up as a sharp increase in the DOS below $\sim-2$~eV, which explains the sharp jump in the tunneling current observed in the experiment at negative biases exceeding that value. For the remainder of this discussion, we focus however on the smooth region dominated by the sp band, which we describe in the FEG model.

\begin{figure*}
\begin{centering} \includegraphics[width=0.9\textwidth]{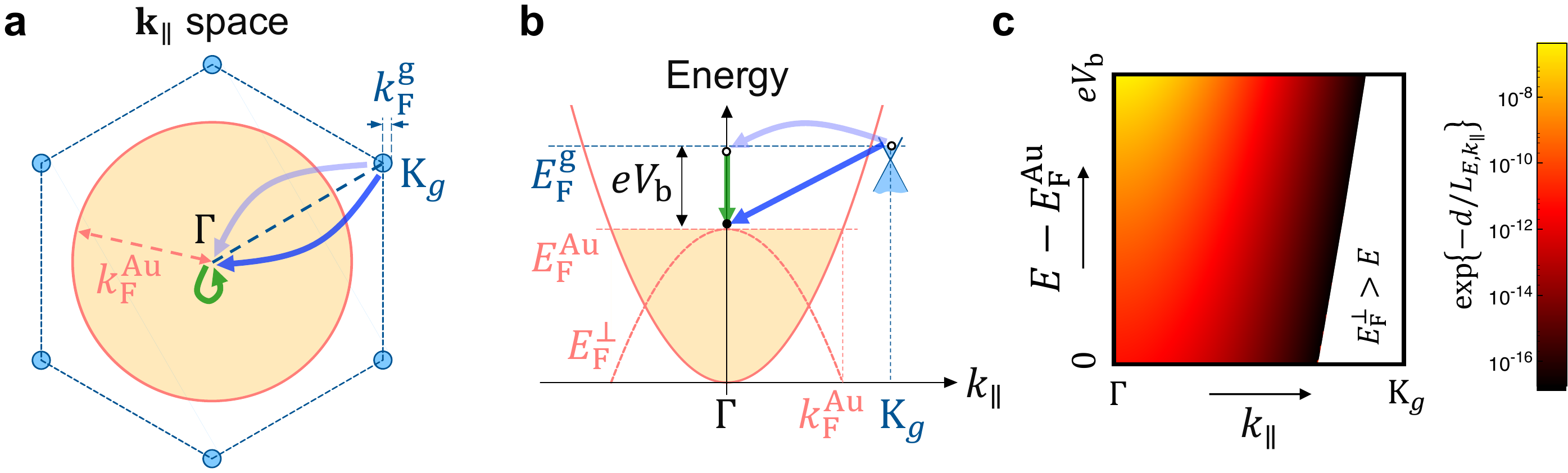} \par\end{centering}
\caption{{\bf Electron tunneling pathways.} ({\bf a}) On-scale representation of the in-plane electron momentum distribution in graphene (light blue) and gold (orange), including their respective Fermi wave vectors. Graphene is taken to be negatively doped to a Fermi energy of 0.5~eV relative to the Dirac points. ({\bf b}) Energy--parallel-momentum diagram corresponding to the configuration in (\textbf{a}) for a graphene--gold bias energy $e\VVb=3$~eV. The downward dashed parabola shows the out-of-plane electron energy in gold at the Fermi level, $E_{\rm F}^\perp=\EFAu-\hbar^2\kpar^2/2m^*_{\rm Au}$, as a function of parallel wave vector $\kpar$. ({\bf c}) Exponential attenuation of gold conduction electrons across the interface with an hBN film (thickness $d=3$~nm) as a function of energy and in-plane wave vector, as determined by the spill-out distance $L_{E,\kpar}\approx\hbar\,[2m^*_{\rm hBN}(\EFAu+E_g/2-E+\hbar^2\kpar^2/2m^*_{\rm Au})]^{-1/2}$, estimated from a one-dimensional potential barrier determined by the hBN gap at the ${\rm \Gamma}$ point~\cite{XC91} $E_g\approx8.9$~eV and the effective mass along the out-of-plane ${\rm \Gamma}$A direction $m^*_{\rm hBN}\approx0.63\,\me$. In (\textbf{a}) and (\textbf{b}), we consider two possible channels for graphene-to-gold tunneling: (1) one-step direct inelastic transitions to near-threshold gold states around the surface projected ${\rm \Gamma}$ point (coinciding with the ${\rm \Gamma}$ point of graphene) assisted by the creation of indirect excitons in a neighboring TMD monolayer (dark blue arrows); and (2) two-step quasi-elastic tunneling from graphene to gold states near the surface-projected ${\rm \Gamma}$ point assisted by large-momentum phonon excitation (purple arrows), followed by quasi-vertical inelastic decay to states in the vicinity of the Fermi level of gold mediated by direct excitons (green arrows).}
\label{FigS10}
\end{figure*}

In a simplified picture, the surface-momentum-projected sp band of gold defines a circle of radius given by the Fermi wave vector $\kFAu$, which we compare to the conduction band of graphene in Fig.~\ref{FigS10}a. This plot shows that graphene electrons near its Fermi level are situated far from the ${\rm \Gamma}$ point, where gold conduction electrons are expected to extend further beyond the metal surface. A band projection on energy and parallel momentum along the ${\rm \Gamma}$K$_g$ direction of n-type doped graphene (Fig.~\ref{FigS10}b) also illustrates how the electrons of this material need to bridge a large momentum mismatch to tunnel to gold states near the ${\rm \Gamma}$ point. This mismatch is smaller for transitions to gold states with high parallel wave vectors, but then, a huge jump in out-of-plane energy is encountered. In this respect, the landscape of the conduction band bottom (CBB) in the hexagonal boron nitride (hBN) layer~\cite{XC91} separating graphene from gold in our samples could play a role, whose analysis would require a realistic description of the electronic states in the system beyond the scope of the present theoretical model. However, a quantitative argument can be constructed upon examination of the exponential attenuation factor displayed by the wave functions of gold conduction electrons across a distance of 3~nm outside the metal in the interface with hBN. The corresponding plot (Fig.~\ref{FigS10}c) clearly indicates that tunneling should be exponentially attenuated for energies and wave vectors deviating from the conditions of both quasi-elastic tunneling ($E-\EFAu\lesssim e\VVb=3$~eV in the figure) and electron momentum in the surface-projected gold ${\rm \Gamma}$ point.

Based on these arguments, we expect tunneling in the TMD/graphene/hBN/Au sandwich for positive bias of the graphene (as considered in the experiment) to involve a momentum transfer to graphene electrons such that they can transition to gold states near the ${\rm \Gamma}$ point (since graphene does not have any states near such point). The required wave vector transfer, $\kpar\sim{\rm \Gamma}$K$_g=4\pi/(3\sqrt{3}a_{\rm CC})\approx17.02$~nm$^{-1}$, dictated by the graphene lattice and the carbon-carbon bond distance $a_{\rm CC}\approx0.1421$~nm, is too high to be supplied by the creation of low-momentum optical modes such as direct excitons. For example, the nonlocal conductivities of the four studied TMDs show a strong attenuation for $\kpar\gtrsim0.5$~nm$^{-1}$ (see Fig.~\ref{FigS18} in Sec.~\ref{TMDsigma}). An additional source of momentum is thus required, for which we postulate two possible channels that can assist graphene-to-gold tunneling:

\begin{itemize}
\item {\it Direct one-step transitions assisted by indirect TMD excitons.} Indirect excitons in the TMD monolayer can be formed by involving an electron and a hole of very different wave vectors. Upon examination of the possible indirect excitons connecting flat points in the band structure of TMD monolayers (i.e., those capable of producing Van Hove singularities), we find KQ$_{\rm TMD}$ excitons (consisting of one hole at the K point and one electron at the Q point of the TMD) as good candidates to contribute to the observed tunneling current because their separation in wave vector is close to the ${\rm \Gamma}$K$_g$ distance in graphene (see Table\ \ref{TableS2}). In addition, our calculated energies for KQ$_{\rm TMD}$ excitons are in relatively good agreement with the observed peaks in our conductance measurements, as shown in Table~\ref{TableS2}, where we also include the corresponding KQ$_{\rm TMD}$ gap energies (again not too far from the observed peaks), as well as information on the A and B excitons. For reference, we show the first Brillouin zone (1BZ)  of MoSe$_2$ (pink) in the following scheme, superimposed on the 1BZ of graphene and the surface-projected gold Fermi surface taken from Fig.~\ref{FigS10}a:
\begin{center} \includegraphics[width=0.2\textwidth]{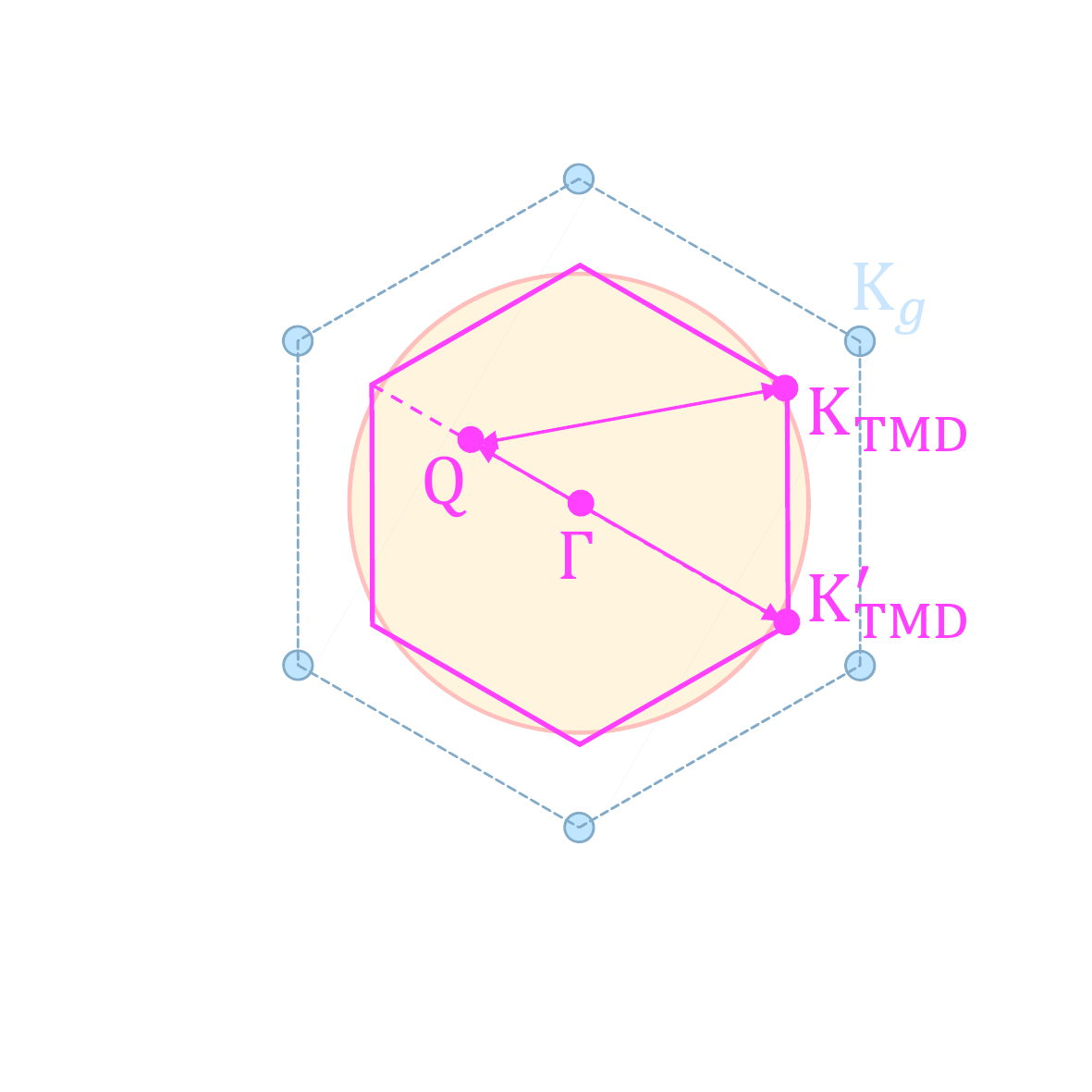} \end{center}
In Table~\ref{TableS2}, we show values of the KQ$_{\rm TMD}$ and K$^\prime$Q$_{\rm TMD}$ distances, indicated by double arrows in this scheme. Although the relative azimuthal orientation between TMD and graphene lattices is not well defined in our samples, the set of KQ$_{\rm TMD}$ distances spanned by the combinations of inequivalent K$_g$ and Q points is dense, and some of them lie close to the required graphene ${\rm \Gamma}$K$_g$ momentum transfer, particularly considering that KQ$_{\rm TMD}$ excitons span a wave vector distribution of finite extension in the $\sim1$~nm$^{-1}$ range (see Fig.~\ref{FigS19} in Sec.~\ref{TMDsigma}).

\item {\it Two-step processes assisted by phonon excitation and direct TMD excitons.} Phonon excitation can bridge the graphene ${\rm \Gamma}$K$_g$ momentum mismatch, as both hBN and graphene have relatively close lattice parameters, and in addition, the electronic and phononic Brillouin zones are identical within each of them. The involved phonon energies are $\sim70$~meV and $\sim120-150$~meV~\cite{MMD07,SSB19}, much smaller than the exciton energies considered in this work. Apart from indirect excitons (see previous point), we cannot find a single source of momentum and energy that can simultaneously break the energy-momentum mismatch between graphene states and gold conduction electrons near the Fermi level at the ${\rm \Gamma}$ point. However, a two-step process can be invoked, involving a first quasi-elastic transition from graphene to gold near the ${\rm \Gamma}$ point, assisted by the excitation of a high-momentum phonon in graphene (or hBN, although proximity favors graphene phonons); and a subsequent inelastic decay from higher- to lower-energy states in gold (both of them near the ${\rm \Gamma}$ point), taking place by investing the transition energy in the creation of an exciton in the TMD monolayer placed on top of graphene. Inverting the order of these two steps would involve inelastic decay from graphene to an intermediate unoccupied gold state of large parallel momentum, and thus, this possibility should produce a negligible contribution because of the marginal spill out of such intermediate state near the Fermi level of gold (see Fig.~\ref{FigS10}c). For negative bias, with p-doped graphene, electrons could still tunnel from gold to graphene following a two-step process, but the intermediate and final states (both in graphene) have a small local DOS compared with gold, and thus, two-step inelastic tunneling should occur with lower probability in this direction.
\end{itemize}

\begin{table*}
\begin{tabular}{c|c|c|c|c|c|c|c|c|c} \hline
\textbf{material} \quad &
$\begin{matrix}\text{\bf A, B}\\ \text{\bf excitons}\\\text{\bf (PL)}\end{matrix}$ &
$\begin{matrix}\text{\bf KQ gap}\\ \text{\bf (experiment)}\end{matrix}$ &
$\begin{matrix}\text{\bf conductance peak}\\ \text{\bf (our experiment)}\end{matrix}$ &
$\begin{matrix}\text{\bf direct excitons}\\ \text{\bf (experiment)}\end{matrix}$ &
$\begin{matrix}\text{\bf A, B excitons} \\ \text{\bf (our theory)}\end{matrix}$ &
$\begin{matrix}\text{\bf ${\rm\bf \Gamma}$K}\\ \text{\bf (nm$^{-1}$)}\end{matrix}$ &
$\begin{matrix}\text{\bf ${\rm\bf \Gamma}$Q}\\ \text{\bf (nm$^{-1}$)}\end{matrix}$ &
$\begin{matrix}\text{\bf KQ}\\ \text{\bf (nm$^{-1}$)}\end{matrix}$ &
$\begin{matrix}\text{\bf K$^\prime$Q}\\ \text{\bf (nm$^{-1}$)}\end{matrix}$ \\
\hline
\textbf{WS$_2$} &1.98, - & 2.00 \cite{WPM21} & 2.07 & 2.00, 2.40 \cite{LCZ14} & 1.92, 2.35 & 13.13 & 6.94 & 17.65 & 20.07 \\
&& 2.05 \cite{ZCJ15} && 2.09, 2.23 \cite{Chernikov2014} &&&&& \\
&&&& 2.06, 2.20 \cite{Goryca2019} &&&&& \\
&&&& \quad\quad\, 2.08 \cite{Hanbicki2015} &&&&& \\ \hline

\textbf{MoS$_2$} &1.87, 2.0 & 2.12 \cite{ZCJ15}& 2.10 & 1.87, 2.03 \cite{LCZ14} & 2.05, 2.19 & 13.13 & 6.46 & 17.28 & 19.58 \\
&&&& 1.92, 2.08 \cite{Robert2018} &&&&& \\
&&&& 1.94, 2.11 \cite{Goryca2019} &&&&& \\ \hline

\textbf{WSe$_2$}&1.64,-  & 1.69 \cite{BEC20} & 1.70 & 1.65, 2.05 \cite{LCZ14} & 1.50, 1.97 & 12.48 & 6.94 & 17.05 & 19.43 \\
&& 2.09\;\; \cite{ZCJ15} && 1.65, 2.08\;\; \cite{He2014} &&&&& \\
&&&& 1.72, 1.85 \cite{Stier2018} &&&&& \\ \hline

\textbf{MoSe$_2$} &1.57, 1.72 & 1.84 \cite{CYJ17_2} & 1.90 & 1.55, 1.75 \cite{LCZ14} & 1.80, 2.00 & 12.62 & 6.60 & 16.92 & 19.23 \\
&& 2.34\;\; \cite{ZCJ15} && 1.64, 1.79 \cite{Han2018} &&&&& \\
&&&& 1.64, 1.79 \cite{Horng2018} &&&&& \\
&&&& 1.64, 1.81 \cite{Goryca2019} &&&&& \\ \hline
\end{tabular}
\caption{{\bf Direct and indirect excitons in TMD monolayers.} We present a compilation of energies for the measured KQ$_{\rm TMD}$ gaps (from photoluminescence \cite{CYJ17_2,BEC20,WPM21} and $I \textendash V$ \cite{ZCJ15} studies), the conductance peaks in our experiment, luminescence measurements from different sources, and our calculated A and B excitons. We show data for the four different TMD monolayer materials under consideration. All energies are in eV. We also provide wave vector distances between singular points of the 1BZ (four rightmost columns). The Q point is defined by an intermediate conduction band minimum along the ${\rm \Gamma}$K$_{\rm TMD}$ line. We consider two of the three distances between K and Q points in the 1BZ of each TMD, whose values lie close to the ${\rm \Gamma}$K$_g$ distance in graphene (17.02\,nm$^{-1}$).}
\label{TableS2}
\end{table*}

We conclude these preliminary considerations by estimating the dependence on the graphene--gold bias voltage $\VVb$ of the shift in the graphene Fermi energy $\EFg$ relative to the Dirac point energy $E_{\rm D}$ (i.e., relative to the graphene Fermi energy at zero bias). Throughout this work, we refer all energies to the CBB of gold, so we write the noted shift as $\DEFg=\EFg-E_{\rm D}$. In addition, we assume hBN to fill the regions both below and above the TMD+graphene heterostructure, and ignore the effect of the TMD monolayer on graphene doping. A variational minimization of the total energy in the system shows that the graphene doping density $n$ (electrons per unit area) is determined by the capacitor formed with gold as $n=(\epsilon_{\rm hBN}/4\pi e\,d)(\VVb-\DEFg/e)$, where $\epsilon_{\rm hBN}\approx3.76$ is the out-of-plane DC permittivity of hBN, $d$ is the hBN spacing layer thickness separating graphene from gold, and the effective bias voltage needs to be reduced by $\DEFg/e$ to account for the change in energy needed to add/subtract every new electron to/from graphene as its Fermi energy is varied. Using the additional relation \cite{CGP09} $\DEFg=\hbar\vFg\sqrt{\pi n}$ for positive bias, where $\vFg\approx10^6$~m/s is the graphene Fermi velocity, we finally obtain the Fermi energy shift as
\begin{align}
\DEFg=\frac{E_0}{2}\bigg(\sqrt{1+\frac{4e\VVb}{E_0}}-1\bigg)
\label{defg}
\end{align}
with $E_0=\epsilon_{\rm hBN}(\hbar\vFg/2e)^2/d$. From this expression, we have, for example, $\DEFg\approx0.49$~eV for $d=3$~nm and $e\VVb=3$~eV ($\gg E_0\approx0.09$~eV). Incidentally, this value is slightly shifted with respect to the large-bias limit $\DEFg\approx\sqrt{e\VVb E_0}\approx0.53$~eV.

\begin{figure}
\begin{centering} \includegraphics[width=0.42\textwidth]{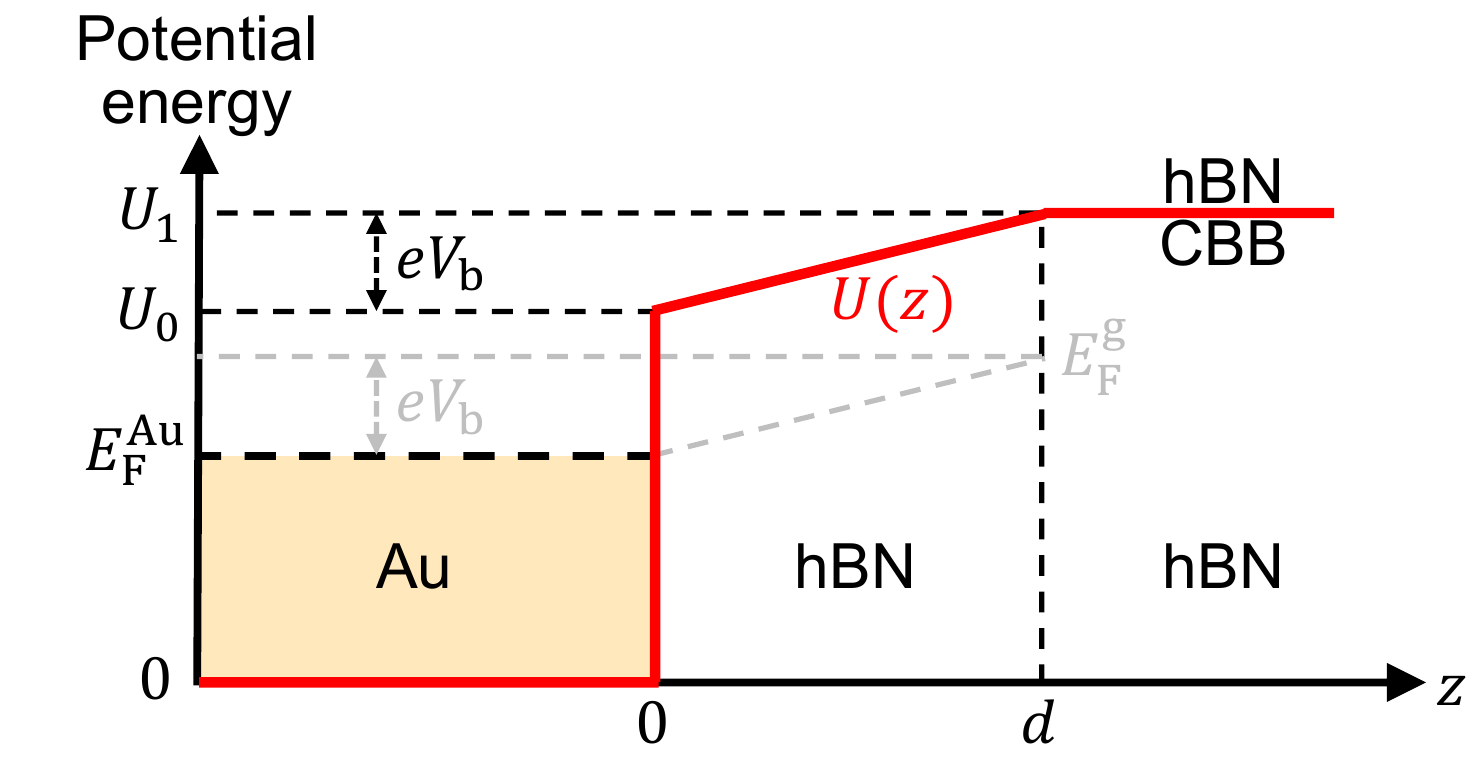} \par\end{centering}
\caption{{\bf Electron potential landscape.} We model the out-of-plane electron dynamics through a one-dimensional potential energy $U(z)$, here sketched for a graphene--gold positive bias potential energy $e\VVb$ and defined by the conduction band bottom (CBB) of gold at $z<0$, as well as the potential energies $U_0$ and $U_1$ of the CBB of hBN at $z=0$ and $z=d$ (the graphene--gold spacing), respectively. The graphene Fermi level $\EFg$ is separated from the gold Fermi level $\EFAu$ by the bias potential energy $\EFg-\EFAu=e\VVb$. We set $U_0-\EFAu\approx4.45$~eV as half of the hBN gap at the ${\rm \Gamma}$ point~\cite{XC91}.}
\label{FigS11}
\end{figure}

\subsubsection{Vertical potential landscape, graphene electrons, and electron Green function}
\label{Greenfunction}

We assume the tunneling electron to evolve in the potential energy landscape sketched in Fig.~\ref{FigS11}, as described by the Hamiltonian
\begin{align}
\mathcal{H}_0=-\frac{\hbar^2\nabla^2}{2\me}+U(z),
\nonumber
\end{align}
where a uniform isotropic electron mass is assumed for simplicity. We note that the strong variation of the electron mass and energy gap within the in-plane 1BZ of hBN~\cite{XC91} can produce substantial corrections to this model, which deserve a future investigation including a more detailed description of the electron states involved in the tunneling process. In addition, we approximate the wave function of each initial graphene electron $i$ as a separable state
\[\varphi_i(\rb)=\varphi_i(z)\ee^{\ii(\Kb_g+\Qb_i)\cdot\Rb}/\sqrt{A}\]
of energy $\hbar\varepsilon_i=E_{\rm D}+s_i\hbar\vFg Q_i$ [with the Dirac points at an energy $E_{\rm D}=\EFg-\DEFg$, see Eq.~(\ref{defg})] relative to the CBB of gold and wave vector $\Qb_i$ relative to one of the K$_g$ points of momentum $\Kb_g$ (see Figs.~\ref{FigS10} and \ref{FigS11}). Here, $s_i=\pm1$ refers to the upper ($+$) and lower ($-$) graphene Dirac cones, $\Kb_g$ and $\Qb_i$ and both 2D vectors, the in-plane position coordinates are $\Rb=(x,y)$, and we introduce the graphene area $A$ for normalization. We also note that $\EFg$ is referred to the CBB of gold. Incidentally, the two inequivalent Dirac cones within the 1BZ should contribute equally to the tunneling current, and therefore, we absorb valley and spin degeneracies in an overall factor of 4.

The evolution of the tunneling electron in the potential landscape of Fig.~\ref{FigS11} can be conveniently described in terms of the three-dimensional (3D) electron Green function $G_0(\rb,\rb',\varepsilon)$ defined by $(\mathcal{H}_0-\hbar\varepsilon)G_0(\rb,\rb',\varepsilon)=-\delta(\rb-\rb')$. Due to the in-plane translational symmetry of the system, the Green function depends on $\Rb$ and $\Rb'$ only through the difference $\Rb-\Rb'$, so it can be written as
\begin{align}
G_0(\rb,\rb',\varepsilon)=\int\frac{d^2\Qb}{(2\pi)^2}\,\ee^{\ii\Qb\cdot(\Rb-\Rb')}\,G_0(z,z',\varepsilon-\hbar Q^2/2\me)
\label{G0G0}
\end{align}
in terms of the one-dimensional Green function $G_0(z,z',\varepsilon)$ satisfying $(\mathcal{H}_0-\hbar\varepsilon)G_0(z,z',\varepsilon)=-\delta(z-z')$, which we correct in Eq.~(\ref{G0G0}) by subtracting the in-plane energy $\hbar^2Q^2/2\me$ from $\hbar\varepsilon$.

In the calculations presented below, we only need the Green function evaluated at $G_0(0,z,\varepsilon)$ and $G_0(z,d,\varepsilon)$, as well as the derivative of the former with respect to the first argument. It is then useful to point out that the spatial arguments of these functions are interchangeable in virtue of the reciprocity property $G_0(z,z',\varepsilon)=G_0(z',z,\varepsilon)$. In addition, we note that the required functions are solutions of $(\mathcal{H}_0-\hbar\varepsilon)G_0(z,z',\varepsilon)=-\delta(z-z')$ for $z'=0$ and $z'=d$. This allows us to express them in terms of analytical functions within each of the three distinct regions of the potential $U(z)$ in Fig.~\ref{FigS11}:
\begin{align}
G_0(z,z',\varepsilon)=\left\{\begin{matrix*}[l]
A(z')\,\ee^{-\ii k z}, & &\quad \text{$z<0$}, \\ \\
B(z')\,\psi_1(z)+C(z')\,\psi_2(z), & &\quad \text{$0\le z\le d$}, \\ \\
D(z')\,\ee^{-\kappa(z-d)}, & &\quad \text{$d<z$},
\end{matrix*}\right.
\label{G0explicit}
\end{align}
where we consider outgoing waves with $k=\sqrt{2\me\varepsilon/\hbar}$ and $\kappa=\sqrt{2\me(U_1/\hbar-\varepsilon)/\hbar}$, which are suitable solutions decaying at $|z|\to\infty$ for a source placed at $z'\in[0,d]$. The functions in the intermediate region can be written as
\begin{align}
&\psi_1(z)={\rm Ai}(\theta), \nonumber \\
&\psi_2(z)={\rm Bi}(\theta)
\nonumber
\end{align}
in terms of the Airy functions Ai and Bi~\cite{AS1972} with an argument $\theta=(2\me e\VVb d^2/\hbar^2)^{1/3}\big[z/d+(U_0-\hbar\varepsilon)/e\VVb\big]$. Finally, the coefficients in Eq.~(\ref{G0explicit}) are obtained from the conditions
\begin{subequations}
\label{cond}
\begin{align}
&A-B\psi_1(0)-C\psi_2(0)=0, \label{cond1} \\
&D-B\psi_1(d)-C\psi_2(d)=0, \label{cond2} \\
&\ii k A+B\psi'_1(0)+C\psi'_2(0)=n_0\times2\me/\hbar^2, \label{cond3} \\
&\kappa D+B\psi'_1(d)+C\psi'_2(d)=n_d\times2\me/\hbar^2. \label{cond4}
\end{align}
\end{subequations}
In particular, Eqs.~(\ref{cond1}) and (\ref{cond2}) guarantee continuity at $z=0$ and $z=d$, respectively, whereas Eqs.~(\ref{cond3}) and (\ref{cond4}) relate to the jump in the derivative ($2\me/\hbar^2$) produced by the inhomogeneous term $-\delta(z-z')$. More precisely, we need to set $n_0=1$ and $n_d=0$ for $z'=0$; and $n_0=0$ and $n_d=-1$ for $z'=d$. The required Green function values are then given by Eq.~(\ref{G0explicit}) with coefficients $A$, $B$, $C$, and $D$ determined by solving the $4\times4$ system of equations (\ref{cond}).

Likewise, we evaluate $\partial_{z'}G_0(z',z,\varepsilon)|_{z'=0^-}$ by writing a solution similar to Eq.~(\ref{G0explicit}),
\begin{align}
\partial_{z'} G_0(z',z,\varepsilon)|_{z'=0}=\left\{\begin{matrix*}[l]
A'\,\ee^{-\ii k z}, & & \text{$z<0$}, \\ \\
B'\,\psi_1(z)+C'\,\psi_2(z), & & \text{$0\le z\le d$}, \\ \\
D'\,\ee^{-\kappa(z-d)}, & & \text{$d<z$},
\end{matrix*}\right.
\nonumber
\end{align}
with coefficients determined by the equations
\begin{subequations}
\label{condbis}
\begin{align}
&A'-B'\psi_1(0)-C'\psi_2(0)=2\me/\hbar^2, \\
&D'-B'\psi_1(d)-C'\psi_2(d)=0, \\
&\ii k A'+B'\psi'_1(0)+C'\psi'_2(0)=0, \\
&\kappa D'+B'\psi'_1(d)+C'\psi'_2(d)=0.
\end{align}
\end{subequations}
This solution if obtained by first writing $G_0(z',z,\varepsilon)$ (with $z'<0$ regarded as a parameter) in terms of $z$-dependent outgoing waves within the $z<z'$ and $z>d$ regions, as well as Airy functions for $0\le z\le d$ and $\ee^{\pm\ii kz}$ waves for $z'\le z<0$. We then take the derivative of the resulting coefficients with respect to $z'$ and calculate the $z'\to0^-$ limit (i.e., approaching $z'=0$ from the left), which leads to Eqs.~(\ref{condbis}).

\subsubsection{Inelastic electron tunneling assisted by excitations in the heterostructure}
\label{generalJ}

The tunneling process can be assisted by the creation of an excitation in the heterostructure, which absorbs the energy and momentum change undergone by the transferred electron. In this section, we derive a general expression for the inelastic tunneling current that encapsulates such excitations $n$ of energies $\hbar\omega_n$ in the dielectric response function. Although we consider electron tunneling from graphene to gold in this work, the formalism presented in this section can be applied to any combination of two conductive materials separated by an insulating layer and subject to a relative bias. Starting from an electron in an initial graphene state $i$ of wave function $\varphi_i(\rb)$ and energy $\hbar\varepsilon_i$, the final electron wave function component of energy $\hbar(\varepsilon_i-\omega_n)$ associated with an additional excitation $n$ is given by
\begin{align}
\psi_{n,i}(\rb)=\int d^3\rb' \; G_0(\rb,\rb',\varepsilon_i-\omega_n)\bra{n} \,\mathcal{H}_1(\rb')\ket{0}{\varphi}_{i}(\rb')
\label{psinip}
\end{align}
within first-order perturbation in the interaction Hamiltonian $\mathcal{H}_1$ of matrix elements
\begin{align}
\bra{n}\mathcal{H}_1(\rb')\ket{0}=-e\int d^3\rb'\,\frac{\rho_{n0}(\rb')}{|\rb-\rb'|},
\label{H1}
\end{align}
sandwiched by the ground and excited states of the heterostructure, $\ket{0}$ and $\ket{n}$, respectively. Here, $\rho_{n0}(\rb)=\bra{n}\hat{\rho}(\rb)\ket{0}$ is the matrix element of the charge density operator $\hat{\rho}(\rb)=\sum_j q_j\,\delta(\rb-\rb_j)$, incorporating all electrons and nuclei of charges $q_j$ at positions $\rb_j$ in the system. The Hamiltonian $\mathcal{H}_1$ thus describes the Coulomb interaction between such charges and the tunneling electron in a completely general fashion. The details of the structure are additionally captured by the unperturbed Green function $G_0(\rb,\rb',\varepsilon)$, which remains general in this section, but we then specify it for the heterostructure under consideration by using the methods described in Sec.~\ref{Greenfunction}.

\begin{figure}
\begin{centering} \includegraphics[width=0.4\textwidth]{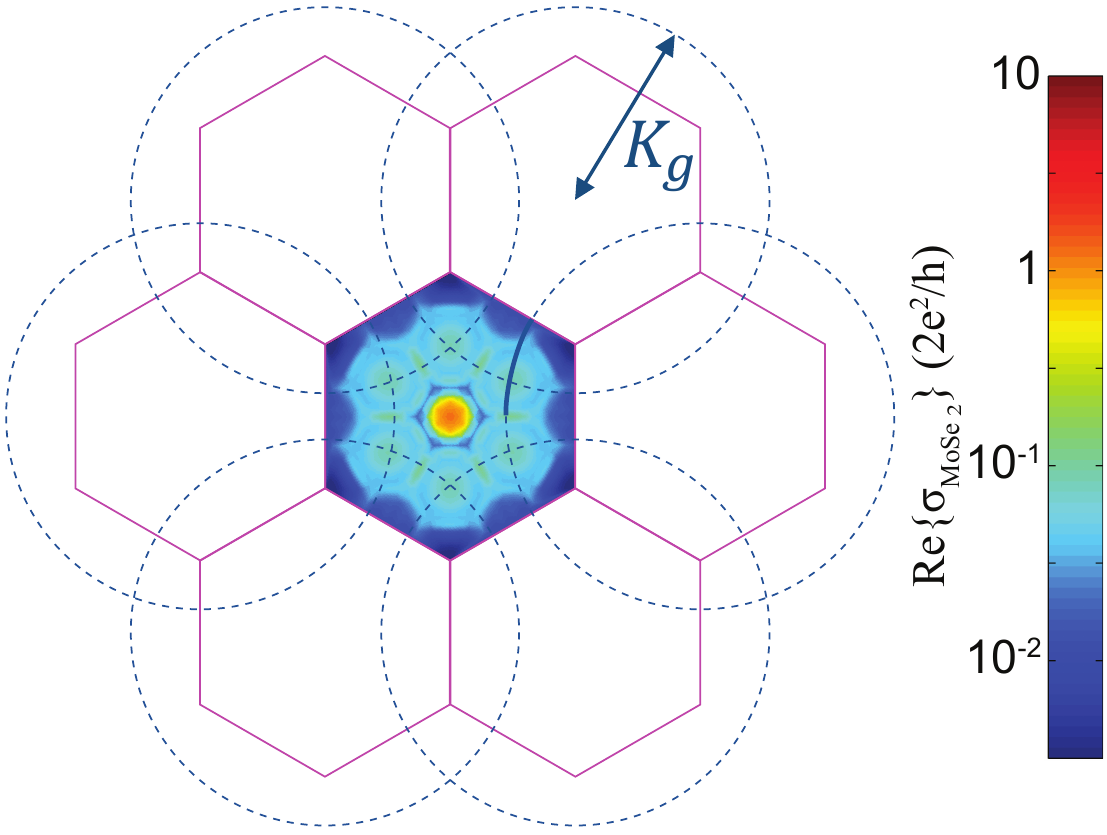} \par\end{centering}
\caption{{\bf Leading exciton contributions to the tunneling current.} We plot ${\rm Re}\{\sigma_{{\rm TMD},00}(\kparb,\omega)\}$ (i.e., the real part of the optical conductivity for $\Gb=\Gb'=0$, see Secs.~\ref{sectionWGG} and \ref{TMDsigma}) within the 1BZ of MoSe$_2$ at a photon energy $\hbar\omega=2.25$~eV, along with the six nearest-neighbor Brillouin zones (hexagons), each of them centered around a $\Gb$ vector. We also show the corresponding circles defined by the conditions $|\Gb-\kparb|=K_g$ (i.e., the ${\rm\Gamma}$K$_g$ distance in graphene).}
\label{FigS12}
\end{figure}

From the wave functions given by Eq.~(\ref{psinip}), we obtain the tunneling current density (charge moving downwards and entering the gold surface per unit time and unit area) as
\begin{widetext}
\begin{align}
J=\frac{e\hbar}{\me}\sum_{n,i}
f_T(\hbar\varepsilon_i)\,\big\{1-f_T[\hbar(\varepsilon_i-\omega_n)]\big\}
\;{\rm Im}\big\{\psi_{n,i}^*(\rb) \partial_z \psi_{n,i}(\rb)\big\}\big|_{z=0^-},
\label{Iphex1}
\end{align}
where we sum over all initial occupied states $i$ and material excitations $n$ that lead to final unoccupied gold states of energy $\hbar(\varepsilon_i-\omega_n)$, as imposed by using the Fermi--Dirac distribution $f_T$ at temperature $T$. The gold surface is placed at $z=0$, so we evaluate the current right below that plane at $z=0^-$ in Eq.~(\ref{Iphex1}). From Eqs.~(\ref{psinip})--(\ref{Iphex1}), we find
\begin{align}
&J=\frac{e^3\hbar}{\me}\sum_{n,i}
f_T(\hbar\varepsilon_i)\,\big\{1-f_T[\hbar(\varepsilon_i-\omega_n)]\big\}
\int d^3\rb'\int d^3\rb''\int d^3\rb_1\int d^3\rb_2 \label{Iphex2}\\
&{\rm Im}\Big\{
G_0^*(\rb,\rb',\varepsilon_i-\omega_n)
\,\partial_z G_0(\rb,\rb'',\varepsilon_i-\omega_n)\big|_{z=0^-}
\;{\varphi}^*_{i}(\rb'){\varphi}_{i}(\rb'')
\,\frac{\rho_{0n}(\rb_1)\rho_{n0}(\rb_2)}{|\rb'-\rb_1||\rb''-\rb_2|}
\Big\}.
\nonumber
\end{align}
Following a method similar to the derivation of the electron energy-loss probability in electron microscopy (see Ref.~\cite{paper149}), this expression can be written in terms of a response function by exploiting the identities \cite{PN1966}
\begin{subequations}
\label{chiW}
\begin{align}
&{\rm Im}\big\{\chi(\rb,\rb',\omega)\big\}=-\frac{\pi}{\hbar}\sum_n
\rho_{0n}(\rb)\rho_{n0}(\rb')\,\delta(\omega_n-\omega), \\
&W^{\rm ind}(\rb,\rb',\omega)=\int d^3\rb_1\int d^3\rb_2
\,\frac{1}{|\rb-\rb_1|}\,\chi(\rb_1,\rb_2,\omega)
\,\frac{1}{|\rb'-\rb_2|},
\end{align}
\end{subequations}
where $\chi(\rb,\rb',\omega)$ is the nonlocal susceptibility and $W^{\rm ind}(\rb,\rb',\omega)$ is the induced part of the screened interaction. The latter is defined as the potential created at a position $\rb$ by a time-dependent unit charge of magnitude $\ee^{-\ii\omega t}$ placed at $\rb'$ (see Secs.~\ref{screening} and \ref{sectionWGG} for explicit calculations of this quantity in the planar heterostructures under consideration). Using Eqs.~(\ref{chiW}) to manipulate Eq.~(\ref{Iphex2}), we obtain
\begin{align}
J=&\frac{e^3\hbar^2}{\pi\me}\sum_{i}\int_0^\infty\!\!\! d\omega
\;f_T(\hbar\varepsilon_i)\,\big\{1-f_T[\hbar(\varepsilon_i-\omega)]\big\}
\int d^3\rb'\int d^3\rb'' \;{\rm Im}\big\{-W(\rb',\rb'',\omega)\big\} \label{Iphex3}\\
&\times{\rm Im}\Big\{
G_0^*(\rb,\rb',\varepsilon_i-\omega)
\,\partial_z G_0(\rb,\rb'',\varepsilon_i-\omega)\big|_{z=0^-}
\;{\varphi}^*_{i}(\rb'){\varphi}_{i}(\rb'')
\Big\},
\nonumber
\end{align}
where we have substituted the induced part of the screened interaction by the full interaction because $W(\rb,\rb',\omega)-W^{\rm ind}(\rb,\rb',\omega)=1/|\rb-\rb'|$ is a real function.

\subsubsection{Indirect exciton-assisted one-step tunneling}
\label{currentindirect}

We now apply Eq.~(\ref{Iphex3}) to discuss one-step transitions from graphene states $i$ assisted by the creation of indirect excitons in the TMD layer. In what follows, the sum over $i$ is understood to be restricted to occupied graphene states and we set $T=0$, such that $1-f_T[\hbar(\varepsilon_i-\omega)]=\Theta[\hbar(\varepsilon_i-\omega)-\EFAu]$ in expressed in terms of the step function $\Theta$ and the gold Fermi energy $\EFAu$. Accordingly, we rewrite Eq.~(\ref{Iphex3}) as
\begin{align}
J^{\rm ind-ex}=&\frac{e^3\hbar^2}{\pi\me}\sum_{i}\int_0^\infty\!\!\! d\omega
\;\Theta\big[\hbar(\varepsilon_i-\omega)-\EFAu\big]
\int d^3\rb'\int d^3\rb'' \;{\rm Im}\big\{-W(\rb',\rb'',\omega)\big\} \label{Iphex1s1}\\
&\times{\rm Im}\Big\{
G_0^*(\rb,\rb',\varepsilon_i-\omega)
\,\partial_z G_0(\rb,\rb'',\varepsilon_i-\omega)\big|_{z=0^-}
\;{\varphi}^*_{i}(\rb'){\varphi}_{i}(\rb'')
\Big\}.
\nonumber
\end{align}
Momentum mismatch is then considered to be bridged by large wave vectors $\kparb$ in the screened interaction, contributed by indirect excitons in the TMD layer.

Assuming isotropy and translational invariance in the in-plane response, one can write the screened interaction as
\begin{align}
W(\rb,\rb',\omega)=\int \frac{d^2\kparb}{(2\pi)^2}\,\ee^{\ii\kparb\cdot(\Rb-\Rb')}\,W(\kpar,z,z',\omega)
\label{Wkpar}
\end{align}
in terms of momentum components $W(\kpar,z,z',\omega)$. However, for the large values of $\kparb$ under consideration, the in-plane atomic lattice can play a substantial role, imprinting a lattice periodicity on the screened interaction, which can be written as
\begin{align}
W(\rb,\rb',\omega)=\sum_{\Gb\Gb'}\int_{\rm 1BZ} \frac{d^2\kparb}{(2\pi)^2}\,\ee^{\ii(\kparb+\Gb)\cdot\Rb}\ee^{-\ii(\kparb+\Gb')\cdot\Rb'}\,W_{\Gb\Gb'}(\kparb,z,z',\omega).
\label{WkparGG}
\end{align}
This expression, which generalizes Eq.~(\ref{Wkpar}) by considering $W_{\Gb,\Gb'}(\kparb,z,z',\omega)$ components labeled by TMD reciprocal lattice vectors $\Gb$ and $\Gb'$ and restricting the $\kparb$ integral to the 1BZ, is general to describe the linear optical response of a TMD layer including atomic periodicity. In our analysis, we neglect lattice contributions to the Green function as a less relevant effect than in the excitonic optical response of the TMD material.

Upon insertion of Eqs.~(\ref{G0G0}) and (\ref{WkparGG}) into Eq.~(\ref{Iphex1s1}), noticing the plane-wave dependence of the wave functions on in-plane coordinates, we can readily carry out the integrals over $\Rb'$ and $\Rb''$. Further averaging the current over in-plane positions $\Rb$, we obtain
\begin{align}
J^{\rm ind-ex}=&\frac{1}{A}\,\frac{e^3\hbar^2}{\pi\me}\sum_{i}\sum_\Gb \int_0^\infty\!\!\! d\omega
\;\Theta\big[\hbar(\varepsilon_i-\omega)-\EFAu\big] \label{Iphex1s2}\\
&\times\int_{\rm 1BZ}\frac{d^2\kparb}{(2\pi)^2}\int dz'\int dz'' \;{\rm Im}\big\{-W_{\Gb\Gb}(\kparb,z',z'',\omega)\big\} \nonumber\\
&\times\,{\rm Im}\Big\{
G_0^*(0,z',\varepsilon_i-\omega-\hbar|\Qb_i-\kparb+\Gb-\Kb_g|^2/2\me) \nonumber\\
&\quad\;\;\;\times\partial_z G_0(z,z'',\varepsilon_i-\omega-\hbar|\Qb_i-\kparb+\Gb-\Kb_g|^2/2\me)\big|_{z=0^-} \;\;{\varphi}^*_{i}(z'){\varphi}_{i}(z'') \Big\} \nonumber,
\nonumber
\end{align}
where $\Kb_g$ is the wave vector at the graphene K$_g$ point and we find contributions from different Brillouin zones labeled by $\Gb$. We note that the average over $\Rb$ has eliminated nondiagonal $W_{\Gb\Gb'}$ components ($\Gb\neq\Gb'$) from this expression. Incidentally, the rightmost argument of the function $G_0(z,z',\varepsilon)$ in Eq.~(\ref{Iphex1s2}) refers to the out-of-plane electron energy $\hbar\varepsilon$ relative to the gold CBB, and consequently, the sums over $i$ and $\Gb$ need to be restricted to only yield positive values of such argument.

We proceed by using the fact that the initial graphene states $i$ are spatially localized near the $z=d$ plane [i.e., the out-of-plane spatial extension of $\varphi_i(z)$ ($\sim\pm1$~{\AA} \cite{paper295}) is small compared with the tunneling distance $d\sim2-3$~nm], and therefore, Eq.~(\ref{Iphex1s2}) can be approximated by setting $z'=z''=d$. In addition, the initial wave vectors $\Qb_i$ relative to the graphene K$_g$ point (see Sec.~\ref{Greenfunction}) are small compared with $\Kb_g$, and their energies $\hbar\varepsilon_i$ close to $E_{\rm F}^{\rm g}$, so we approximate $\varepsilon_i-\omega-\hbar|\Qb_i-\kparb+\Gb-\Kb_g|^2/2\me\approx E_{\rm F}^{\rm g}/\hbar-\omega-\hbar|\Gb-\Kb_g-\kparb|^2/2\me$. The sum over $i$ is then trivially yielding an overall constant. Putting these elements together, we transform Eq.~(\ref{Iphex1s2}) into
\begin{align}
J^{\rm ind-ex}\propto&\sum_\Gb \int_0^{e\VVb/\hbar}\!\!\! d\omega
\int_{\rm 1BZ}d^2\kparb \;{\rm Im}\big\{-W_{\Gb\Gb}(\kparb,d,d,\omega)\big\} \label{Jindexinter}\\
&\times\,{\rm Im}\Big\{
G_0^*(0,d,E_{\rm F}^{\rm g}/\hbar-\omega-\hbar|\Gb-\Kb_g-\kparb|^2/2\me) \nonumber\\
&\quad\;\;\;\times\partial_z G_0(z,d,E_{\rm F}^{\rm g}/\hbar-\omega-\hbar|\Gb-\Kb_g-\kparb|^2/2\me)\big|_{z=0^-} \Big\}.
\nonumber
\end{align}
Finally, we argue that $G_0(0,d,\varepsilon)$ is a strongly decreasing function of $\varepsilon$, and therefore, the lowest values of the argument $\varepsilon$ should contribute maximally to the tunneling current. Such values are encountered at $\hbar\omega=e\VVb$ in Eq.~(\ref{Jindexinter}). Using an analogous argument, the region near the point $\kparb=\Gb-\Kb_g$ for each reciprocal lattice vector $\Gb$ should produce a dominant contribution to the $\kparb$ integral. In addition, because the relative orientation of the TMD and graphene lattices is undefined, we average over the azimuthal angle of $\Kb_g$, which we denote as $\varphi_{\Kb_g}$. By doing so, the tunneling current is found to depend on bias voltage $\VVb$ roughly as
\begin{align}
J^{\rm ind-ex}\propto&\sum_\Gb
\int_{\rm 1BZ}d^2\kparb \;{\rm Im}\big\{-W_{\Gb\Gb}(\kparb,d,d,e\VVb/\hbar)\big\}
\; \int d\varphi_{\Kb_g}\;\delta(\Gb-\Kb_g-\kparb)  \nonumber\\
\propto&\sum_\Gb
\int_{\rm 1BZ}d^2\kparb \;{\rm Im}\big\{-W_{\Gb\Gb}(\kparb,d,d,e\VVb/\hbar)\big\}
\; \delta\big(K_g-|\Gb-\kparb|\big). \label{Jindexfinal}
\end{align}
\end{widetext}
The 1BZ integration region of Eq.~(\ref{Jindexfinal}) is represented in $\kparb$ space in Fig.~\ref{FigS12} for MoSe$_2$, while similar results are found for the other TMDs. The six nearest-neighbor TMD Brillouin zones are indicated by hexagons in this figure, each of them centered around its corresponding $\Gb$ vector. We also show the circles defined by the respective conditions $|\Gb-\kparb|=K_g$. It is clear that only the six smallest non-vanishing TMD reciprocal lattice vectors $\Gb$ can satisfy such conditions for $\kparb$ within the 1BZ. Furthermore, as we have averaged over the orientation of ${\rm\Gamma}$K$_g$ in graphene, and in virtue of symmetry, all of those $\Gb$'s should make equal contributions. Consequently, we only need to consider a single $\Gb=\Gb_0\equiv G_0\,\xx$ with $G_0=4\pi/(\sqrt{3}a_{\rm TMD})$, where $a_{\rm TMD}$ is the in-plane lattice constant of the TMD. This allows us to rewrite Eq.~(\ref{Jindexfinal}) as
\begin{align}
J^{\rm ind-ex}\!\!\propto\!&
\int_{G_0-K_g}^{G_0-\sqrt{3}K/2} \!\frac{dk_x}{k_y} {\rm Im}\big\{-W_{\Gb_0\Gb_0}(\kparb,d,d,e\VVb/\hbar)\big\},
\label{Jindexfinalbis}
\end{align}
where $\kparb=(k_x,k_y)$ with $k_y=\sqrt{K_g^2-(G_0-k_x)^2}$, and we recall that $K_g\approx17.02$~nm$^{-1}$ is the ${\rm\Gamma}$K$_g$ distance in graphene. The integration contour in Eq.~(\ref{Jindexfinalbis}) is indicated by a thick circular segment inside the 1BZ of Fig.~\ref{FigS12}.

\begin{figure}
\begin{centering} \includegraphics[width=0.45\textwidth]{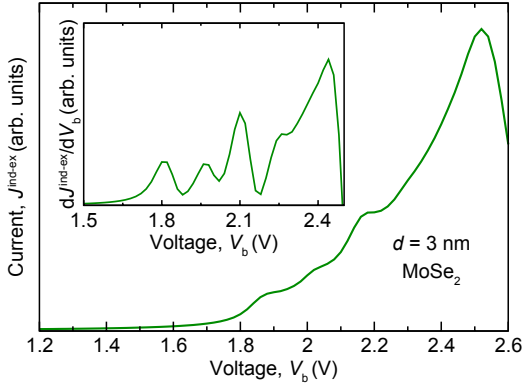} \par\end{centering}
\caption{{\bf Calculated $I \textendash V$ curve for indirect exciton-assisted one-step tunneling.} We plot the current obtained by using Eq.~(\ref{Jindexfinalfinal}) for MoSe$_2$ (see Fig.~\ref{fig:fig4} in the main text for other TMDs) with a hBN tunnel barrier distance $d=3$~nm. The inset shows the corresponding $dI/dV$ curve. Incidentally, the current drops beyond $\sim2.5$~eV because of the reduced number of bands used to calculate the response associated with indirect excitons (see Sec.~\ref{TMDsigma}).}
\label{FigS13}
\end{figure}

The optical conductivity of the TMD, illustrated for MoSe$_2$ at $\hbar\omega=2.25$~eV in Fig.~\ref{FigS12} through a color plot of ${\rm Re}\{\sigma_{{\rm TMD},00}(\kparb,\omega)\}$ (i.e., with $\Gb=\Gb'=0$, see Secs.~\ref{sectionWGG} and \ref{TMDsigma}), shows that the response is dominated by localized momentum regions corresponding to the KQ$_{\rm TMD}$ excitons (see Table~\ref{TableS2}), inside of which the wave vector component $k_y$ does not vary significantly, so we approximate the current as
\begin{align}
J^{\rm ind-ex}\propto {\rm Im}\big\{-W_{\Gb_0\Gb_0}(\kparb,d,d,e\VVb/\hbar)\big\}
\label{Jindexfinalfinal}
\end{align}
for the sake of the discussion of one-step tunneling processes in the main text. [Nevertheless, the results presented in Fig.~\ref{FigS13} and Fig.~\ref{fig:fig4}c in the main text are obtained by evaluating Eq.~(\ref{Jindexfinalbis}).] We remind that $\kparb$ is understood to be determined by the condition $|G_0\,\xx-\kparb|=K_g$ in Eq.~(\ref{Jindexfinalfinal}).

We evaluate Eq.~(\ref{Jindexfinalbis}) from the diagonal elements of the screened interaction $W_{\Gb\Gb}(\kparb,d,d,\omega)$, which are in turn calculated as explained in Sec.~\ref{sectionWGG} below. A result for the $I \textendash V$ curve is presented in Fig.~\ref{FigS13} for MoSe$_2$ with a hBN tunnel barrier distance $d=3$~nm. Spectral features are clearly visible in the 1.9--2.2~eV range, corresponding to indirect excitons that are also observed in our first-principles calculations of the TMD conductivity (see Fig.~\ref{FigS19} in Sec.~\ref{TMDsigma}). The spectral range in which these features show up is in good correspondence with experiments, although the detailed spectral shape presents differences that are possibly connected to the approximations adopted within the present theoretical model.

\subsubsection{Two-step tunneling assisted by phonon and exciton creation}

We now analyze two-step processes involving phonon excitation followed by the creation of nearly direct excitons as an additional channel contributing to electron tunneling from graphene to gold and also producing an excitonic signature in the $dI/dV$ curves.

{\bf Quasi-elastic phonon-assisted transitions.} The phonon energies at the K point are large compared with $\kB T$ at room temperature $T$, and therefore, we can safely neglect phonon absorption processes. Starting from a graphene state $i$ (see Sec.~\ref{Greenfunction}), within first-order perturbation theory, the excitation of a phonon $p$ of wave vector $\Kb_g+\Qb_p$ (i.e., with $\Qb_i$ relative to the graphene K$_g$ point) and frequency $\omega_p$ contributes with an electron wave function $\tilde{\varphi}_{ip}(\rb)=\varphi_i(z)\,\ee^{\ii(\Qb_i-\Qb_p)\cdot\Rb}/\sqrt{A}$ of energy $\hbar(\varepsilon_i-\omega_p)$, where
\begin{align}
\tilde{\varphi}_{ip}(\rb)=\int d^3\rb' G_0(\rb,\rb',\varepsilon_i-\omega_p)
\,\mathcal{H}_p^{\rm ph}(\rb') \,\varphi_i(\rb')
\label{G0Hphir}
\end{align}
is expressed in terms of the electron-phonon coupling Hamiltonian $\mathcal{H}^{\rm ph}_p(\rb')$ and the 3D electron Green function $G_0(\rb,\rb',\varepsilon)=(2\pi)^{-1}\int d^2\Qb\,\ee^{\ii\Qb\cdot(\Rb-\Rb')}\,G_0(z,z',\varepsilon-\hbar Q^2/2\me)$ [see Sec.~\ref{Greenfunction} for details of this function and the definitions of $\varphi_i(z)$, $\Qb_i$, and $A$]. Separating the in- and out-of-plane dependence of the phonon Hamiltonian as $\mathcal{H}^{\rm ph}_p(\rb')=\mathcal{H}^{\rm ph}_p(z')\ee^{-\ii(\Kb_g+\Qb_p)\cdot\Rb}$, we can recast Eq.~(\ref{G0Hphir}) into
\begin{align}
\tilde{\varphi}_{ip}(z)=\int dz' \;&G_0\big(z,z',\varepsilon_i-\omega_p-\hbar|\Qb_i-\Qb_p|^2/2\me\big) \nonumber\\
&\times \mathcal{H}_p^{\rm ph}(z') \,\varphi_i(z'),
\nonumber
\end{align}
and noticing again that $\varphi_i(z)$ is tightly localized around $z=d$, we can approximate this expression as
\begin{align}
\tilde{\varphi}_{ip}(z)\approx C_p\, G_0\big(z,d,\varepsilon_i-\omega_p-\hbar|\Qb_i-\Qb_p|^2/2\me\big),
\label{phiip}
\end{align}
where the constant $C_p$ depends on the specific phonon mode $p$ under consideration. In addition, as tunneling is expected to involve large phonon momentum transfers that place the electron near the ${\rm \Gamma}$ point, only a reduced region in the phonon bands contributes to the process, and therefore, we assume $C_1\equiv|C_p|$ to be roughly independent of $p$. The tunneling current per unit area traversing the hBN/Au interface is finally obtained as
\begin{widetext}
\begin{align}
J^{\rm ph}&=-\frac{e\hbar}{\me}\sum_{ip}
\Theta\big[\hbar(\varepsilon_i-\omega_p)-\EFAu\big]
\;{\rm Im}\big\{\tilde{\varphi}_{ip}^*(\rb) \partial_z \tilde{\varphi}_{ip}(\rb)\big\}\big|_{z=0^-} \nonumber\\
&\approx -\frac{C_1^2}{A}\,\frac{e\hbar}{\me}\sum_{ip}
\Theta\big[\hbar(\varepsilon_i-\omega_p)-\EFAu\big]
\;{\rm Im}\Big\{G_0^*\big(z,d,\varepsilon_i-\omega_p-\hbar|\Qb_i-\Qb_p|^2/2\me\big) \nonumber\\
&\quad\quad\quad\quad\quad\quad\quad\quad\quad\quad\quad\quad\quad\quad\quad\quad\;
\times\partial_z G_0\big(z,d,\varepsilon_i-\omega_p-\hbar|\Qb_i-\Qb_p|^2/2\me\big)\Big\}\Big|_{z=0^-},
\nonumber
\end{align}
\end{widetext}
which is evaluated at a position $z=0^-$ right inside the metal (see Fig.~\ref{FigS11} and Sec.~\ref{generalJ}), with the step function restricting the sum over $i$ and $p$ such that only final unoccupied electron gold states above the Fermi level $\EFAu$ are included.

{\bf Phonon+exciton-assisted tunneling.} In the two-step channel, after a phonon has been created, a direct exciton absorbs most of the energy lost by the tunneling electron. This second step is described by Eq.~(\ref{Iphex3}) in Sec.~\ref{generalJ}, with $\varphi_i$ substituted by the phonon-scattered wave function $\tilde{\varphi}_{ip}$ given in Eq.~(\ref{phiip}). Following a similar procedure as in Sec.~\ref{currentindirect}, we insert Eqs.~(\ref{G0G0}) and (\ref{Wkpar}) into Eq.~(\ref{Iphex3}) and carry out the integrals over $\Rb$, $\Rb'$, and $\Rb''$ to find
\begin{widetext}
\begin{align}
J^{\rm ph+ex}=\frac{1}{A}\,\frac{e^3\hbar^2}{\pi\me}\sum_{ip}&\int_0^\infty\!\!\! d\omega
\;\Theta\big[\hbar(\varepsilon_i-\omega_p-\omega)-\EFAu\big]
\int\frac{d^2\kparb}{(2\pi)^2}\int dz'\int dz'' \;{\rm Im}\big\{-W(\kpar,z',z'',\omega)\big\} \nonumber\\
&\times\Theta(\Omega_{ip\omega\kparb})
\;{\rm Im}\Big\{
G_0^*(0,z',\Omega_{ip\omega\kparb}) \;\partial_z G_0(z,z'',\Omega_{ip\omega\kparb})\big|_{z=0^-}
\;\tilde{\varphi}^*_{ip}(z')\tilde{\varphi}_{ip}(z'') \Big\},  \label{Iphex40}
\end{align}
where we have assumed translational invariance in the screened interaction [Eq.~(\ref{Wkpar})] and defined $\Omega_{ip\omega\kparb}=\varepsilon_i-\omega_p-\omega-\hbar|\Qb_i-\Qb_p-\kparb|^2/2\me$.

We note that the intermediate phonon-assisted state $\tilde{\varphi}_{ip}(z)$ is generated from the graphene region (at $z=d$), and therefore, we expect it to be exposed to further inelastic interactions from that region, including TMD excitons. Rather than propagating it elastically (once the phonon has been created) to tunnel into the metal, we are interested in the immediate interaction with the excitons. Therefore, proceeding as we did in Sec.~\ref{currentindirect}, we approximate Eq.~(\ref{Iphex40}) by setting $z'=z''=d$ and encapsulating the dependence on the intermediate states in an overall constant factor, so we write
\begin{align}
J^{\rm ph+ex}\propto\sum_{ip}&\int_0^\infty\!\!\! d\omega
\;\Theta\big[\hbar(\varepsilon_i-\omega_p-\omega)-\EFAu\big]
\int d^2\kparb\;{\rm Im}\big\{-W(\kpar,d,d,\omega)\big\} \label{Iphex4}\\
&\times\,{\rm Im}\Big\{
G_0^*(0,d,\varepsilon_i-\omega_p-\omega-\hbar|\Qb_i-\Qb_p-\kparb|^2/2\me) \nonumber\\
&\quad\;\;\;\times\partial_z G_0(z,d,\varepsilon_i-\omega_p-\omega-\hbar|\Qb_i-\Qb_p-\kparb|^2/2\me)\big|_{z=0^-} \Big\}. \nonumber
\nonumber
\end{align}
We further simplify the evaluation of Eq.~(\ref{Iphex4}) by arguing that the response function dies out quickly as $\kpar$ increases to values beyond 1~nm$^{-1}$ (see Sec.~\ref{screening} and Fig.~\ref{FigS18}), which is small compared with both $K_g\approx17.02$~nm$^{-1}$ and $\kFAu\approx12$~nm$^{-1}$, so we can dismiss the $\kparb$ dependence in front of $\Qb_p$ in the last argument of the Green function $G_0$. Likewise, the wave vectors of graphene electrons relative to the Dirac point are also small and can be neglected in $G_0$ when compared with $\Qb_p$. We thus approximate $|\Qb_i-\Qb_p-\kparb|\approx Q_p$. In contrast, the phonon wave vector $\Qb_p$ is large, so it has to be retained. We replace the sum over phonons $p$ by $C_2\int_0^\infty QdQ$, where $Q\equiv Q_p$ and the constant $C_2$ accounts for further details of the electron-phonon coupling. Analogously, we replace the sum over graphene electrons $i$ by $C_3\sum_{s=\pm1}\int_0^{Q_s} Q'dQ'$, where $Q'\equiv Q_i$ and the constant $C_3$ incorporates further graphene band details. We also sum over $s=+1$ and $s=-1$, referring to the upper (with $Q'$ running up to the graphene Fermi wave vector $Q_1=\kFg$) and lower (with $Q_{-1}=\infty$) Dirac cones, respectively. In addition, we assume $\omega_p$ to be independent of $p$. These considerations allow us to reduce Eq.~(\ref{Iphex4}) to
\begin{align}
J^{\rm ph+ex}\propto\sum_{s=\pm1} &\int_0^\infty\!\!\! d\omega
\int_0^{Q_s}\!\!\! Q'dQ'
\;\Theta\big[\hbar(\varepsilon_0+s\vFg Q'-\omega_p-\omega)-\EFAu\big]
\int_0^\infty\!\!\! \kpar d\kpar\;{\rm Im}\big\{-W(\kpar,d,d,\omega)\big\} \nonumber\\
&\times\int_0^{Q_c}\!\!\! QdQ\;{\rm Im}\Big\{
G_0^*(0,d,\varepsilon_0+s\vFg Q'-\omega_p-\omega-\hbar Q^2/2\me) \nonumber\\
&\quad\quad\quad\quad\quad\times\partial_z G_0(z,d,\varepsilon_0+s\vFg Q'-\omega_p-\omega-\hbar Q^2/2\me)\big|_{z=0^-}
\Big\} \label{Iphex5},
\end{align}
where $\hbar\varepsilon_0\equiv E_{\rm D}=\EFg-\DEFg$ is the Dirac point energy relative to the CBB of gold (see Figs.~\ref{FigS10} and \ref{FigS11}), and a cutoff wave vector $Q_c=\big[2\me(\varepsilon_0+s\vFg Q'-\omega_p-\omega)/\hbar\big]^{1/2}$ is introduced to guarantee that the final out-of-plane electron energy is above the CBB of gold.

We are interested in examining the tunneling current for bias potential energies close to the TMD exciton energies, which are large compared with the graphene electron energies relative to the Dirac point. Therefore, we further approximate Eq.~(\ref{Iphex5}) by setting $\varepsilon_0+s\vFg Q'\approx\EFg/\hbar=(\EFAu+e\VVb)/\hbar\equiv\varepsilon_1$ and treating $\sum_s\int Q'dQ'$ as an additional overall constant. This leads to
\begin{align}
J^{\rm ph+ex}\propto&\int_0^{e\VVb/\hbar-\omega_p}\!\!\! d\omega \;
\int_0^\infty\!\!\! \kpar d\kpar\;{\rm Im}\big\{-W(\kpar,d,d,\omega)\big\} \label{Iphexfinal}\\
&\times\int_0^{Q_c}\!\!\! QdQ\;{\rm Im}\Big\{
G_0^*(0,d,\varepsilon_1-\omega_p-\omega-\hbar Q^2/2\me)
\;\partial_z G_0(z,d,\varepsilon_1-\omega_p-\omega-\hbar Q^2/2\me)\big|_{z=0^-} \Big\}, \nonumber
\end{align}
where we need to redefine $Q_c=\big[2\me(\varepsilon_1-\omega_p-\omega)/\hbar\big]^{1/2}$.
\end{widetext}

\begin{figure}
\begin{centering} \includegraphics[width=0.45\textwidth]{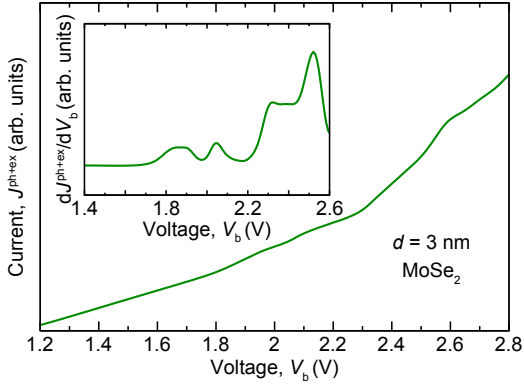} \par\end{centering}
\caption{{\bf Calculated $I \textendash V$ curves for phonon+exciton-assisted two-step tunneling.} We plot the current obtained by using Eq.~(\ref{Iphexfinalfinal}) for MoSe$_2$ (see Fig.~\ref{fig:fig4} in the main text for other TMDs) with a hBN tunnel barrier distance $d=3$~nm. The inset shows the corresponding $dI/dV$ curve.}
\label{FigS14}
\end{figure}

Finally, we argue that $G_0(0,d,\varepsilon)$ is a strongly decreasing function of $\varepsilon$, and therefore, the lowest values of the argument $\varepsilon$ should contribute maximally to the tunneling current. Such values are encountered at $\hbar\omega=e\VVb$ according to Eq.~(\ref{Iphexfinal}), from which the phonon+exciton-assisted tunneling current is found to depend on bias voltage $\VVb$ roughly as
\begin{align}
J^{\rm ph+ex}\propto&\int_0^\infty\!\!\! \kpar d\kpar\;{\rm Im}\big\{-W(\kpar,d,d,e\VVb/\hbar-\omega_p)\big\}. \label{Iphexfinalfinal}
\end{align}
This result relates the phonon+exciton-assisted tunneling current to the screened interaction in the heterostructure. The latter is studied in detail in Sec.~\ref{screening}.

In Fig.~\ref{FigS14}, we plot the $I \textendash V$ curves obtained from Eq.~(\ref{Iphexfinalfinal}) for $d=3$~nm, with the screened interaction obtained by describing gold in the specular-reflection model, hBN through a local anisotropic permittivity, graphene in the random-phase approximation (RPA), and the TMD monolayer through a nonlocal surface conductivity (see Sec.~\ref{TMDsigma}). The tunneling current shows dominant features associated with C excitons, while A and B excitons emerge as weak shoulders at lower biases.

Upon examination of the $\kpar$ dependence of ${\rm Im}\big\{-W(\kpar,d,d,e\VVb/\hbar-\omega_p)\big\}$ (see Fig.~\ref{FigS18}), we conclude that the integral in Eq.~(\ref{Iphexfinalfinal}) yields a result similar to ${\rm Im}\big\{-W(0,d,d,e\VVb/\hbar-\omega_p)\big\}$ (i.e., at $\kpar=0$), so we write
\begin{align}
J^{\rm ph+ex}\propto\;&{\rm Im}\big\{-W_{00}(0,d,d,e\VVb/\hbar-\omega_p)\big\}, \nonumber
\end{align}
where we insert subindices indicating that this part involves $\Gb=0$ (i.e., no umklapp processes), for the sake of the discussion of two-step tunneling in the main text. [Nevertheless, the results presented in Fig.~\ref{FigS14} and Fig.~\ref{fig:fig4}d in the main text are obtained by evaluating Eq.~(\ref{Iphexfinalfinal}).]

\subsection{Screened interaction under the in-plane isotropic and translational-invariant approximations}
\label{screening}
\begin{figure}
\begin{centering} \includegraphics[width=0.4\textwidth]{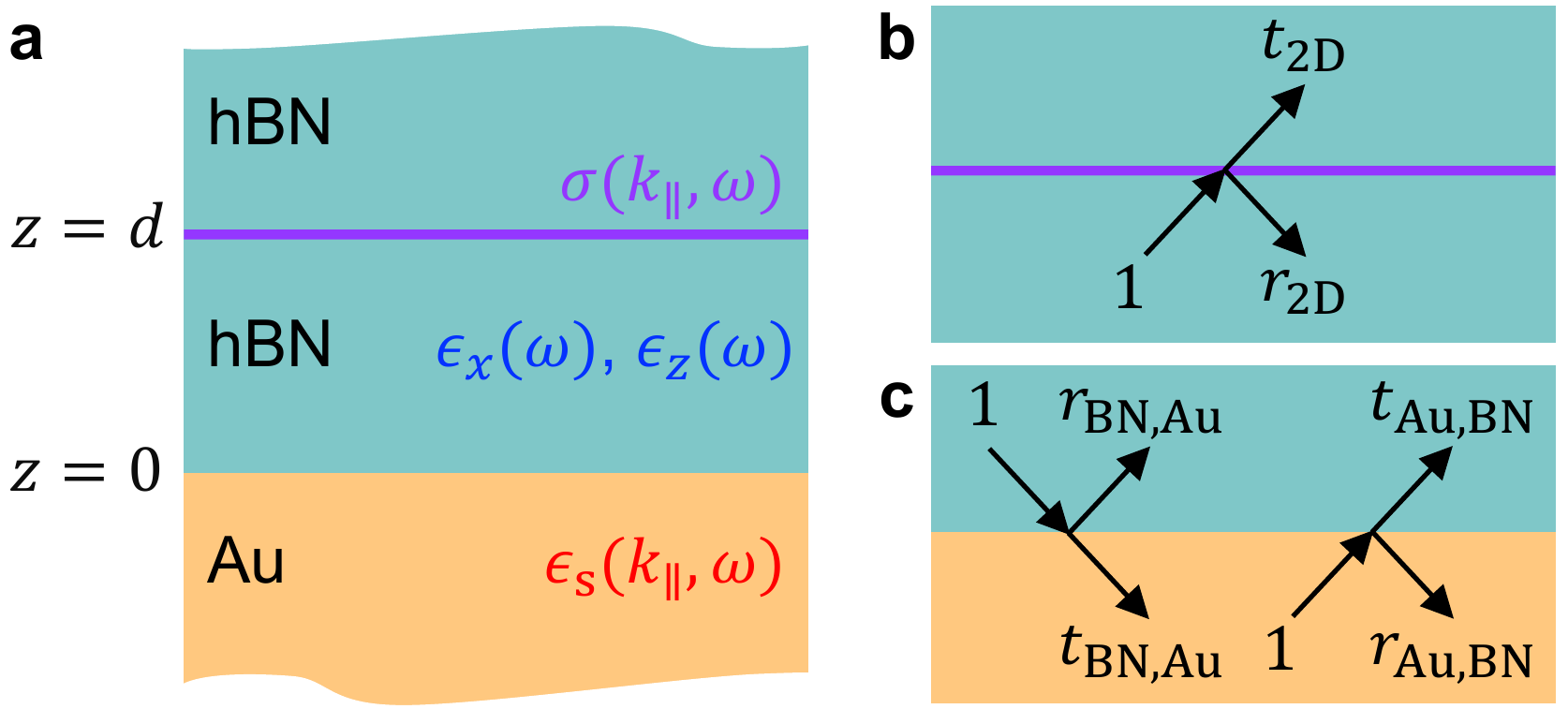} \par\end{centering}
\caption{{\bf Analysis of the dielectric response of the studied multilayer structure.} ({\bf a}) We consider a 2D material described by a nonlocal surface conductivity $\sigma(\kpar,\omega)$, separated from a gold surface by an hBN tunnel barrier of thickness $d$. The structure is covered by an upper hBN layer. Gold is described through a surface-corrected nonlocal dielectric function $\epsilon_{\rm s}(\kpar,\omega)$, whereas hBN is modeled through its anisotropic frequency-dependent permittivity~\cite{GPR1966} of in-plane and out-of-plane components $\epsilon_x(\omega)$ and $\epsilon_z(\omega)$, respectively. The gold and upper hBN media are assumed to have an infinite thickness, while the 2D material is treated in the zero-thickness limit. ({\bf b,c}) The response of the system is described in terms of Fresnel transmission and reflection coefficients for each of the two interfaces.}
\label{FigS15}
\end{figure}

Tunneling electrons couple to excitations in the structure through the screened interaction $W(\rb,\rb',\omega)$. Because the distances involved are small compared with the optical wavelength at frequency $\omega$, we can safely neglect retardation effects and treat the interaction in the electrostatic limit. The structure under consideration is sketched in Fig.~\ref{FigS15}a and consists of a layer of 2D materials separated by a distance $d$ from a gold surface. The region in between the two of them is filled with hBN, and the medium above the 2D materials is also hBN.

The layer of 2D materials here considered consists of a TMD monolayer on top of graphene. We describe it in the zero-thickness approximation through a nonlocal wave-vector- and frequency-dependent surface conductivity $\sigma(\kpar,\omega)$. This approximation simplifies the algebra considerably and yields reasonable results in comparison to finite-thickness layers~\cite{paper329}. We thus set $\sigma(\kpar,\omega)=\sigma_{\rm TMD}(\kpar,\omega)+\sigma_{\rm g}(\kpar,\omega)$ as the sum of the surface conductivities of the TMD and graphene monolayers. A nonlocal version of the former is calculated from first principles (see Sec.~\ref{TMDsigma}), whereas the nonlocal conductivity of graphene is modeled in the RPA~\cite{WSS06}.

The interfaces in our structure are described by first defining Fresnel's reflection coefficients, as shown in Fig.~\ref{FigS15}b,c. More precisely, only p-polarization coefficients survive in the electrostatic limit. For convenience, we define in-plane coordinates $\Rb=(x,y)$ and work in the space of in-plane wave vectors $\kparb=(k_x,k_y)$, with electrostatic quantities such as the electric potential written as $\phi(\rb,\omega)=(2\pi)^{-2}\int d^2\kparb\,\ee^{\ii\kparb\cdot\Rb}\,\phi(\kparb,z,\omega)$.

We start by considering the electrostatic potential created by a point charge of time-dependent magnitude $\ee^{-\ii\omega t}$ placed at the origin in bulk hBN. The momentum components of such potential are given by $[2\pi/\kpar\bar{\epsilon}(\omega)]\,\ee^{-q|z|}$, where $\bar{\epsilon}(\omega)=\sqrt{\epsilon_x(\omega)\epsilon_z(\omega)}$ and $q=\kpar\sqrt{\epsilon_x(\omega)/\epsilon_z(\omega)}$, with the square root sign chosen to yield positive real parts. Now, with the 2D material layer placed at $z=d$, scattering of an {\it incident} potential component $\ee^{-q(z-d)}$ coming from the $z<d$ region gives rise to a {\it transmitted} potential $t_{\rm 2D}\ee^{-q(z-d)}$ at $z>d$ and an {\it incident}+{\it reflected} potential $\ee^{-q(z-d)}-r_{\rm 2D}\ee^{q(z-d)}$ at $z<d$. (Incidentally, note that the reflection coefficient is preceded by a $-$ sign in this expression for the potential \cite{paper329}.) The corresponding Fresnel coefficients are determined by imposing the continuity of the potential at $z=d$, as well as the condition that the jump in the normal electric displacement is $4\pi$ times the surface charge produced through the conductivity $\sigma(\kpar,\omega)$. Writing such charge in terms of the current through the continuity equation, and this in turn as the conductivity times the in-plane electric field, we are left with a set of two linear equations from which we obtain
\begin{subequations}
\label{r2Dt2D}
\begin{align}
r_{\rm 2D}&=\frac{1}{1-\ii\omega\bar{\epsilon}(\omega)/[2\pi\kpar\sigma(\kpar,\omega)]}, \\
t_{\rm 2D}&=1-r_{\rm 2D}
\end{align}
\end{subequations}
for the Fresnel coefficients of the 2D layer, which generalize previous results \cite{paper235} to deal with an anisotropic host medium. We note that the $z$ dependence of the potential components inside hBN is simply given by $\ee^{\mp q(z-d)}$ for upward ($-$ sign) and downward ($+$ sign) components (see Fig.~\ref{FigS15}b).

Given the small thickness of the hBN tunnel layer, nonlocal effects in the response of gold can play an important role, and therefore, we use the nonlocal bulk permittivity of this material $\epsilon_{\rm Au}(k,\omega)$. As a reasonably accurate prescription, we express $\epsilon_{\rm Au}(k,\omega)$ as the sum of a conduction-electron component, treated in the RPA, and a local contribution associated with interband transitions, as described in detail elsewhere~\cite{paper119}. In addition, we adopt the specular-reflection model~\cite{RM1966,paper119,paper149} to incorporate surface effects through a surface response function
\begin{align}
\epsilon_{\rm s}(\kpar,z,\omega)
&=\frac{\kpar}{\pi}\int\frac{dk_z}{k^2}\frac{\ee^{\ii k_zz}}{\epsilon_{\rm Au}(k,\omega)} \nonumber\\
&=\frac{2\kpar}{\pi}\int_0^\infty\frac{dk_z}{k^2}\frac{\cos(k_zz)}{\epsilon_{\rm Au}(k,\omega)},
\nonumber
\end{align}
where $k^2=\kpar^2+k_z^2$. Incidentally, this quantity becomes $\epsilon_{\rm s}(\kpar,z,\omega)=\ee^{-\kpar|z|}/\epsilon_{\rm Au}(\omega)$ in the local limit (i.e., when dismissing the $k$ dependence of the gold permittivity). We also define $\epsilon_{\rm s}(\kpar,\omega)\equiv\epsilon_{\rm s}(\kpar,0,\omega)$, in terms of which we find the Fresnel coefficients
\begin{subequations}
\begin{align}
r_{\rm Au,BN}&=-r_{\rm BN,Au}=\frac{\bar{\epsilon}(\omega)\epsilon_{\rm s}(\kpar,\omega)-1}{\bar{\epsilon}(\omega)\epsilon_{\rm s}(\kpar,\omega)+1}, \label{rBAu}\\
t_{\rm Au,BN}&=\frac{2}{\bar{\epsilon}(\omega)\epsilon_{\rm s}(\kpar,\omega)+1}, \\
t_{\rm BN,Au}&=\frac{2\,\bar{\epsilon}(\omega)\epsilon_{\rm s}(\kpar,\omega)}{\bar{\epsilon}(\omega)\epsilon_{\rm s}(\kpar,\omega)+1}
\end{align}
\end{subequations}
for the hBN/Au interface by imposing the continuity of both the potential and the normal electric displacement. Incidentally, the latter involves $\epsilon_{\rm Au}\cdot\epsilon_{\rm s}$ (in operator notation), which reduces to $\ee^{-\kpar |z|}$ in $(\kpar,z)$ space. With the hBN/Au interface placed at $z=0$, the $z$ dependence of the potential indicated by the arrows in Fig.~\ref{FigS15}c is given by $\ee^{\mp qz}$ for $z>0$. However, the dependence inside gold ($z<0$) is more complex~\cite{paper149}: it is partially encapsulated in a factor $\epsilon_{\rm s}(\kpar,z,\omega)/\epsilon_{\rm s}(\kpar,\omega)$ multiplying the transmission coefficients; in addition, an involved dependence is encountered for internal reflection from the gold side (see more details in Ref.~\cite{paper149}).

\begin{figure}[b]
\begin{centering} \includegraphics[width=0.42\textwidth]{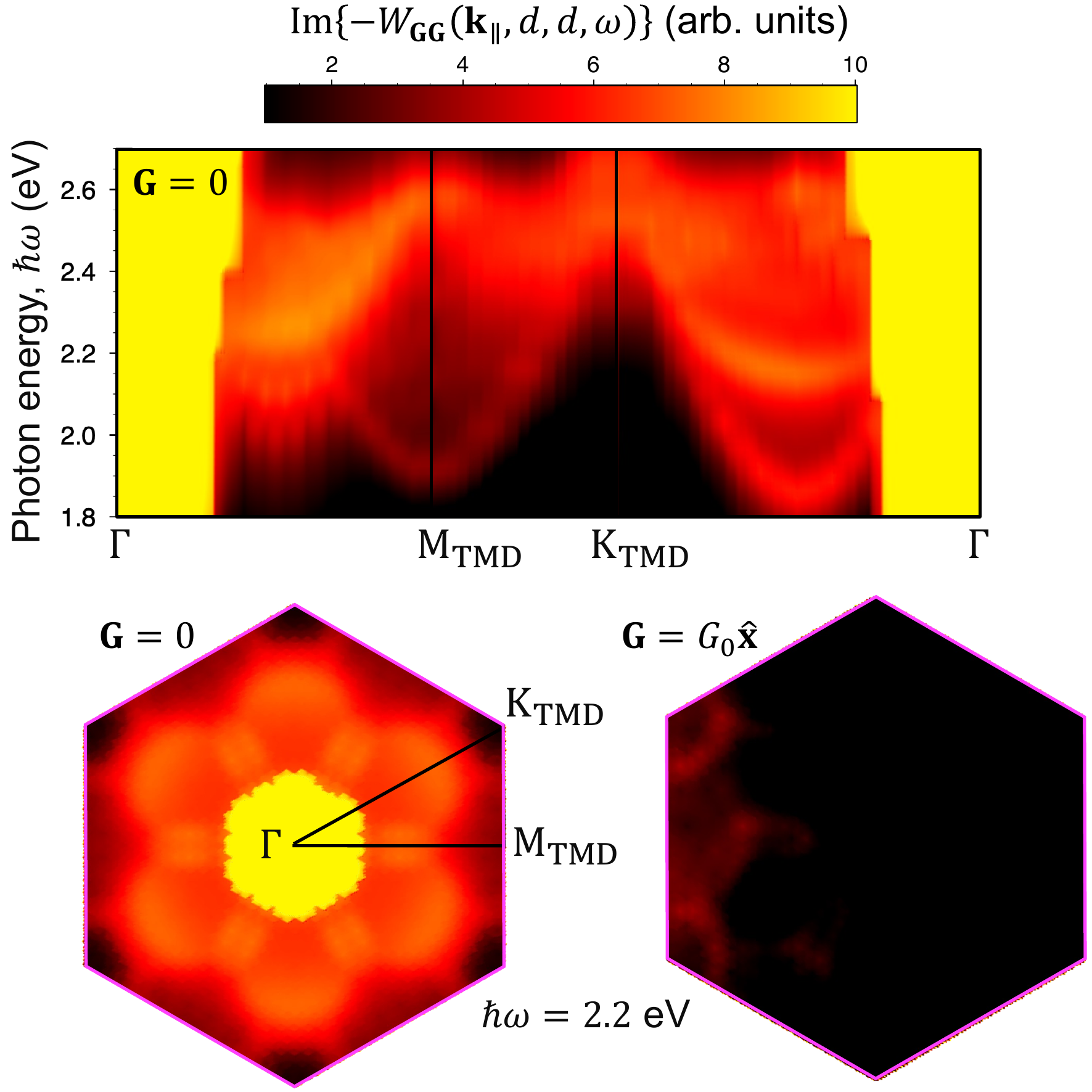} \par\end{centering}
\caption{{\bf Screened interaction in MoSe$_2$.} We plot ${\rm Im}\{-W_{\Gb\Gb}(\kparb,d,d,\omega)\}$ for a heterostructure like that in Fig.~\ref{FigS15} with $d=3$~nm and the 2D layer consisting of monolayer MoSe$_2$ and graphene. The upper plot shows the dependence on photon energy and momentum along the indicated excursion within the 1BZ of the TMD for $\Gb=0$. The lower plots show the $\kparb$ dependence within the 1BZ for $\hbar\omega=2.2$~eV and two values of $\Gb$ [0 and $G_0\xx$, with $G_0=4\pi/(\sqrt{3}a_{\rm TMD})$, where $a_{\rm TMD}$ is the in-plane lattice constant of the TMD]. The color scale is shared by all plots and saturated outside the displayed range.}
\label{FigS16}
\end{figure}

\begin{widetext}
Finally, the screened interaction can also be decomposed in parallel wave vectors as shown in Eq.~(\ref{Wkpar}), where $W(\kpar,z,z',\omega)=W^{\rm dir}(\kpar,z,z',\omega)+W^{\rm ref}(\kpar,z,z',\omega)$ is in turn separated into direct and reflected components. The former corresponds to the direct Coulomb interaction when $z$ and $z'$ lie in the same medium (see Fig.~\ref{FigS15}a):
\begin{align}
W^{\rm dir}(\kpar,z,z',\omega)=\frac{2\pi}{\kpar}\times\left\{\begin{matrix*}[l]
\dfrac{1}{\bar{\epsilon}(\omega)}\ee^{-q|z-z'|}, &\quad\quad \text{$z,z'\ge0$ inside hBN}, \\
\epsilon_{\rm s}(\kpar,z-z',\omega), &\quad\quad \text{$z,z'<0$ inside Au}, \\
0, &\quad\quad \text{otherwise}, \end{matrix*}\right. \label{Wdir}
\end{align}
where the part inside gold incorporates nonlocal bulk corrections. The remaining surface contribution is given by
\begin{align}
W^{\rm ref}(\kpar,z,z',\omega)=\frac{2\pi}{\kpar}\times\left\{\begin{matrix*}[l]
-\dfrac{1}{\bar{\epsilon}(\omega)}\ee^{-q(z+z'-2d)} \Big[r_{\rm 2D}+\ee^{-2qd}\Delta\,t_{\rm 2D}^2r_{\rm BN,Au}\Big], &\quad\quad \text{$d\le z,z'$}, \\
\dfrac{1}{\bar{\epsilon}(\omega)}\Delta\,t_{\rm 2D}\Big[\ee^{-q(z-z')}-\ee^{-q(z+z')}r_{\rm BN,Au}\Big]-\dfrac{\ee^{-q(z-z')}}{\bar{\epsilon}(\omega)}, &\quad\quad \text{$0\le z'<d\le z$}, \\
\ee^{-qz}\Delta\,t_{\rm 2D}t_{\rm Au,BN}\epsilon_{\rm s}(\kpar,z',\omega), &\quad\quad \text{$d\le z$, $z'<0$}, \\
\dfrac{1}{\bar{\epsilon}(\omega)}\Delta\,\Big[\big(\ee^{-q(2d+z-z')}+\ee^{-q(2d-z+z')}\big)r_{\rm 2D}r_{\rm BN,Au} & \\
\quad\quad\quad\quad-\ee^{-q(2d-z-z')}r_{\rm 2D}-\ee^{-q(z+z')}r_{\rm BN,Au}\Big], &\quad\quad \text{$0\le z,z'<d$}, \\
\Delta\,t_{\rm Au,BN}\Big[\ee^{-qz}-\ee^{-q(2d-z)}r_{\rm 2D}\Big]\epsilon_{\rm s}(\kpar,z',\omega), &\quad\quad \text{$z'<0\le z<d$}, \\
\epsilon_{\rm s}(\kpar,z+z',\omega)-\bar{\epsilon}(\omega)\Delta\,t_{\rm Au,BN}\Big[1+\ee^{-2qd}r_{\rm 2D}\Big]& \\
\quad\quad\quad\quad\quad\quad\quad\quad\times\;\epsilon_{\rm s}(\kpar,z,\omega)\epsilon_{\rm s}(\kpar,z',\omega), & \quad\quad \text{$z,z'<0$},
\end{matrix*}\right. \label{Wref}
\end{align}
\end{widetext}
for $z\ge z'$, while the reciprocity property $W^{\rm dir}(\kpar,z,z',\omega)=W^{\rm dir}(\kpar,z',z,\omega)$ can be used to compute it for $z<z'$. In Eq.~(\ref{Wref}), we use the Fabry--Perot factor $\Delta=\big(1-r_{\rm 2D}r_{\rm BN,Au}\ee^{-2qd}\big)^{-1}$ to account for multiple round trips within the 2D/hBN/Au cavity. We have explicitly verified that the sum of Eqs.~(\ref{Wdir}) and (\ref{Wref}) is continuous in both $z$ and $z'$ across the $z=0$ and $z=d$ interfaces, while the normal displacement is continuous at $z=0$ and undergoes a jump by $2\,\bar{\epsilon}\,\kpar/(1-1/r_{\rm 2D})=\ii\omega\bar{\epsilon}^2/\pi\sigma(\kpar,\omega)$ at $z=d$ due to the presence of the surface conductivity associated with the 2D materials.

\begin{figure}
\begin{centering} \includegraphics[width=0.45\textwidth]{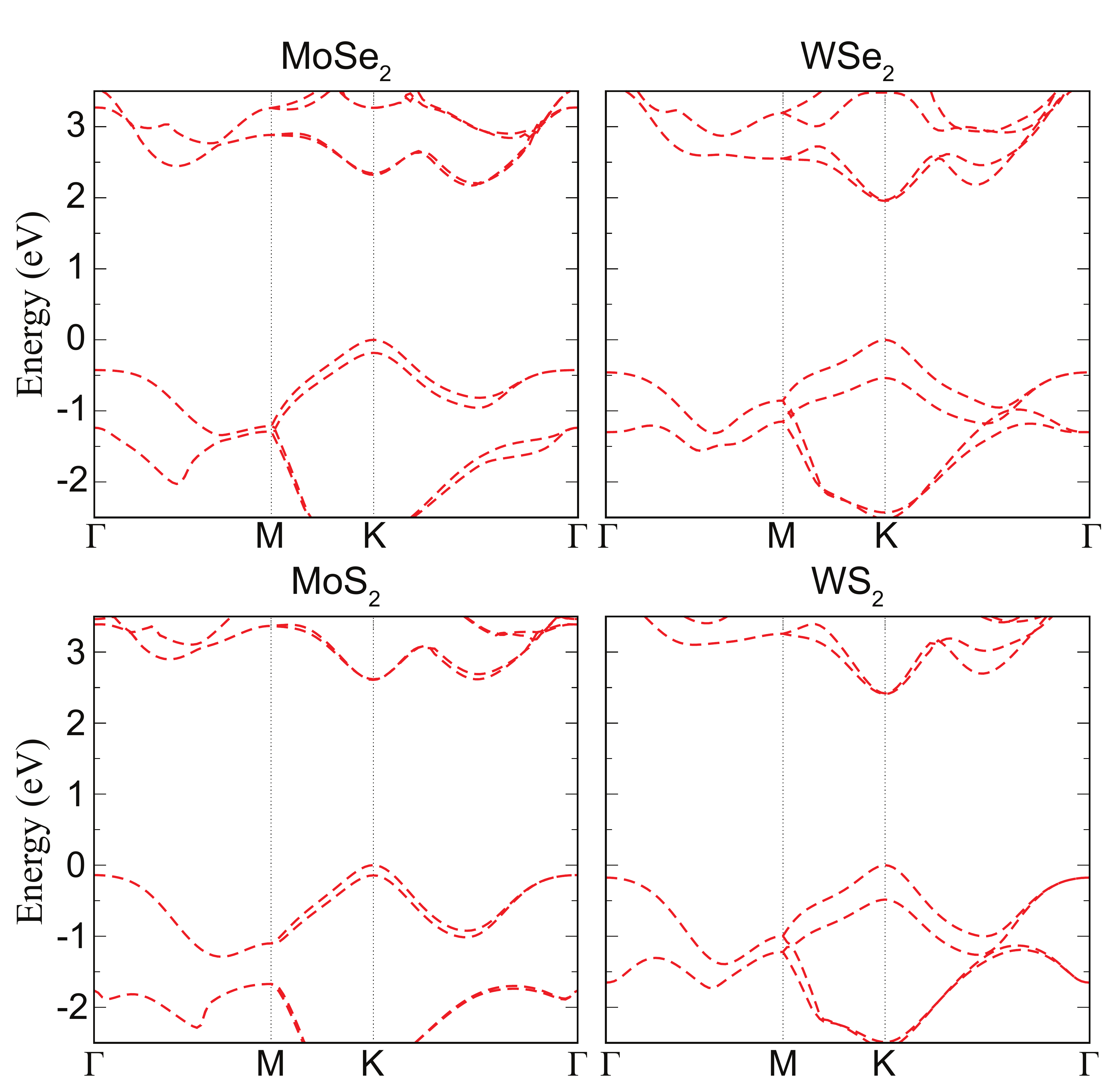} \par\end{centering}
\caption{{\bf Electronic bands of TMD monolayers.} We show DFT calculations for MoSe$_2$, WSe$_2$, MoS$_2$, and WS$_2$, obtained within the G$_{0}$W$_{0}$ approximation.}
\label{FigS17}
\end{figure}

\begin{figure}
\begin{centering} \includegraphics[width=0.45\textwidth]{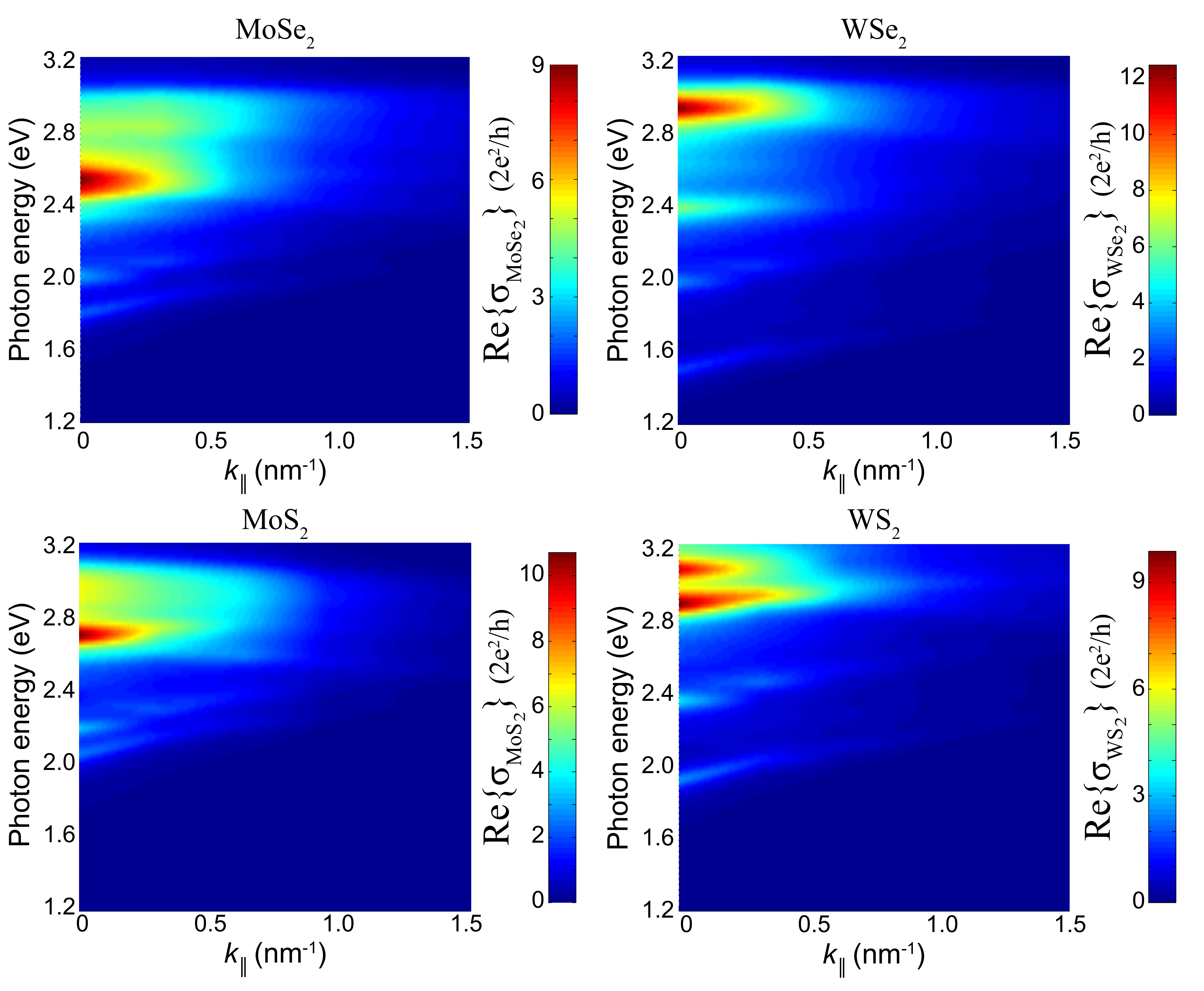} \par\end{centering}
\begin{centering} \includegraphics[width=0.3\textwidth]{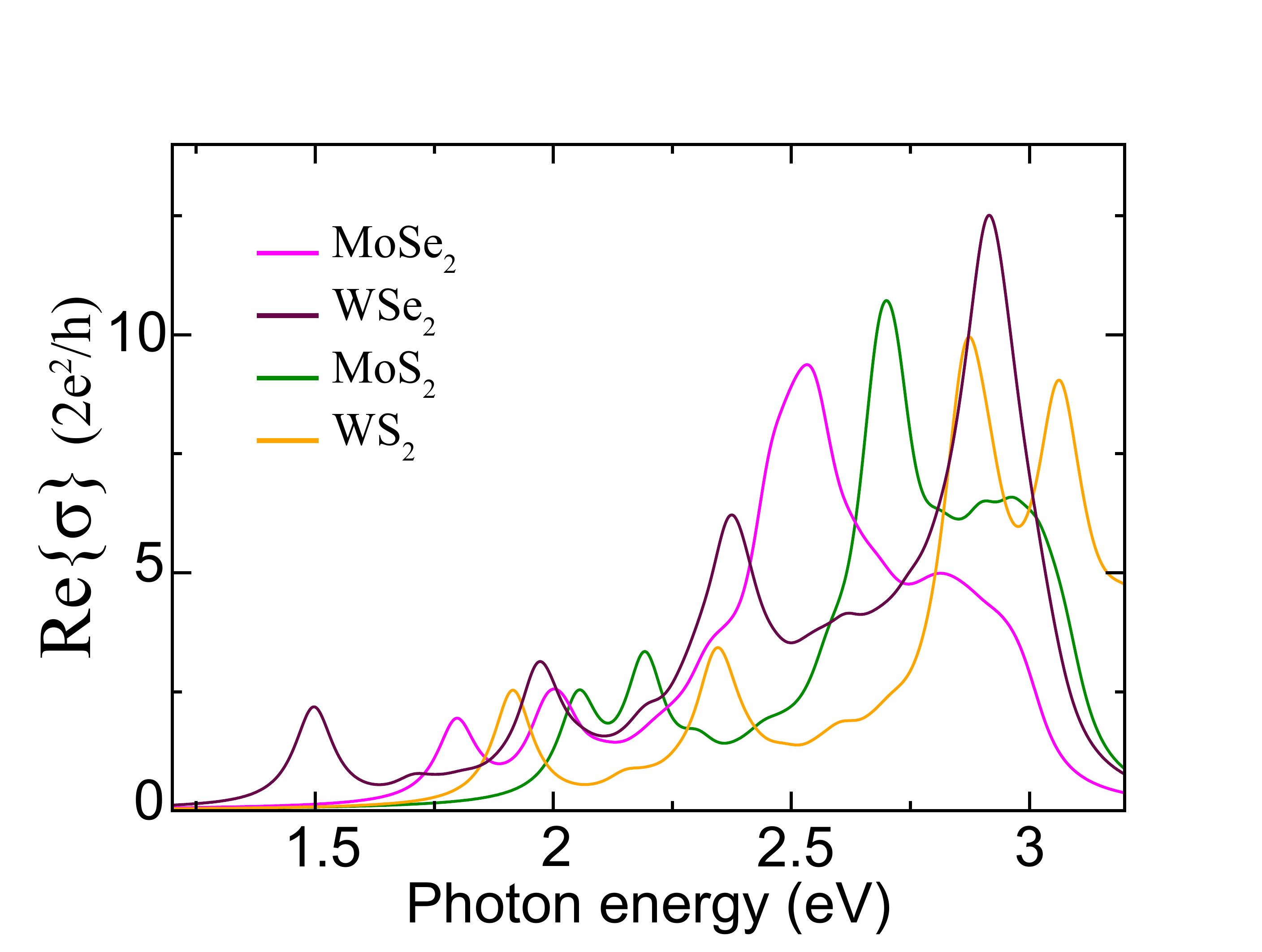} \par\end{centering}
\caption{{\bf Nonlocal conductivity of TMD monolayers.} We plot the calculated real part of the 2D conductivities of MoSe$_2$, WSe$_2$, MoS$_2$, and WS$_2$ monolayers, expressed in units of $2e^2/h$, for relatively low values of the in-plane wave vector $\kpar$ and ignoring umklapp processes. The lower panel shows ${\rm Re}\{\sigma_{\rm TMD}(\kpar,\omega)\}$ in the $\kpar\rightarrow0$ limit.}
\label{FigS18}
\end{figure}

\begin{figure}
\begin{centering} \includegraphics[width=0.45\textwidth]{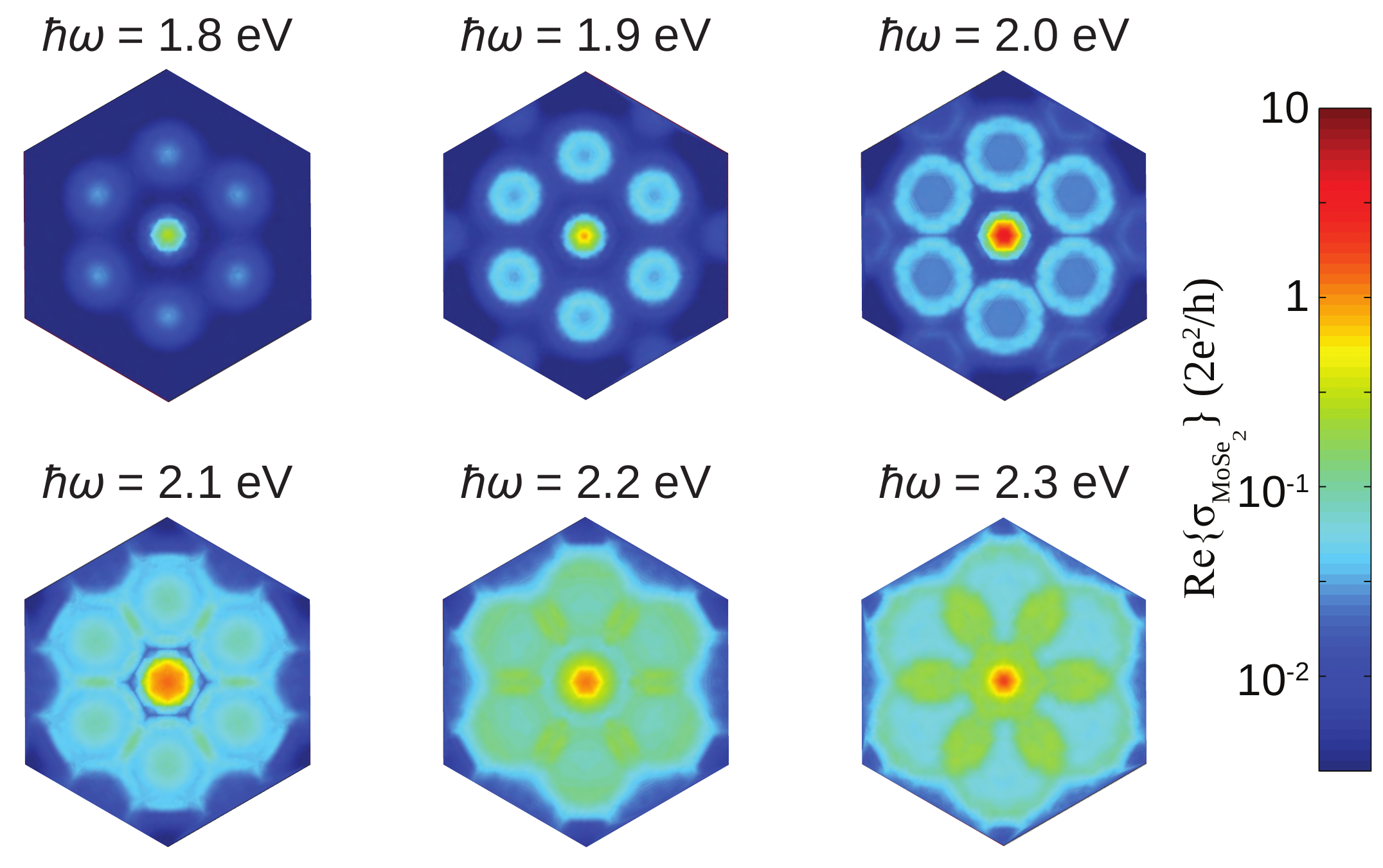} \par\end{centering}
\caption{{\bf Nonlocal response within the 1BZ of MoSe$_2$.} We plot ${\rm Re}\{\sigma_{{\rm TMD},00}(\kparb,\omega)\}$ (i.e., the real part of the optical conductivity for $\Gb=\Gb'=0$, see Secs.~\ref{sectionWGG} and \ref{TMDsigma}) within the 1BZ of MoSe$_2$ at different photon energies $\hbar\omega$.}
\label{FigS19}
\end{figure}

\subsection{Screened interaction including the atomic lattice periodicity}
\label{sectionWGG}

A more rigorous description of the screened interaction incorporating the in-plane atomic corrugation of the TMD layer may become necessary when large wave vectors are involved, such as in the current experiment (i.e., for momentum transfers $\sim{\rm\Gamma}$K$_g$ in graphene). This is the case of the one-step tunneling channel mediated by the creation of indirect excitons (see Sec.~\ref{currentindirect}), as described by Eq.~(\ref{Jindexinter}), in which the screened interaction enters through the components $W_{\Gb\Gb'}(\kparb,d,d,\omega)$ [see Eq.~(\ref{WkparGG})], with the source and probing positions both lying on the $z=d$ plane (i.e., where the TMD--graphene heterostructure is located, assuming again a zero-thickness limit in the description of its optical response).

We obtain $W_{\Gb\Gb'}(\kparb,d,d,\omega)$ directly from a Fabry--Perot (FP) description of the pathways followed by the electric potential generated by a time-dependent unit charge of magnitude $\ee^{-\ii\omega t}$ placed at $\rb'$ with $z'=d^+$ taken right above the 2D material layer. Using the notation and methods of Sec.~\ref{screening}, the direct potential in an infinite hBN host medium can be written in a form analogous to Eq.~(\ref{WkparGG}):
\begin{subequations}
\begin{align}
\phi^{\rm dir}(\rb,\omega)&=\sum_{\Gb}\int_{\rm 1BZ} \frac{d^2\kparb}{(2\pi)^2}\,\ee^{\ii(\kparb+\Gb)\cdot\Rb}\phi^{\rm dir}_\Gb(\kparb,z,\omega), \nonumber
\end{align}
where
\begin{align}
\phi^{\rm dir}_\Gb(\kparb,z,\omega)=\frac{2\pi}{|\kparb+\Gb|\bar{\epsilon}}\,\ee^{-q_\Gb|z-d|} \nonumber
\end{align}
\end{subequations}
and we define
$q_\Gb=|\kparb+\Gb|\sqrt{\epsilon_x(\omega)/\epsilon_z(\omega)}$ similar to $q$ in Sec.~\ref{screening}, but with $\kpar$ substituted by $|\kparb+\Gb|$. We now substitute the reflection and transmission coefficients $r_{\rm 2D}$ and $t_{\rm 2D}$ of the 2D material layer by reflection and transmission tensors of components $r_{{\rm 2D},\Gb\Gb'}$ and $t_{{\rm 2D},\Gb\Gb'}$, labelled by incident ($\Gb'$) and reflected/transmitted ($\Gb$) reciprocal lattice vectors (see below). In addition, the reflection coefficient at the BN--Au interface is still described by Eq.~(\ref{rBAu}), assuming isotropy and translational invariance (i.e., $\kparb$ and $\Gb$ are both conserved upon reflection at this interface), but introducing a $\Gb$ dependence by also replacing $\kpar$ by $|\kparb+\Gb|$ in such an equation, which we indicate by writing the corresponding reflection coefficient as $r_{{\rm BN,Au},\Gb}$. Using these elements, a straightforward FP analysis allows us to write
\begin{align}
&W_{\Gb\Gb'}(\kparb,d,d,\omega)=\frac{2\pi}{|\kparb+\Gb'|\bar{\epsilon}} \bigg[
\delta_{\Gb\Gb'}-r_{{\rm 2D},\Gb\Gb'} \label{WkparGGexplicit}\\
&-\!\!\sum_{\Gb_1\Gb_2}\! t_{{\rm 2D},\Gb\Gb_2} \,\ee^{-q_{\Gb_2}d} \,\Delta_{\Gb_2\Gb_1}
\,r_{{\rm BN,Au},\Gb_1} \,\ee^{-q_{\Gb_1}d} \,t_{{\rm 2D},\Gb_1\Gb'}
\bigg], \nonumber
\end{align}
where $\Delta$ is a tensor of inverse components $\big[\Delta^{-1}\big]_{\Gb\Gb'}=\delta_{\Gb\Gb'}-r_{{\rm BN,Au},\Gb}\,r_{{\rm 2D},\Gb\Gb'}\,\ee^{-(q_{\Gb}+q_{\Gb'})d}$.

The reflection and transmission tensors of the 2D layer entering Eq.~(\ref{WkparGGexplicit}) can be in turn obtained from the conductivity tensor $\sigma_{\Gb\Gb'}(\kparb,\omega)$ by applying the same boundary conditions as invoked in the derivation of Eqs.~(\ref{r2Dt2D}), but taking into account the expansion of different quantities in reciprocal-lattice-vector components. A detailed analysis leads to the result
\begin{subequations}
\begin{align}
r_{{\rm 2D},\Gb\Gb'}&=\Big[\frac{\mathcal{I}}{\mathcal{I}+2\bar{\epsilon}\;\mathcal{A}^{-1}}\Big]_{\Gb\Gb'}, \nonumber\\
t_{{\rm 2D},\Gb\Gb'}&=\delta_{\Gb\Gb'}-r_{{\rm 2D},\Gb\Gb'}, \nonumber
\end{align}
\end{subequations}
where $\mathcal{I}_{\Gb\Gb'}=\delta_{\Gb\Gb'}$ is the identity matrix and we introduce a tensor of components
\begin{align}
\mathcal{A}_{\Gb\Gb'}=\frac{4\pi\ii}{\omega}\,\frac{(\kparb+\Gb)\cdot(\kparb+\Gb')}{|\kparb+\Gb|}
\,\sigma_{\Gb\Gb'}(\kparb,\omega).  \label{AGG}
\end{align}
Obviously, the $\Gb=\Gb'=0$ elements $r_{{\rm 2D},00}$ and $t_{{\rm 2D},00}$ reproduce Eqs.~(\ref{r2Dt2D}).

Finally, we express the conductivity tensor $\sigma_{\Gb\Gb'}(\kparb,\omega)$ in terms of the non-interacting susceptibility tensor $\chi_{\Gb\Gb'}^0(\kparb,\omega)$ by writing the charge $\rho^{\rm ind}$ induced in the 2D layer in response to a total potential $\phi$ expressed in $(\rb,\omega)$ space as $\rho^{\rm ind}=(\ii\omega)^{-1}\nabla\cdot\jb^{\rm ind}=(\ii/\omega)\nabla\cdot\sigma\nabla*\phi$, where $\jb^{\rm ind}$ is the induced current, given in turn by the conductivity times the electric field $-\nabla\phi$. Here, the dot stands for the scalar product of two 2D vectors and the star indicates a functional linear operation. By definition, the non-interacting susceptibility relates the total potential and the induced charge as $\rho^{\rm ind}=\chi^0*\phi$, so we have the relation $\chi^0=(\ii/\omega)\nabla\cdot\sigma$. Projecting on $\kparb$ (within the 1BZ) and $\Gb$ components, we obtain
\begin{align}
\sigma_{\Gb\Gb'}(\kparb,\omega)=\frac{\ii\omega}{(\kparb+\Gb)\cdot(\kparb+\Gb')}\;\chi_{\Gb\Gb'}^0(\kparb,\omega).  \label{sigmaGG}
\end{align}
Incidentally, using Eq.~(\ref{sigmaGG}), we can write the matrix in Eq.~(\ref{AGG}) as $\mathcal{A}_{\Gb\Gb'}=-4\pi\,|\kparb+\Gb|^{-1}\,\chi_{\Gb\Gb'}^0(\kparb,\omega)$.

As explained in Sec.~\ref{screening}, the conductivity of the TMD--graphene 2D heterostructure is taken as the sum $\sigma(\kparb,\omega)=\sigma_{\rm TMD}(\kparb,\omega)+\sigma_{\rm g}(\kparb,\omega)$. We apply Eq.~(\ref{sigmaGG}) to obtain the conductivity of the TMD monolayer from the susceptibility calculated from first principles in Sec.~\ref{TMDsigma}. In addition, because the excitons involved in the tunneling process are supported by the TMD layer, and the 1BZ of graphene is substantially larger than that of the TMD, we treat the conductivity of graphene in the isotropic and translational-invariant approximations [i.e., $\sigma_{{\rm g}}(\kpar,\omega)$ is calculated in the RPA~\cite{WSS06}, as described in Sec.~\ref{screening}], with tensor components given by $\sigma_{{\rm g},\Gb\Gb'}(\kparb,\omega)=\delta_{\Gb\Gb'}\;\sigma_{{\rm g}}(|\kparb+\Gb|,\omega)$ for $\kparb$ within the 1BZ of the TMD and the $\Gb$ vectors running over the reciprocal lattice of this material.

An example of the obtained response function is shown in Fig.~\ref{FigS16} for a heterostructure containing MoSe$_2$. We remark that $\kparb$ refers to the optical wave vector, which is associated with the difference between electron and hole wave vectors of the contributing excitons. The response is dominated by nearly direct excitons at $\kparb\approx0$ as well as weaker features corresponding to ${\rm\Gamma}$Q$_{\rm TMD}$ excitons away from that region.

\subsection{First-principles nonlocal optical response of TMD monolayers}
\label{TMDsigma}
The exciton-dominated nonlocal optical conductivities of monolayer TMDs constitute a central ingredient in our theory. We perform first-principles calculations in the framework of many-body perturbation theory to appropriately describe the excited-state properties of the considered materials. The ground-state wave functions and one-electron eigenenergies are obtained by performing density-functional-theory (DFT) calculations with the Perdew--Burke--Ernzerhof (PBE)~\cite{PBE96} exchange--correlation functional as implemented in the Quantum Espresso code~\cite{GBB09}. In addition, the spin-orbit interaction is included using optimized norm-conserving pseudopotentials~\cite{H13_3,SG15} with a kinetic energy cutoff of 80~Ry. We employ the Coulomb-cutoff technique~\cite{RVM06} at the edges of the unit cells in the out-of-plane direction with a vacuum layer of 37 atomic units.

Using the DFT output, we perform Bethe--Salpeter-equation (BSE) calculations within the YAMBO code~\cite{MHG09,SFM19}. The G$_{0}$W$_{0}$ and plasmon-pole approximations~\cite{H1965,ORR02} are employed to correct the Kohn--Sham eigenenergies. In Fig.~\ref{FigS17}, we present the so-obtained quasiparticle electronic band dispersions for the TMDs under consideration. The direct electronic band gap at the K point is found to be 2.32, 1.95, 2.61, and 2.41~eV for MoSe$_2$, WSe$_2$, MoS$_2$, and WS$_2$, respectively. We solve the BSE~\cite{RL00,PPD04} to calculate the eigenenergies and eigenvectors of the excitons supported by these materials, from which we obtain the dielectric function. More precisely, we write the BSE as
\begin{widetext}
\begin{align}\label{eq:BSE}
    (\varepsilon_{c\qparb-\kparb} - \varepsilon_{v\qparb})A^{j}_{vc\qparb\kparb} + \hbar^{-1}\sum_{v'c'\qparb'}\langle v\qparb;c\qparb-\kparb|\mathcal{K}^{\rm eh}|v'\qparb';c'\qparb'-\kparb\rangle A^{j}_{v'c'\qparb'\kparb}=\varepsilon^{j}_{\kparb} A^{j}_{vc\qparb\kparb},
\end{align}
\end{widetext}
where a combination of the index $j$ and the in-plane wave vector $\kparb$ are used to label each exciton state, $\hbar\varepsilon^{j}_{\kparb}$ is the corresponding exciton energy, $\hbar\varepsilon_{v\qparb}$ and $\hbar\varepsilon_{c\qparb-\kparb}$ denote valence- and conduction-band quasiparticle energies obtained from G$_{0}$W$_{0}$ simulations, $|v\qparb;c\qparb-\kparb\rangle$ are pure electron--hole excitonic states, $A^{j}_{vc\qparb\kparb}$ are the exciton eigenvector coefficients, $\mathcal{K}^{\rm eh}$ is the electron--hole interaction kernel, and $\qb$ are 3D one-electron wave vectors. The generalized oscillator strengths of an excitonic transition can be defined as $\Omega^{j}_{\kparb\Gb} = \sum_{vc\qparb}A^{j}_{vc\qparb\kparb}\langle v\qparb|\ee^{-\ii(\kparb+\Gb)\cdot\Rb}|c\qparb-\kparb\rangle$, from which we calculate the non-interacting susceptibility as
\begin{align} \label{eq:Chi-eps}
   \chi^{0}_{\Gb\Gb'}(\kparb,\omega) = \frac{f_{s}e^2t}{N_{\qb}V\hbar}\sum_{j} \frac{\Omega^{j\ast}_{\kparb\Gb}\Omega^{j}_{\kparb\Gb'}}{\omega-\varepsilon^{j}_{\kparb} + i\gamma}.
\end{align}
Here, we set $f_{s}=1\;(2)$ for spin-polarized (-unpolarized) calculations, $N_{\qb}$ is the number of $\qb$ points in the 1BZ, $V$ denotes the 3D unit-cell volume, and we multiply by the out-of-plane lattice constant of the simulation unit cell $t$ to obtain a 2D surface response function.

For the two-step tunneling process, we describe direct excitonic transitions using the dipole matrix elements $\langle v\qparb|\ee^{-\ii\kparb\cdot\Rb}|c\qparb-\kparb\rangle$, calculated from the YAMBO code for small values of $\kpar$, in which the in-plane response is isotropic. To obtain an optical spectrum comparable to experimental results, we use a dense and uniform grid of electron wave vectors $\qb$ of size 72$\times$72$\times$1 in the G$_{0}$W$_{0}$ and BSE calculations. We ignore umklapp processes (i.e., we only use the $\Gb=\Gb'=0$ element of the $\chi^0_{\Gb\Gb'}$ matrix of components labeled by the 2D reciprocal lattice vectors $\Gb$ and $\Gb'$ \cite{MHG09}).   Transitions between the four highest-valence and the four lowest-conduction bands are incorporated in the BSE calculations. We employ the Scalable Library for Eigenvalue Problem Computations (SLEPC) \cite{HRV05} to obtain detailed information about each BSE eigenmode and validate our results with the full Lanczos--Haydock solution~\cite{H1980_2}. Since the matrix that we need to diagonalize is very large, we only extract the BSE eigenmodes for transitions below 3~eV. The 2D conductivities $\sigma_{\rm TMD}(\kpar,\omega)$ of the studied TMD monolayers are obtained from the 2D non-interacting susceptibility by using the relation $\sigma_{\rm TMD}(\kpar,\omega)=\ii\,\omega\chi^{0}_{00}(\kpar,\omega)/\kpar^{2}$. The calculated conductivities are presented in Fig.~\ref{FigS18} as a function of photon energy and parallel wave vector.

In the one-step tunneling process, we consider indirect excitonic transitions by calculating the matrix elements $\langle c\qparb-\kparb| \ee^{\ii(\kparb+\Gb)\cdot\textbf{r}} |v\qparb \rangle$ for $\kparb$ running over the full 1BZ. We also include the first six neighboring Brouillin zones corresponding to the six smallest non-vanishing reciprocal lattice vectors $\Gb$. To reduce the increased computational burden for such a large set of $\kparb$ and $\Gb$ vectors, we decrease the $\qparb$-point grid size to 30$\times$30$\times$1 and do not include spin-orbit coupling in this part of our calculations. In addition, since we focus on momentum-transfer-induced changes in the excitonic spectra, we solve the BSE only for the degenerate highest-valence and lowest-conduction bands. We present examples of the so-obtained conductivity of MoSe$_2$ in Fig.~\ref{FigS19}, showing features associated with indirect excitons of high momentum within the 1BZ in optical wave vector space $\kparb$.


%

\end{document}